\documentclass[journal,12pt,onecolumn]{IEEEtran}

\ifCLASSINFOpdf
\else
\fi

\usepackage{cite}
\usepackage{setspace}
\usepackage{graphicx}
\usepackage{graphics}
\usepackage{subfigure}
\usepackage{url}
\usepackage{verbatim} 
\usepackage{multicol}
\usepackage{array}
\usepackage{margins}
\usepackage{psfrag}
\usepackage{xcolor}
\usepackage{enumerate}
\usepackage{amsmath, amssymb}
\usepackage{amsfonts}

 {\everymath{\scriptscriptstyle\everymath{}}\array}%
 {\endarray}
\newcolumntype{x}[1]{>{\centering\arraybackslash\hspace{0pt}}p{#1}}

\newcommand{\cS}{{\mathcal S}}
\newcommand{\cX}{{\mathcal X}}

\newcommand{\cC}{{\mathcal C}}
\newcommand{\g}{\gamma}
\newcommand{\al}{\alpha}

\newcommand{\tl}{\tilde}
\newcommand{\bx}{\mathbf{x}}
\newcommand{\by}{\mathbf{y}}
\newcommand{\bp}{\mathbf{p}}
\newcommand{\bs}{\mathbf{s}}
\newcommand{\bq}{\mathbf{q}}
\newcommand{\bd}{\mathbf{d}}
\newcommand{\bw}{\mathbf{w}}

\newcommand{\sqstack}[2]{\Bigl[\genfrac{}{}{0pt}{1}{#2}{#1}\Bigr]}
\newcommand{\braceup}[4]{\draw[decorate, decoration={brace, amplitude=5pt},thick] ([xshift=0.5mm,yshift=#3]#1.north west)--([xshift=-0.5mm,yshift=#3]#2.north east) node[midway,anchor=south,outer sep=2mm] {#4}}
\newcommand{\bracedn}[4]{\draw[decorate, decoration={brace, amplitude=5pt},thick] ([xshift=-0.5mm,yshift=#3]#2.south east)--([xshift=0.5mm,yshift=#3]#1.south west) node[midway,anchor=north,outer sep=2mm] {#4}}
\newcommand{\dimup}[4]{\draw[<->,thick] ([xshift=0.5mm,yshift=#3]#1.north west)--([xshift=-0.5mm,yshift=#3]#2.north east) node[midway,anchor=south] {#4}}
\newcommand{\dimdn}[4]{\draw[<->,thick] ([xshift=-0.5mm,yshift=#3]#2.south east)--([xshift=0.5mm,yshift=#3]#1.south west) node[midway,anchor=north] {#4}}

\definecolor{light-gray}{gray}{0.75}
\usepackage{tikz}
\usetikzlibrary{shapes,arrows,shadows,positioning,decorations.pathreplacing,trees,calc}
\tikzstyle{sym} = [draw, thick, rectangle, font=\small, minimum width=4mm, minimum height=4mm, text centered]
\tikzstyle{esym} = [sym, fill=light-gray]
\tikzstyle{usym} = [sym, fill=white]
\tikzstyle{diagbox} = [draw, rectangle, font=\footnotesize, fill=white, text centered, rounded corners]
\tikzstyle{codebox} = [draw, rectangle, font=\footnotesize, minimum height=7mm, fill=white, text centered]

\newcommand{\Am}{\mathcal{A}}
\newcommand{\Bm}{\mathcal{B}}
\newcommand{\Bbm}{\bar{\mathcal{B}}}
\newcommand{\Cm}{\mathcal{C}}
\newcommand{\Dm}{\mathcal{D}}

\newtheorem{thm}{Theorem}
\newtheorem{lem}{Lemma}
\newtheorem{prop}{Proposition}
\newtheorem{remark}{Remark}

\begin{document}
\title{Streaming-Codes for Multicast over \\ Burst Erasure Channels}

\author{\IEEEauthorblockN{Ahmed Badr,   Devin Lui and Ashish Khisti\\}
\IEEEauthorblockA{School of Electrical and Computer Engineering\\
University of Toronto\\
Toronto, ON, M5S 3G4, Canada\\
Email: \{abadr, dlui, akhisti\}@comm.utoronto.ca}
\thanks{The corresponding author is Ashish Khisti (ashish.khisti@gmail.com). This work was supported by a discovery grant from National Science and Engineering Research Council of Canada, Hewlett Packard through a HP-IRP Award and an Ontario Early Researcher Award. Preliminary results of this work were presented at the 2010  Allerton Conference on Communication, Control, and Computing, Montecillo, IL.}}

\maketitle

\vspace{-3em}

\begin{abstract}
We study the capacity limits of real-time streaming over burst-erasure channels.
A  stream of source packets must be sequentially encoded and the resulting channel packets must be transmitted over a
two-receiver burst-erasure broadcast channel. The source packets must be sequentially reconstructed at each receiver 
with a possibly different reconstruction deadline.  We study the associated capacity as a function of burst-lengths and delays at the two receivers.

We establish that  the operation of the system can be divided into two main regimes: a {\em low-delay regime} and a {\em large-delay regime}.
We fully characterize the capacity in the large delay regime. The key to this characterization is 
an inherent {\em{slackness}} in the delay of one of the receivers. At every point in this regime we can reduce the delay of at-least one of the
users until a certain critical value and thus it suffices to obtain code constructions for certain critical delays.  We partially characterize the capacity
in the low-delay regime. Our capacity results involve  code constructions and converse techniques that appear to be novel.
We also provide a rigorous information theoretic converse theorem in the  point-to-point setting which was studied by 
Martinian in an earlier work.
\end{abstract}


\IEEEpeerreviewmaketitle

\section{Introduction}

\IEEEPARstart{D}{}elay is often ignored in the analysis of classical communication systems.
Traditional error correction codes are designed to operate on message blocks, and can incur arbitrarily
long encoding and decoding delays. In contrast several emerging applications are highly delay-sensitive.
Both the fundamental limits and error correction techniques in such systems can be very different, see e.g.,
~\cite{ftn1, MartinianDebt,Leith,Sahai,teja,neuhoff,JayKumar,effros,yeung}, and references therein.

An  information theoretic framework for the study of low-delay streaming codes has been introduced in
~\cite{Martinian_Thesis,Martinian_Sundberg,Martinian_Trott}.
The encoder observes a stream of source packets, and sequentially encodes it into a a stream of channel packets.
The decoder is required to reconstruct each source packet with a maximum delay of $T$ units.
The proposed channel is a burst erasure channel.
It can erase up to $B$ packets in a single burst, but otherwise transmits each packet instantaneously.
The maximum possible rate $C_0(B,T)$  is characterized by proposing a coding scheme and proving a converse.
We  refer to  this class of codes as streaming codes (SCo) throughout this paper.


From a practical point of view, the $(B,T)$ SCo code should be used over  a
burst-erasure channel, where the maximum length of any single burst is $B$ and the guard interval
separating multiple bursts is at-least $T$.
Extensions of SCo codes thats correct both burst and isolated erasures have been recently developed~\cite{BKTA, BKTA-2}.
Such codes were demonstrated to exhibit significant performance gains over the Gilbert-Elliott channel and Fritchman channel models, thus opening up the exciting possibility of developing structured codes for delay-constrained streaming communication in practical wireless networks.

In this paper we are interested in a different extension of the SCo constructions. Instead of committing to a particular burst length $B$
and delay $T$, our constructions adapt to the burst-length introduced by the channel. When the channel introduces an erasure-burst
of length up to $B_1$, the reconstruction-delay must be no greater than $T_1$, whereas if the burst-length is larger, say $B_2$, the reconstruction delay
can be increased to $T_2$. Such constructions can be relevant for error concealment techniques such as adaptive media playback\cite{Girod}.
Such methods adjust the play-out rate as a function of the receiver buffer size, so that  a temporary increase in delay can be naturally accommodated.
A natural way to study such constructions is to consider a multicast setup involving one sender and two receivers.
The two receivers are connected to the sender over a burst-erasure broadcast channel and both the receivers are interested in reconstructing
the same source stream, but with different delays.
One receiver's channel introduces a burst of length $B_1$ and the required reconstruction delay is $T_1$. The second receiver's channel introduces
a burst of length $B_2$ and the associated reconstruction delay is $T_2$.
We seek to characterize the multicast streaming capacity $C(B_1, T_1, B_2, T_2)$ in this paper.

In an earlier work~\cite{Our_JSAC}, we investigate the necessary and sufficient conditions under which $C(B_1, T_1, B_2, T_2)=C_0(B_1, T_1)$ (with $B_2 > B_1$).  In particular, we show that if the delay $T_2$ of the weaker user is larger than a certain threshold, then the multicast capacity reduces to the single-user capacity of the stronger receiver. A particular code construction --- {\em diversity embedded streaming erasure codes} (DE-SCo) --- is proposed to achieve this capacity. In the present paper we obtain several new results. First, we observe that system performance can be
divided into two operating regimes.  When both the delays $T_1$ and $T_2$ are smaller than certain thresholds the system operates in a low-delay
regime. Otherwise it operates in a large-delay regime. In the latter case we identify a surprising {\em{slackness}} property and use it in our code constructions. The slackness property enables us to reduce the delay of either receiver $1$ or receiver $2$  to a certain minimum threshold without reducing the capacity. In the low-delay regime the characterization of the capacity is more challenging. We characterize the capacity for a subset of this region by proposing a new coding scheme and a matching converse. For the remainder of this region we propose an upper bound on the capacity,
but leave open whether this bound is the true capacity. Preliminary results of our work appeared in~\cite{Mu-SCo-Allerton}. For related work see e.g.,~\cite{Khisti,Vyetrenko,Li,Deng2,Li2,Wornell,Soljanin4,Tracey,Tracey-2,Tracey-3} and references therein.

\begin{figure}
\centering
 \resizebox{\columnwidth}{!}{\includegraphics{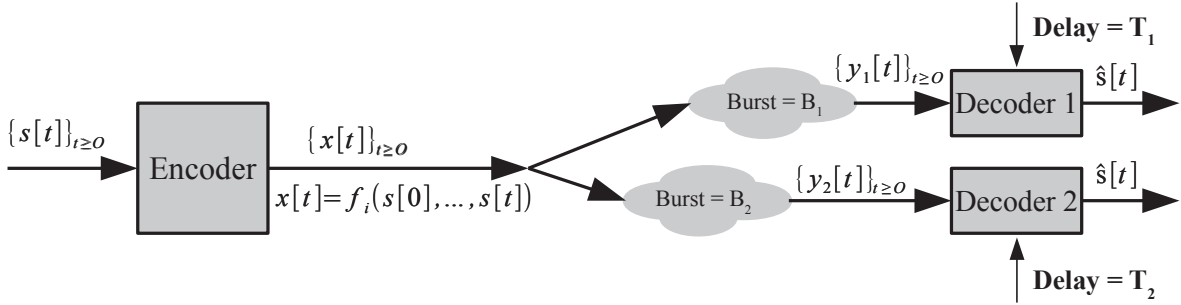}}
\caption{The source stream $\{\bs[i]\}$ is causally mapped into an output stream $\{\bx[i]\}$. Both the receivers observe these packets via their channels. The channel introduces an erasure-burst of length $B_i$, and each receiver tolerates a delay of $T_i$, for $i=1,2$.}
\label{fig:ProblemFormulationFigure}
\vspace{-1 em}
\end{figure}

\section{System Model}

\label{sec:SystemModel}
Fig.~\ref{fig:ProblemFormulationFigure} shows the proposed system model.
The transmitter encodes a stream of source packets $\{ \bs[t] \}_{t \geq 0}$ intended to be received at two receivers.
The channel packets $\{ \bx[t] \}_{t \geq 0}$ are produced causally from the source stream, i.e.,

\begin{equation}
\label{eq:Code_Function}
\bx[t] = f_t (\bs[0],\dots,\bs[t]).
\end{equation}


The channel of receiver $i$ introduces an erasure-burst of length $B_i$ i.e., the channel output at receiver $i$ at time $t$ is given by

\vspace{-1em}

\begin{equation}
\label{eq:Channel_Function_Multi}
\by_i[t] =
\left\{
\begin{array}{ll}
\star & t \in [j_i,j_i + B_i - 1] \\
\bx[t] & \text{otherwise}
\end{array}
\right.
\end{equation}
for ${i=1,2},$ and for some ${j_i \ge 0}$. Furthermore, user $i$ tolerates a delay of $T_i$, i.e., there exists a sequence of decoding functions $\gamma_{1t}(.)$ and $\gamma_{2t}(.)$ such that
\begin{equation}
\label{eq:Decoders}
\hat{\bs}[t] = \gamma_{it} (\by_i[0],\by_i[1],\dots,\by_i[t+T_i]), \qquad i=1,2,
\end{equation}
and $\text{Pr} (\bs[t] \neq \hat{\bs}[t]) = 0, \; \; \; \;	\forall t \geq 0, \; \;$.

The source stream is an i.i.d.\ process; each source symbol is sampled from a distribution $p_\bs(\cdot)$ over some finite alphabet $\cS$. The channel symbols $\bx[t]$ belong to some alphabet $\cX$. The rate of the multicast code is defined as the ratio of the (marginal) entropy of the source symbol to the alphabet size i.e., $R = H(\bs)/{\log_2|\cX|}$ and the multicast streaming capacity, $C(B_1,T_1,B_2,T_2)$ is the maximum achievable rate. An \emph{optimal multicast streaming erasure code (Mu-SCo)} achieves such capacity for a given choice of $(B_1,  T_1, B_2, T_2)$. Without loss of generality, we assume throughout the paper that $B_2 \geq B_1$.

Note that our model only considers a single erasure burst on each channel. As is the case with (single user) SCo, our constructions correct multiple erasure-bursts separated sufficiently apart. Also we only consider the erasure channel model. More general channel models can be transformed into an erasure model by applying an appropriate inner code~\cite[Chapter 7]{Martinian_Thesis}.


\section{Main Results}
\label{sec:MainResults}
To keep the paper self contained, we first briefly review the single user scenario~\cite{Martinian_Thesis, Martinian_Trott, Martinian_Sundberg}. We point the reader to these references
as well as a summary in~\cite{Our_JSAC} for a   more exhaustive treatment. 
\subsection{Single-User Capacity}
\begin{thm}[Point-to-Point Capacity:{\cite{Martinian_Thesis}}]
\label{thm:SCo_Capacity}
The capacity of a point-to-point system described by~\eqref{eq:Code_Function},~\eqref{eq:Channel_Function_Multi} and~\eqref{eq:Decoders} (with $i=1$)  is
\begin{equation}
\label{eq:SCo_Capacity}
C =
\left\{
\begin{array}{ll}
\frac{T}{T+B} & T \geq B \\
0 & T < B,
\end{array}
\right.
\end{equation}
where $T_1$ and $B_1$ are replaced by $T$ and $B$ for simplicity.
\end{thm}
The associated code construction involves a two step approach.
\begin{itemize}
\item Construct a low-delay burst-erasure block code (LD-BEBC) that takes $T$ source symbols, say $(s_0, \ldots, s_{T-1})$ and generates $B$ parity checks, say $(p_0,\ldots, p_{B-1})$. The resulting
codeword ${\bx = (s_0,\ldots, s_{T-1}, p_0,\ldots, p_{B-1})}$ has the property that it can fully recover all erased symbols from any erasure burst of length $B$.  Furthermore each of the erased source symbols $s_i$ for $i \in \{0,\ldots, T-1\}$  is recovered by time $\min\left\{i+T, T+B\right\}$.
An explicit construction of such a code  is proposed in~\cite{Martinian_Thesis, Martinian_Trott, Martinian_Sundberg}.
\item Apply diagonal-interleaving to the LD-BEBC code to construct the streaming code.
 \end{itemize}
The resulting streaming code is a time-invariant, systematic convolutional code of memory $T,$ that takes in  $T$ source symbols at any given time and outputs $T+B$
symbols\footnote{In this work will not be using any special properties of convolutional codes~\cite{Forney2} and the reader is not assumed to  have familiarity with this topic. Some properties of SCo codes from the context of convolutional codes are discussed in~\cite{Martinian_Sundberg,BKTA}.}.
The converse is based on a periodic erasure channel argument, similar to the upper bounding technique used in classical burst-noise channels~\cite[Section 6.10]{Gallager}.
The basic idea is to amplify the effect of a single isolated erasure burst into a periodic erasure channel and use the capacity of such a channel as an upper bound.
We  compliment this argument with a rigorous information theoretic proof for~\eqref{eq:SCo_Capacity} in Section~\ref{sec:SCo_Converse}.
The information theoretic proof is more general and provides a tighter upper bound when we consider the multicast setup.


\subsection{DE-SCo Construction}
In earlier work~\cite{Our_JSAC} Badr et.\ al consider the proposed multicast setup
when the delay of the weaker user i.e., user 2 is sufficiently large.
\begin{thm}[Badr et.\ al~\cite{Our_JSAC}]
\label{thm:DESCo}
The  multicast streaming capacity  $C(B_1, T_1, B_2, T_2)$  in the regime where ${B_2 > B_1}$ and  $T_2 \geq \al T_1 + B_1$ (with $\al = \frac{B_2}{B_1}$) is given by:
\begin{equation}
\label{eq:Ca}
 C_1 = \frac{T_1}{T_1+B_1}.
\end{equation}
\end{thm}

The associated code construction ---  Diversity Embedded  Streaming Codes (DE-SCo) ---
involves constructing two groups of parity checks: one along the main
diagonal and the other along the off-diagonal and then combining these parity checks in a suitable manner.
We refer the reader to~\cite{Our_JSAC} for the detailed construction. A converse argument
is also provided in~\cite{Our_JSAC} to establish that  $T_2$ is indeed the smallest
possible threshold to achieve the rate of $C_1$.

\begin{figure}
  \begin{minipage}[b]{0.47\linewidth}
    \centering
   \resizebox{\columnwidth}{!}{\includegraphics{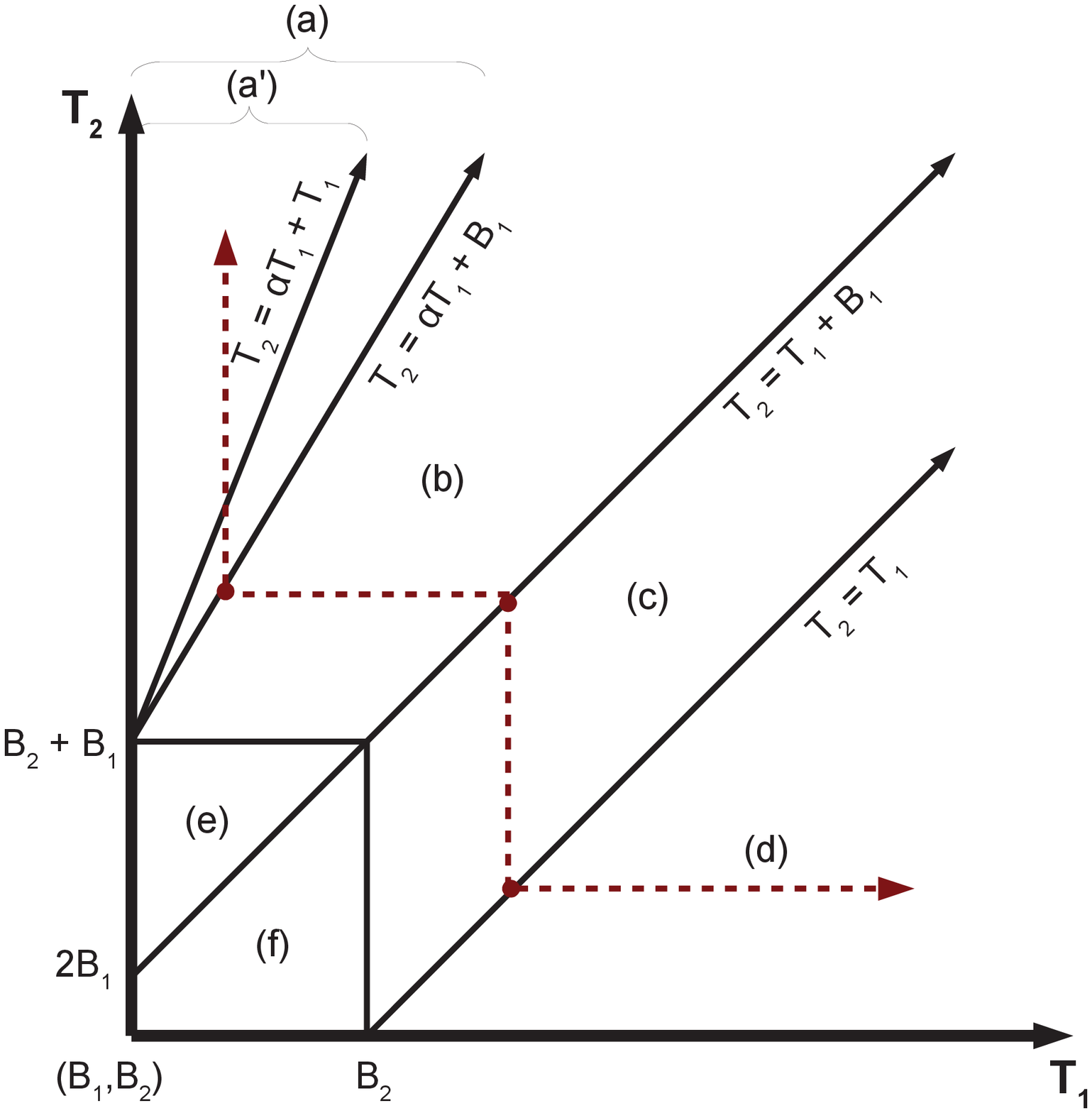}}
		\caption{Capacity behavior in the $(T_1, T_2)$ plane. We hold $B_1$ and $B_2$ as constants with ($B_2>B_1$), so the regions depend on the relation between $T_1$ and $T_2$ only.
		The red dashed line shows the contour of constant capacity in regions (a), (b), (c) and (d).}
		\label{fig:Regions}
  \end{minipage}
  \hspace{0.5cm}
  \begin{minipage}[b]{0.47\linewidth}
    \centering
    \resizebox{\columnwidth}{!}{\includegraphics{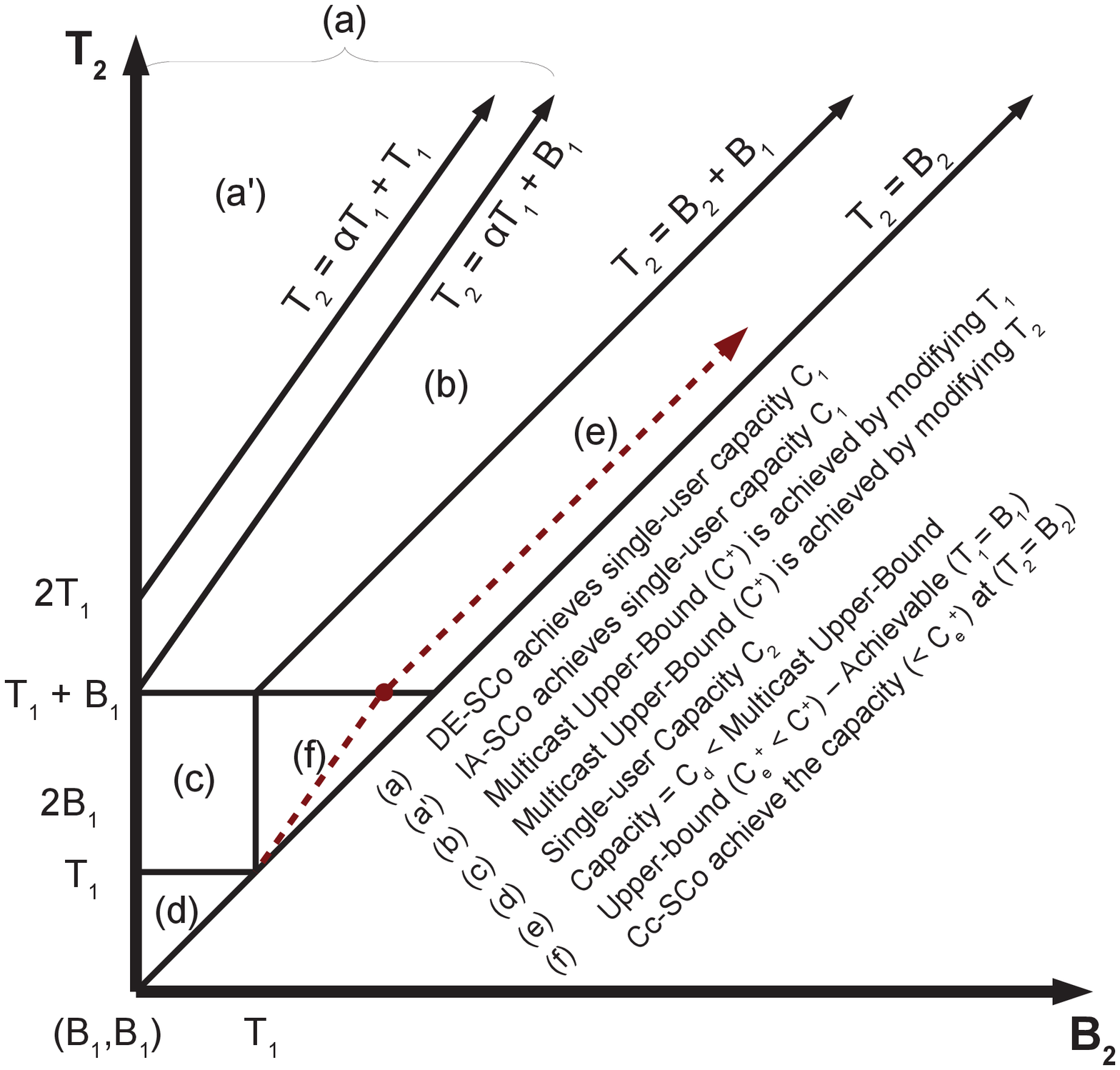}}
		\caption{Capacity behaviour in the $(B_2, T_2)$ plane. We hold $B_1$ and $T_1$ as constants, so the regions depend on the relation between $T_2$ and $B_2$ only. The dashed line gives the contour of constant capacity in region (e) as well as in the special case of $T_1=B_1$ in region (f).}
		\label{fig:Regions2}
  \end{minipage}
\end{figure}

\subsection{Large Delay Regime}
The parameters of the DE-SCo construction in Theorem~\ref{thm:DESCo}  fall within a larger class which
we refer to as the {\em large-delay regime}. In particular if at-least one of $T_1$ and $T_2$ is larger than a certain threshold:
\begin{align}
T_1 \ge B_2, \quad \mathrm(or) \quad
T_2 \ge B_1+B_2. \label{eq:delay-2}
\end{align}
we have been able to determine the multicast capacity.
In Fig.~\ref{fig:Regions} this regime consists of all pairs $(T_1, T_2)$ outside the rectangular box $[B_1, B_2] \times [B_2, B_1+B_2]$.

\begin{thm}
\label{thm:MU-SCo}
When the delays $T_1$ and $T_2$ satisfy~\eqref{eq:delay-2} and $B_2 > B_1$ the multicast capacity is given by
\begin{align}
C =	\left\{\begin{array}{ll}
									C_1, & T_2 \geq \al T_1 + B_1, \\
									\frac{T_2 - B_1}{T_2 - B_1 + B_2}, 	& T_1 + B_1 \le T_2 \le \al T_1 + B_1,\\
									\frac{T_1}{T_1 + B_2}, & T_1 \le T_2 \leq T_1 + B_1, \\
									C_2,  & T_2 \leq T_1.
									\end{array}\right.
\label{eq:Mu-SCo}\end{align}
where $C_i = \frac{T_i}{T_i+B_i}$ is the single user capacity of user $i = 1,2$ and we have defined $\al = \frac{B_2}{B_1}$.
$\hfill\Box$
\end{thm}

The proof of Theorem~\ref{thm:MU-SCo} appears in section~\ref{sec:MU-SCo}.

\begin{remark}[Delay-Slackness Property]
A closer look at~\eqref{eq:Mu-SCo} shows that in each of the four cases the capacity only depends on either $T_1$ or $T_2$,
but not on both of them simultaneously. In particular as shown in Fig.~\ref{fig:Regions}  the contour of constant capacity
is a piecewise constant line.   On the horizontal portions, the delay $T_1$ can be reduced without reducing the capacity
whereas on the vertical portions the delay $T_2$ can be reduced without reducing the capacity. This {\em slackness}
in the delay of the receivers is rather unexpected and one of the surprises in this work.
\end{remark}

We next comment of the key ingredients in the proof of  Theorem~\ref{thm:MU-SCo}. The converse  is obtained by combining  the following upper bound with single user capacity bounds.
\begin{thm}
\label{thm:Multicast_UB}
\textbf{(Upper-Bounds via Periodic Erasure Channel)}
For any two receivers with burst-delay parameters of $(B_1,T_1)$ and $(B_2,T_2)$, the multicast streaming capacity is upper-bounded by $C \leq C^+$, where
\begin{equation}
\label{eq:Multicast_UB}
C^+ =
\left\{
\begin{array}{ll}
\frac{T_2 - B_1}{T_2 - B_1 + B_2} & T_2 > T_1 + B_1, \\
\frac{T_1}{T_1 + B_2} & T_2 \leq T_1 + B_1,
\end{array}
\right.
\end{equation}
\end{thm}
The proof of Theorem~\ref{thm:Multicast_UB} is provided in section~\ref{sec:Multicast_UB}.
It involves simultaneously
using the decoding constraints of both the receivers to obtain a tighter upper bound than a simple point-to-point bound.

We next discuss the achievability part of Theorem~\ref{thm:MU-SCo}.
The first case in~\eqref{eq:Mu-SCo}, i.e., when $T_2 \ge \al T_1 + B_1$ coincides with the condition of DE-SCo codes in Theorem~\ref{thm:DESCo} and thus~\eqref{eq:Ca} applies.
The code construction associated with this region appears in~\cite{Our_JSAC}.
The construction of the remainder of the cases in~\eqref{eq:Mu-SCo} exploits the delay-slackness property.
The second case corresponds to region (b) in Fig.~\ref{fig:Regions}, where user $1$ experiences slackness in its delay. We can reduce $T_1$ so that we just hit the boundary of region (a) and then use the DE-SCo code construction. In contrast, region (c) in Fig.~\ref{fig:Regions} which corresponds to the third case  in~\eqref{eq:Mu-SCo} involves slackness in the delay of user $2$.  We can reduce $T_2$ till we just hit the boundary of region (d). For region (d) it can be easily seen that a single-user code for user $2$ is optimal. The details of the above reductions are presented in section~\ref{sec:MU-SCo}.

For a subset of  region (a), where DE-SCo codes are optimal,
 we also propose a  simpler construction --- Interference Avoidance Streaming Codes (IA-SCo) that only requires us to
construct two single user codes and combine the associated parity checks to avoid mutual interference.
\begin{prop}
\label{prop:IASCo}
\textbf{(Interference-Avoidance SCo)}
An IA-SCo construction achieves a rate of $C_1 = \frac{T_1}{T_1 + B_1}$ when  $B_2 = \al B_1$ (with $\al>1$ an integer) and
\begin{equation}
T_2 \ge \al T_1 + T_1.
\label{eq:necc}
\end{equation}
\end{prop}
The region associated with~\eqref{eq:necc} is marked  by $(a')$  in Fig.~\ref{fig:Regions}.

The proposed scheme involves starting with single user streaming codes $\cC_1: (B_1,T_1)$ and $\cC_2: (B_2,T_2)$, delaying the parity checks of $\cC_2$ by $T_1$ units and then directly combining them with the parity checks of $\cC_1$ such that they do not interfere with one another.
The  complete IA-SCo construction is provided in Section.~\ref{sec:IASCo}.

\subsection{Low Delay Regime}
We next consider the case when the delay pair $(T_1, T_2)$ falls in the box $[B_1, B_2] \times [B_2, B_1 + B_2]$ i.e.,
\begin{align}
B_1 \le T_1 \le B_2, \quad \mathrm(and) \quad
B_2 \le T_2 \le B_1+B_2. \label{eq:delay}
\end{align}

This regime appears to be more challenging and the capacity has only been established in some special cases.

\begin{thm}
\textbf{(Capacity in Region (e))}
\label{thm:Region_e}
The multicast streaming capacity in region (e) defined by $T_1+B_1 \le T_2 \le B_2+B_1$  and $B_1 \le T_1 < B_2$ is given by,

\begin{equation}
\label{eq:Ce}
C_e = \frac{T_1}{2T_1 + B_1 + B_2 - T_2}.
\end{equation}
\end{thm}

Note that the capacity expression $C_e$ only depends on $B_2$ and $T_2$ via the difference $B_2-T_2$. To identify the contour of constant capacity in the (e) region it is natural to fix $B_1$ and $T_1$
and classify the various regions as shown in Fig.~\ref{fig:Regions2}. Observe that the streaming capacity for any point in region (e) is constant across the 45-degrees line and is equal to the multicast upper-bound at the lowest point on the line separating regions (e) and (f) in Fig.~\ref{fig:Regions2}.

The complete proof for Theorem~\ref{thm:Region_e} is divided into two main parts. The achievability scheme is provided in Section~\ref{sec:Region_e_CC} while the converse is given in Section~\ref{sec:Region_e_UB}. The achievability involves first constructing a single user $(B_1, T_1)$ SCo code for the first user and then carefully embedding additional parity checks to satisfy the decoding constraint of user $2$.  The converse too involves a new insight of revealing some of the source symbols to a virtual decoder to obtain a tighter bound than a periodic erasure channel argument.

The remainder of the low delay regime is called region (f). The capacity remains open except
in the special cases of either $T_1=B_1$ or $T_2=B_2$.


\begin{thm}
\textbf{(Upper-bound in Region (f))}
\label{thm:Region_f1}
An upper-bound on the multicast streaming capacity in region (f) defined by $T_2 <T_1 + B_1$ and $T_1 \in [B_1, B_2]$ is given by,
\begin{equation}
\label{eq:C_f}
C_f \leq C_f^+ = \frac{T_2 - B_1}{2(T_2 - B_1) + (B_2 - T_1)}.
\end{equation}
The above expression equals the streaming capacity if  $T_1 = B_1$.
\end{thm}

The proof of the upper bound is given in Section~\ref{sec:Region_f_UB1}. The code construction for $T_1 = B_1$ case appears in Section.~\ref{sec:Region_f_CC}.

The capacity has also been obtained when $T_2 = B_2$ for any $T_1 \in [B_1, B_2]$.

\begin{thm}
\label{thm:Region_f_T2B2}
\textbf{(Capacity in Region (f) at $(T_2 = B_2)$)}
The multicast streaming capacity in region (f) defined by $T_2 < T_1 + B_1$ and $ T_1 \in [B_1, B_2] $ at the minimum delay case for user 2 $(T_2 = B_2)$ is given by,
\begin{equation}
\label{C_e_T2B2}
C_{f(T_2 = B_2)} = \frac{T_1}{2T_1 + B_1}.
\end{equation}
\end{thm}

The achievability scheme is based on  concatenation of the parity checks of suitably constructed single-user codes and appears in Section~\ref{sec:Region_f_CC2}. The proof of the converse part for Theorem~\ref{thm:Region_f_T2B2} is provided in Section~\ref{sec:Region_f_UB}. The technique is significantly different than earlier converses and involves carefully double-counting the redundancy arising from the recovery of certain source symbols.

This concludes the main results of the paper.

\section{Converse Proof of Theorem~\ref{thm:SCo_Capacity}}
\label{sec:SCo_Converse}
In this section we provide an information theoretic converse to Theorem~\ref{thm:SCo_Capacity}.
While the capacity of the point-to-point case was established in~\cite{Martinian_Thesis, Martinian_Trott, Martinian_Sundberg},
the converse argument was based on a somewhat informal use a periodic erasure channel (PEC). Our information theoretic approach is not
only more rigorous but also generalizes to the multicast setting in subsequent sections. 
Furthermore it has the following advantages over the PEC approach which might also be of interest.
\begin{enumerate}
\item The PEC approach
requires the channel packet $\bx[t]$ is a deterministic function of the past source packets i.e., $\bx[t]$ must be exactly computed given $\bs[0],\ldots, \bs[t]$.
The information theoretic converse does not impose this restriction and allows for e.g., stochastic encoders.
\item The PEC approach as presented in~\cite{Martinian_Thesis} requires the code to be systematic. The information theoretic approach does not impose this restriction.
\item The PEC approach requires zero error in the recovery of each source symbol. The information theoretic approach can remove this restriction
by suitably invoking Fano's inequality.
\end{enumerate}
Let us use the following notation:
\begin{align}
	\bs\sqstack{a}{b} = \left\{
	\begin{array}{ll}
		\bs[a], \bs[a+1],\ldots,\bs[b-1],\bs[b], & a \leq b \\
		\emptyset, & \mathrm{otherwise}
	\end{array}
	\right.
\end{align}
\begin{align}
	W_a^b = \left\{
	\begin{array}{ll}
		W_a, W_{a+1},\ldots,W_{b-1},W_{b}, & a \leq b \\
		\emptyset, & \mathrm{otherwise}
	\end{array}
	\right.
\end{align}

To aid us in our proof, let us introduce the terms
\begin{align}
	V_i = \bs\sqstack{i(T+B)}{(i+1)(T+B)-1}, \quad \quad W_i &= \bx\sqstack{i(T+B)+B}{(i+1)(T+B)-1}
\end{align}
where $i = 0,1,2,\ldots$. Note that $V_i$ refers to a group of source packets, whereas $W_i$ is a group of channel packets. Fig.~\ref{fig:PEC_SCo} shows the time slots that the packets come from while Fig.~\ref{fig:OnePeriod_SCo} shows the size of $V_i$ and $W_i$.

\begin{figure}[htbp]
	\centering
	\resizebox{\columnwidth}{!}{
	\begin{tikzpicture}[node distance=0mm]
		\node                       (x1start) {Link:};
		\node[esym, right = of x1start] (x100) {};
		\node[esym, right = of x100]     (x101) {};
		\node[esym, right = of x101]     (x102) {};
		\node[esym, right = of x102]     (x103) {};
		\node[usym, right = of x103]     (x104) {};
		\node[usym, right = of x104]     (x105) {};
		\node[usym, right = of x105]     (x106) {};
		\node[usym, right = of x106]     (x107) {};
		\node[usym, right = of x107]     (x108) {};
		\node[esym, right = of x108]     (x109) {};
		\node[esym, right = of x109]     (x110) {};
		\node[esym, right = of x110]     (x111) {};
		\node[esym, right = of x111]     (x112) {};
		\node[usym, right = of x112]     (x113) {};
		\node[usym, right = of x113]     (x114) {};
		\node[usym, right = of x114]     (x115) {};
		\node[usym, right = of x115]     (x116) {};
		\node[usym, right = of x116]     (x117) {};
		\node[esym, right = of x117]     (x118) {};
		\node[esym, right = of x118]     (x119) {};
		\node[esym, right = of x119]     (x120) {};
		\node[esym, right = of x120]     (x121) {};
		\node[usym, right = of x121]     (x122) {};
		\node[usym, right = of x122]     (x123) {};
		\node[usym, right = of x123]     (x124) {};
		\node[usym, right = of x124]     (x125) {};
		\node[usym, right = of x125]     (x126) {};
		\node[esym, right = of x126]     (x127) {};
		\node[esym, right = of x127]     (x128) {};
		\node[esym, right = of x128]     (x129) {};
		\node[esym, right = of x129]     (x130) {};
		\node[usym, right = of x130]     (x131) {};
		\node[usym, right = of x131]     (x132) {};
		\node[usym, right = of x132]     (x133) {};
		\node[usym, right = of x133]     (x134) {};
		\node[usym, right = of x134]     (x135) {};
		\node      [right = of x135]   (x1end) {$\cdots$};

		\braceup{x100}{x108}{1mm}{$V_0$};
		\braceup{x109}{x117}{1mm}{$V_1$};
		\braceup{x118}{x126}{1mm}{$V_2$};
		\braceup{x127}{x135}{1mm}{$V_3$};

		\bracedn{x104}{x108}{-1mm}{$W_0$};
		\bracedn{x113}{x117}{-1mm}{$W_1$};
		\bracedn{x122}{x126}{-1mm}{$W_2$};
		\bracedn{x131}{x135}{-1mm}{$W_3$};
	\end{tikzpicture}}
	\caption{The periodic erasure channel used in proving the upper-bound of the single user scenario in Theorem~\ref{thm:SCo_Capacity}, but with indication of which packets are in the groups $V_i$ and $W_i$. Grey and white squares resemble erased and unerased symbols respectively.}
	\label{fig:PEC_SCo}
\end{figure}
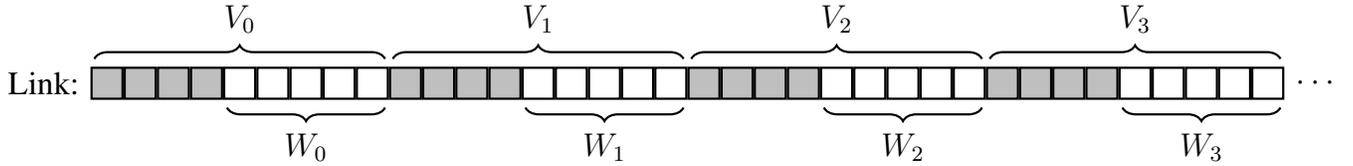

\begin{figure}[htbp]
	\centering
	\begin{tikzpicture}[node distance=0mm]
		\node                       (x1start) {Link:};
		\node[esym, right = of x1start]  (x100) {};
		\node[esym, right = of x100]     (x101) {};
		\node[esym, right = of x101]     (x102) {};
		\node[esym, right = of x102]     (x103) {};
		\node[usym, right = of x103]     (x104) {};
		\node[usym, right = of x104]     (x105) {};
		\node[usym, right = of x105]     (x106) {};
		\node[usym, right = of x106]     (x107) {};
		\node[usym, right = of x107]     (x108) {};
		\node      [right = of x108]     (x1end) {$\cdots$};

		\dimup{x100}{x103}{3mm}{$B$};
		\dimdn{x100}{x108}{-3mm}{$B+T$};
	\end{tikzpicture}
	\caption{One period of the periodic erasure channel in Fig.~\ref{fig:PEC_SCo}, with labels.}
	\label{fig:OnePeriod_SCo}
\end{figure}
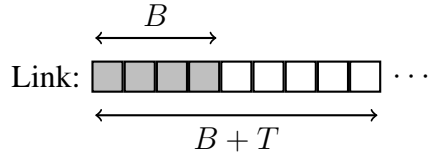

We start with the following equations, which are a result of the $(B,T)$ code. If the first $B$ channel packets are erased, and then the next $T$ channel packets are received perfectly, the $(B,T)$ code can be used to recover the source packets $\bs[0],\ldots,\bs[B-1]$. Using the conditional entropy notation, this can be written as:
\begin{equation}
	H\Bigl(\bs\sqstack{0}{B-1} \Big| W_0\Bigr) = 0. \label{eq:slsr_first_relation}
\end{equation}
Although the next $T$ channel packets $\bx\sqstack{B}{B+T-1}$ are received, we cannot assume that the corresponding source packets $\bs\sqstack{B}{B+T-1}$ are able to be decoded because the code may not be systematic. To recover those source packets, we can use the next group of $T$ unerased packets in $\bx\sqstack{(B+T)+B}{2(B+T)-1}$. In general, we may not need all of these channel packets, but the proof is simpler if we have it all available. We can then write the relation using Fano's Inequality\footnote{The conditional entropy is zero because the receiver needs to perfectly recover each of the source packets with zero error. While we do not allow vanishingly small error probabilities at the decoder, the setup can be easily generalized in this case.}
\begin{equation}
	H\Bigl(\bs\sqstack{B}{B+T-1} \Big| \bx\sqstack{0}{B-1} W_0^1\Bigr) = 0. \label{eq:slsr_second_relation}
\end{equation}
The equations \eqref{eq:slsr_first_relation} and \eqref{eq:slsr_second_relation} can be generalized to
\begin{align}
	H\Bigl(\bs\sqstack{i(B+T)}{i(B+T)+B-1} \Big| \bx\sqstack{0}{i(B+T)-1} W_i) &= 0 \label{eq:slsr_gen_first_relation} \\
	H\Bigl(\bs\sqstack{i(B+T)+B}{(i+1)(B+T)-1} \Big| \bx\sqstack{0}{i(B+T)+B-1} W_i^{i+1}) &= 0 \label{eq:slsr_gen_second_relation}
\end{align}
Note that the above expressions still only assume that there was one burst erasure of length $B$. For instance, in \eqref{eq:slsr_gen_first_relation}, we assume that the packets $\bx\sqstack{i(B+T)}{i(B+T)+B-1}$ were erased so they are not used in the expression.

Next, we will prove the following relation for $n \geq 0$:
\begin{equation}
	H(W_0^n) \geq H(V_0^{n-1}) + H\Bigl(W_n \Big| V_0^{n-1} \bx\sqstack{0}{n(B+T)-1}\Bigr). \label{eq:slsr_target_form}
\end{equation}
For the base case, substitute $n=0$ into \eqref{eq:slsr_target_form}. This gives 
\begin{align}
	H(W_0) \geq H(V_0^{-1}) + H\Bigl(W_0 \Big| V_0^{-1} \bx\sqstack{0}{-1}) = H(W_0)
\end{align}
which is clearly true. We assume that \eqref{eq:slsr_target_form} is true for $n=k$ in the induction step,
\begin{equation}
\label{eq:SCo_nk}
	H(W_0^k) \geq H(V_0^{k-1}) + H\Bigl(W_k \Big| V_0^{k-1} \bx\sqstack{0}{k(B+T)-1}\Bigr).
\end{equation}

With the availability of $W_k$, one can use~\eqref{eq:slsr_gen_first_relation} to recover $\bs\sqstack{k(B+T)}{k(B+T)+B-1}$ and~\eqref{eq:SCo_nk} can be re-written as follows,
\begin{align}
\label{eq:SCo_a}
H(W_0^k) \geq H\Bigl(V_0^{k-1} \bs\sqstack{k(B+T)}{k(B+T)+B-1} \Bigr) + H\Bigl(W_k \Big| V_0^{k-1} \bs\sqstack{k(B+T)}{k(B+T)+B-1} \bx\sqstack{0}{k(B+T)+B-1}\Bigr).
\end{align}
The detailed steps from~\eqref{eq:SCo_nk} to~\eqref{eq:SCo_a} is shown in Appendix.~\ref{app:SCo}.

Next, we add $H(W_{k+1} | W_0^k)$ to both sides and then use~\eqref{eq:slsr_gen_second_relation} to recover the source symbols corresponding to $W_k$, $\bs\sqstack{k(B+T)+B}{(k+1)(B+T)-1}$ and the following can be written (c.f. Appendix.~\ref{app:SCo}),
\begin{align}
\label{eq:SCo_b}
	H&(W_0^{k+1}) \geq H(V_0^{k}) + H\Bigl(W_{k+1} \Big| V_0^{k} \bx\sqstack{0}{(k+1)(B+T)-1} \Bigr),
\end{align}

From~\eqref{eq:SCo_nk} to~\eqref{eq:SCo_b} and passing by~\eqref{eq:SCo_a}, we have shown that if \eqref{eq:slsr_target_form} is true for $n=k\geq0$, then it is also true for $n=k+1$. Thus, by induction~\eqref{eq:slsr_target_form} is true for $n \geq0$. We take \eqref{eq:slsr_target_form} and finalize it as
\begin{align}
	H(W_0^n) \geq H(V_0^{n-1}) + H\Bigl(W_n \Big| V_0^{n-1} \bx\sqstack{0}{n(B+T)-1}\Bigr) \geq H(V_0^{n-1}). \label{eq:slsr_final_form}
\end{align}

Next, we expand the groups of channel packets
\begin{align}
	H(W_0^n) &= H\Bigl(\bx\sqstack{B}{T+B-1} \bx\sqstack{(T+B)+B}{2(T+B)-1} \cdots \bx\sqstack{n(T+B)+B}{(n+1)(T+B)-1}\Bigr) \nonumber \\
	&\leq \sum_{i=0}^n{\sum_{j=B}^{T+B-1}{H(\bx[i(T+B)+j])}} = (n+1)\cdot T \cdot H(\bx)
\end{align}
and also expand the groups of source packets
\begin{align}
	H(V_0^{n-1}) &= H\Bigl(\bs\sqstack{0}{n(T+B)-1}\Bigr) = \sum_{i=0}^{n(T+B)-1}{H\Bigl(\bs[i] \Big| \bs\sqstack{0}{i-1}\Bigr)} \nonumber \\
	&\overset{(a)}{=} \sum_{i=0}^{n(T+B)-1}{H(\bs[i])} = n \cdot (T+B) \cdot H(\bs)
\end{align}
where step (a) is because the source packets are independent. Then we can take \eqref{eq:slsr_final_form} and write it as
\begin{align}
	H(W_0^n) &\geq H(V_0^{n-1}) \nonumber \\
	(n+1)\cdot T \cdot H(\bx) &\geq n \cdot (T+B) \cdot H(\bs) \nonumber \\
	\frac{(n+1)}{n} \cdot \frac{T}{T+B} &\geq \frac{H(\bs)}{H(\bx)}. \nonumber
\end{align}
Finally, we conclude that any $(B,T)$ streaming erasure code must satisfy
\begin{align}
	R = \frac{H(\bs)}{H(\bx)} &\leq \frac{T}{T+B} \; (\mathrm{as} \; n \rightarrow \infty)
\end{align}
which gives our upper bound of the rate.



\section{Proof of Theorem~\ref{thm:MU-SCo}}
\label{sec:MU-SCo}

For  the converse we start with the upper bound in~\eqref{eq:Multicast_UB} in Theorem~\ref{thm:Multicast_UB}
which we reproduce below for convenience. The proof of Theorem~\ref{thm:Multicast_UB} appears section~\ref{sec:Multicast_UB}.
\begin{equation}
\label{eq:Multicast_UB2}
C^+ =
\left\{
\begin{array}{ll}
\frac{T_2 - B_1}{T_2 - B_1 + B_2} & T_2 > T_1 + B_1, \\
\frac{T_1}{T_1 + B_2} & T_2 \leq T_1 + B_1.
\end{array}
\right.
\end{equation}

We further tighten the upper-bound in \eqref{eq:Multicast_UB2} as follows,

\begin{align}
\label{eq:CUB}
C^\mathrm{U} 	&=	\min\left\{C^+, C_1, C_2 \right\} =	\left\{
									\begin{array}{ll}
									\min\left\{\frac{T_2 - B_1}{T_2 - B_1 + B_2}, C_1, C_2 \right\}, & T_2 > T_1 + B_1 \\
									\min\left\{\frac{T_1}{T_1 + B_2}, C_1, C_2 \right\}, & T_2 \leq T_1 + B_1.
									\end{array}
									\right. \end{align}
									
Through straightforward calculations one can further simplify:
									\begin{align}
C^\mathrm{U}	
							&=	\left\{
									\begin{array}{ll}
									\min\left\{\frac{T_2 - B_1}{T_2 - B_1 + B_2}, C_1 \right\}, & T_2 > T_1 + B_1 \\
									\min\left\{\frac{T_1}{T_1 + B_2}, C_2 \right\}, & T_2 \leq T_1 + B_1
									\end{array}
									\right. \notag\\
 							&=	\left\{
									\begin{array}{ll}
									C_1 \triangleq C_a, 																& T_2 \geq \al T_1 + B_1 \\
									\frac{T_2 - B_1}{T_2 - B_1 + B_2} \triangleq C_b, 	& T_1 + B_1 < T_2 < \al T_1 + B_1\\
									\frac{T_1}{T_1 + B_2} \triangleq C_c, 							& T_1 < T_2 \leq T_1 + B_1 \\
									C_2 \triangleq C_d, 																& T_2 \leq T_1.
									\end{array}
									\right. \label{eq:CUB3}
									\end{align}

 where recall that $\al=\frac{B_2}{B_1}$.  This completes the proof of the converse.

We discuss the code constructions for each of these regions below.

\subsection{Region (a)}

The code-construction achieving $C_1$  in region (a) appeared in~\cite{Our_JSAC}. We summarize the key-steps for completeness and provide an example with $\{(B_1,T_1)-(B_2,T_2)\} = \{(2,3)-(4,8)\}$ in Table~\ref{table:Code2348} which we will require in a subsequent example. We will assume for simplicity that $B_2 = \al B_1$ where $\al$ is an integer.

\begingroup
\singlespacing
\everymath{\scriptstyle}
\begin{table}[t]
\begin{tabular}{x{65pt}|x{65pt}|x{68pt}|x{65pt}|x{65pt}|x{65pt}}
$\displaystyle [i-1]$ & $\displaystyle [i]$ & $\displaystyle [i+1]$ & $\displaystyle [i+2]$ & $\displaystyle [i+3]$ & $\displaystyle [i+4]$ \\
\hline
\fbox{$s_0[i-1]$} & $s_0[i]$ & \fboxrule=1pt \fbox{$s_0[i+1]$} & $s_0[i+2]$ & $s_0[i+3]$ & $s_0[i+4]$ \\
$s_1[i-1]$ & \fboxrule=1pt \fbox{\fboxrule=0.25pt \fbox{$s_1[i]$}} & $s_1[i+1]$ & $s_1[i+2]$ & $s_1[i+3]$ & $s_1[i+4]$ \\
\fboxrule=1pt \fbox{$s_2[i-1]$} & $s_2[i]$ & \fbox{$s_2[i+1]$} & $s_2[i+2]$ & $s_2[i+3]$ & $s_2[i+4]$ \\
\hline
$s_0[i-4] \oplus s_2[i-2]$ & $s_0[i-3] \oplus s_2[i-1]$ & $s_0[i-2] \oplus s_2[i]$ & \fbox{$s_0[i-1] \oplus s_2[i+1]$} & $s_0[i] \oplus s_2[i+2]$ & $s_0[i+1] \oplus s_2[i+3]$ \\
$\oplus$ & $\oplus$ & $\oplus$ & $\oplus$ & $\oplus$ & $\oplus$ \\
$s_2[i-9] \oplus s_0[i-7]$ & $s_2[i-8] \oplus s_0[i-6]$ & $s_2[i-7] \oplus s_0[i-5]$ & $s_2[i-6] \oplus s_0[i-4]$ & $s_2[i-5] \oplus s_0[i-3]$ & $s_2[i-4] \oplus s_0[i-2]$ \\
\hline
$s_1[i-4] \oplus s_2[i-3]$ & $s_1[i-3] \oplus s_2[i-2]$ & $s_1[i-2] \oplus s_2[i-1]$ & $s_1[i-1] \oplus s_2[i]$ & \fbox{$s_1[i] \oplus s_2[i+1]$} & $s_1[i+1] \oplus s_2[i+2]$ \\
$\oplus$ & $\oplus$ & $\oplus$ & $\oplus$ & $\oplus$ & $\oplus$ \\
$s_1[i-9] \oplus s_0[i-8]$ & $s_1[i-8] \oplus s_0[i-7]$ & $s_1[i-7] \oplus s_0[i-6]$ & $s_1[i-6] \oplus s_0[i-5]$ & $s_1[i-5] \oplus s_0[i-4]$ & $s_1[i-4] \oplus s_0[i-3]$ \\
\hline
\hline
$\displaystyle [i+5]$ & $\displaystyle [i+6]$ & $\displaystyle [i+7]$ & $\displaystyle [i+8]$ & $\displaystyle [i+9]$ & $\displaystyle [i+10]$\\
\hline
$s_0[i+5]$ & $s_0[i+6]$ & $s_0[i+7]$ & $s_0[i+8]$ & $s_0[i+9]$ & $s_0[i+10]$ \\
$s_1[i+5]$ & $s_1[i+6]$ & $s_1[i+7]$ & $s_1[i+8]$ & $s_1[i+9]$ & $s_1[i+10]$ \\
$s_2[i+5]$ & $s_2[i+6]$ & $s_2[i+7]$ & $s_2[i+8]$ & $s_1[i+9]$ & $s_1[i+10]$ \\
\hline
$s_0[i+2] \oplus s_2[i+4]$ & $s_0[i+3] \oplus s_2[i+5]$ & $s_0[i+4] \oplus s_2[i+6]$ & $s_0[i+5] \oplus s_2[i+7]$ & $s_0[i+6] \oplus s_2[i+8]$ & $s_0[i+7] \oplus s_2[i+9]$ \\
$\oplus$ & $\oplus$ & $\oplus$ & $\oplus$ & $\oplus$ & $\oplus$ \\
$s_2[i-3] \oplus s_0[i-1]$ & $s_2[i-2] \oplus s_0[i]$ & \fboxrule=1pt \fbox{$s_2[i-1] \oplus s_0[i+1]$} & $s_2[i] \oplus s_0[i+2]$ & $s_2[i+1] \oplus s_0[i+3]$ & $s_2[i+2] \oplus s_0[i+4]$ \\
\hline
$s_1[i+2] \oplus s_2[i+3]$ & $s_1[i+3] \oplus s_2[i+4]$ & $s_1[i+4] \oplus s_2[i+5]$ & $s_1[i+5] \oplus s_2[i+6]$ & $s_1[i+6] \oplus s_2[i+7]$ & $s_1[i+7] \oplus s_2[i+8]$ \\
$\oplus$ & $\oplus$ & $\oplus$ & $\oplus$ & $\oplus$ & $\oplus$ \\
$s_1[i-3] \oplus s_0[i-2]$ & $s_1[i-2] \oplus s_0[i-1]$ & $s_1[i-1] \oplus s_0[i]$ & \fboxrule=1pt \fbox{$s_1[i] \oplus s_0[i+1]$} & $s_1[i+1] \oplus s_0[i+2]$ & $s_1[i+2] \oplus s_0[i+3]$ \\
\hline
\hline
\end{tabular}
\everymath{\displaystyle}
\caption{Rate $3/5$ DE-SCo construction that satisfy the region (a) point described by user 1 with $(B_1,T_1) = (2,3)$ and user $2$ with $(B_2,T_2) = (2B_1,2T_1+B_1) = (4,8)$.}
\label{table:Code2348}
\end{table}
\endgroup

\begin{itemize}
\item Generate a $(B_1, T_1)$ SCo code $(\bs[i], \bp[i])$ consisting of $T_1$ source sub-symbols and $B_1$
parity check sub-symbols.  Recall that the parity check sub-symbols are generated by applying a low-delay burst-erasure block codes (LD-BEBC)  across the main diagonal of the stream of source sub-symbols.
\item Generate a $(\al B_1, \al T_1)$ SCo code $(\bs[i], \bq[i])$ consisting of ${T_1 + B_1}$ sub-symbols where the parity check symbols $\bq[i]$
are generated by applying a LD-BEBC across the opposite diagonal of the stream of source sub-symbols and with a interleaving factor of $(\al-1)$.
\item The transmitted packet at time $i$ is given by $\bx[i] = (\bs[i], \bp[i] + \bq[i-T_1])$.
\end{itemize}

We omit the steps in decoding as they are rather involved and refer to~\cite{Our_JSAC}.


\subsection{Region (b)}
In region (b) in Fig.~\ref{fig:Regions} we show that the rate
$$C_b = \frac{T_2 - B_1}{T_2 - B_1 + B_2}, \quad \quad \quad T_1 + B_1 \le T_2 \le \al T_1 + B_1,$$
is achievable.

Since the capacity does not depend on $T_1$, we can reduce the value of $T_1$ to $\tl{T}_1$ such that
we meet the left hand side with equality i.e., we select
$$T_2 = \alpha \tl{T}_1 + B_1,$$
which in turn implies that
\begin{equation}
\label{eq:newT1}
\tilde{T}_1 = \frac{B_1}{B_2}(T_2 - B_1).
\end{equation}

Provided that $\tl{T}_1 \ge B_1$  and furthermore $\tl{T}_1$ is an integer we can use a
$\{ (B_1,\tilde{T}_1)-(B_2,T_2) \}$ DE-SCo code~\cite{Our_JSAC} to achieve $\frac{\tl{T_1}}{\tl{T_1}+B_1} = C_b$  and hence for the original point in region (b).  The former condition is equivalent to $T_2 \ge B_2 + B_1$  which naturally holds in region (b).
If $\tl{T}_1$  it is not an integer a suitable expansion of every source symbol  is needed as discussed below.

\begin{itemize}
\item Split each source symbol into $n^2 \tilde{T}_1$ sub-symbols $s_0[i],\dots,s_{n^2 \tilde{T}_1 -1}[i]$ where $n$ is the smallest integer such that $n \tilde{T}_1$ is an integer.
\item Construct an expanded source sequence $\tilde{\bs}[\cdot]$ such that $\tilde{\bs}[ni+r] = (s_{r n \tilde{T}_1}[i],\dots,s_{(r+1) n \tilde{T}_1 - 1}[i])$ where $r \in \{ 0,\dots,n-1\}$.
\item We apply a DESCo code with parameters $\{ (n B,n \tilde{T}_1)-(n \alpha B, n (\alpha \tilde{T}_1 + B)) \}$ to $\tilde{\bs}[\cdot]$ using the earlier construction.
\end{itemize}

With the channel of user 2 introducing $B_2$ erasures on the original input stream, there will be $n B_2$ erasures on the expanded stream. These will be decoded with a delay of $n (\alpha \tilde{T}_1 + B) = n T_2$ on the expanded stream, which can be easily verified to incur a delay of $T_2$ on the original stream.

For user 1 the expanded source stream incurs a delay of  $n \tilde{T}_1$, which reduces to a delay of $\lceil \tilde{T}_1 \rceil$ on the original stream. This suffices the requirements of user 1 as by construction $T_1 \geq \lceil \tilde{T}_1 \rceil$.

We provide a numerical example below.

\subsubsection{Example --- Source Expansion}
Consider a Mu-SCo with parameters $\{(1,2),(2,4)\}$ which falls in the (b) region. The capacity is given by $R = 3/5$.  The construction is provided in Table~\ref{table:Code1224_Optimal}. Through direct calculation note that $\tilde{T}_1 = 1.5$. Hence we implement a source expansion technique with $n=2$ as follows.

We split each source symbol $\bs[i]$ into six sub-symbols $s_0[i],\ldots, s_5[i]$ and construct an expanded source sequence $\tl{\bs}[\cdot]$ such that $\tl{\bs}[2i] = (s_0[i],s_1[i],s_2[i])$ and $\tl{\bs}[2i+1]=(s_3[i],s_4[i],s_5[i])$. We use the $\{(2,3),(4,8)\}$ DE-SCo code (see Table~\ref{table:Code2348}) that we apply to $\tl{\bs}[\cdot]$ to produce the parity checks $\tl{\bp}[\cdot]$ and transmit $\bp[i] = (\tl{\bp}[2i],\tl{\bp}[2i+1])$ along with $\bs[i]$ at time $i$. It can be verified directly that the resulting code corrects a single erasure with a delay of $2$ symbols and an erasure-burst of length $2$ with a delay of $4$.

\begingroup
\singlespacing
\everymath{\scriptstyle}
\begin{table}[t]
\begin{tabular}{x{65pt}|x{65pt}||x{65pt}|x{65pt}||x{65pt}|x{65pt}}
\multicolumn{2}{c||}{$\displaystyle [i-1]$} & \multicolumn{2}{c||}{$\displaystyle [i]$} & \multicolumn{2}{c}{$\displaystyle [i+1]$} \\
\hline
\fbox{$s_0[i-1]$} & $s_3[i-1]$ & \fboxrule=1pt \fbox{$s_0[i]$} & $s_3[i]$ & $s_0[i+1]$ & $s_3[i+1]$ \\
$s_1[i-1]$ & \fboxrule=1pt \fbox{\fboxrule=0.5pt \fbox{$s_4[i-1]$}} & $s_1[i]$ & $s_4[i]$ & $s_1[i+1]$ & $s_4[i+1]$ \\
\fboxrule=1pt \fbox{$s_2[i-1]$} & $s_5[i-1]$ & \fbox{$s_2[i]$} & $s_5[i]$ & $s_2[i+1]$ & $s_5[i+1]$ \\
\hline
$s_3[i-3] \oplus s_5[i-2]$ & $s_0[i-2] \oplus s_2[i-1]$ & $s_3[i-2] \oplus s_5[i-1]$ & \fbox{$s_0[i-1] \oplus s_2[i]$} & $s_3[i-1] \oplus s_5[i]$ & $s_0[i] \oplus s_2[i+1]$ \\
$\oplus$ & $\oplus$ & $\oplus$ & $\oplus$ & $\oplus$ & $\oplus$ \\
$s_2[i-5] \oplus s_0[i-4]$ & $s_5[i-5] \oplus s_3[i-4]$ & $s_2[i-4] \oplus s_0[i-3]$ & $s_5[i-4] \oplus s_3[i-3]$ & $s_2[i-3] \oplus s_0[i-2]$ & $s_5[i-3] \oplus s_3[i-2]$ \\
\hline
$s_4[i-3] \oplus s_2[i-2]$ & $s_1[i-2] \oplus s_5[i-2]$ & $s_4[i-2] \oplus s_2[i-1]$ & $s_1[i-1] \oplus s_5[i-1]$ & \fbox{$s_4[i-1] \oplus s_2[i]$} & $s_1[i] \oplus s_5[i]$ \\
$\oplus$ & $\oplus$ & $\oplus$ & $\oplus$ & $\oplus$ & $\oplus$ \\
$s_1[i-5] \oplus s_3[i-5]$ & $s_4[i-5] \oplus s_0[i-4]$ & $s_1[i-4] \oplus s_3[i-4]$ & $s_4[i-4] \oplus s_0[i-3]$ & $s_1[i-3] \oplus s_3[i-3]$ & $s_4[i-3] \oplus s_0[i-2]$ \\
\hline
\hline
\multicolumn{2}{c||}{$\displaystyle [i+2]$} & \multicolumn{2}{c||}{$\displaystyle [i+3]$} & \multicolumn{2}{c}{$\displaystyle [i+4]$} \\
\hline
$s_0[i+2]$ & $s_3[i+2]$ & $s_0[i+3]$ & $s_3[i+3]$ & $s_0[i+4]$ & $s_3[i+4]$ \\
$s_1[i+2]$ & $s_4[i+2]$ & $s_1[i+3]$ & $s_4[i+3]$ & $s_1[i+4]$ & $s_4[i+4]$ \\
$s_2[i+2]$ & $s_5[i+2]$ & $s_2[i+3]$ & $s_5[i+3]$ & $s_1[i+4]$ & $s_5[i+4]$ \\
\hline
$s_3[i] \oplus s_5[i+1]$ & $s_0[i+1] \oplus s_2[i+2]$ & $s_3[i+1] \oplus s_5[i+2]$ & $s_0[i+2] \oplus s_2[i+3]$ & $s_3[i+2] \oplus s_5[i+3]$ & $s_0[i+3] \oplus s_2[i+4]$ \\
$\oplus$ & $\oplus$ & $\oplus$ & $\oplus$ & $\oplus$ & $\oplus$ \\
$s_2[i-2] \oplus s_0[i-1]$ & $s_5[i-2] \oplus s_3[i-1]$ & \fboxrule=1pt \fbox{$s_2[i-1] \oplus s_0[i]$} & $s_5[i-1] \oplus s_3[i]$ & $s_2[i] \oplus s_0[i+1]$ & $s_5[i] \oplus s_3[i+1]$ \\
\hline
$s_4[i] \oplus s_2[i+1]$ & $s_1[i+1] \oplus s_5[i+1]$ & $s_4[i+1] \oplus s_2[i+2]$ & $s_1[i+2] \oplus s_5[i+2]$ & $s_4[i+2] \oplus s_2[i+3]$ & $s_1[i+3] \oplus s_5[i+3]$ \\
$\oplus$ & $\oplus$ & $\oplus$ & $\oplus$ & $\oplus$ & $\oplus$ \\
$s_1[i-2] \oplus s_3[i-2]$ & $s_4[i-2] \oplus s_0[i-1]$ & $s_1[i-1] \oplus s_3[i-1]$ & \fboxrule=1pt \fbox{$s_4[i-1] \oplus s_0[i]$} & $s_1[i] \oplus s_3[i]$ & $s_4[i] \oplus s_0[i+1]$ \\
\hline
\hline
\end{tabular}
\everymath{\displaystyle}
\caption{Rate $3/5$ Mu-SCo construction that satisfy the region (b) point described by user 1 with $(B_1,T_1) = (1,2)$ and user $2$ with $(B_2,T_2) = (2,4)$.}
\label{table:Code1224_Optimal}
\end{table}
\endgroup

\subsection{Region (c)}
Region (c) is sandwiched between $T_1 \le T_2 \le T_1  + B_1$ and also satisfies $T_1 \ge B_2$ in Fig.~\ref{fig:Regions}. The capacity is given by
\begin{equation}
\label{eq:Cc}
C_c = \frac{T_1}{T_1 + B_2}.
\end{equation}

For the achievability scheme, we use an  approach similar to region (b). We  can reduce the delay $T_2$ of user $2$ in region (c) so that it meets the $T_1 = T_2$ line without changing
the capacity $C_c$.  We can then apply a single user $(B_2, T_1)$ code that simultaneously satisfies both the users. Clearly this code is feasible since $T_1  \ge B_2$. The rate of this SCo
code meets the capacity. Note also that we do not require {\em source expansion} in this step.

\subsection{Region (d)}
In this region $T_2 \le T_1$ and $B_2 \ge B_1$. It suffices to serve user $2$ and the upper bound shows that
the capacity $C_d = C_2$ is also achieved using a single user SCo of parameters $(B_2,T_2)$.


\section{IA-SCo Construction (Proposition~\ref{prop:IASCo})}
\label{sec:IASCo}

We first provide a simple example to illustrate the main idea behind IA-SCo  and then provide the general construction.
We note that the IA-SCo codes achieve the capacity in a subset of region (a) in Fig.~\ref{fig:Regions}.
While IA-SCo codes do not provide any new capacity results, their construction is much simpler than DE-SCo and perhaps easier to generalize when there are more than two receivers.

\begingroup
\singlespacing
\everymath{\scriptstyle}
\begin{table}[t]
\centering
\subfigure[IA-SCo Code Construction for $(B_1,T_1) = (1,2)$ and $(B_2,T_2) = (2,6)$]{
\begin{tabular}{x{65pt}|x{65pt}|x{65pt}|x{65pt}|x{65pt}|x{65pt}}
$\displaystyle [i-1]$ & $\displaystyle [i]$ & $\displaystyle [i+1]$ & $\displaystyle [i+2]$ & $\displaystyle [i+3]$ & $\displaystyle [i+4]$ \\
\hline
\fboxrule=1pt \fbox{$s_0[i-1]$} & $s_0[i]$ & $s_0[i+1]$ & $s_0[i+2]$ & $s_0[i+3]$ & $s_0[i+4]$ \\
$s_1[i-1]$ & \fboxrule=1pt \fbox{$s_1[i]$} & $s_1[i+1]$ & $s_1[i+2]$ & $s_1[i+3]$ & $s_1[i+4]$ \\
\hline
$s_0[i-3] \oplus s_1[i-2]$ & $s_0[i-2] \oplus s_1[i-1]$ & \fboxrule=1pt \fbox{$s_0[i-1] \oplus s_1[i]$} & $s_0[i] \oplus s_1[i+1]$ & $s_0[i+1] \oplus s_1[i+2]$ & $s_0[i+2] \oplus s_1[i+3]$ \\
$\oplus$ & $\oplus$ & $\oplus$ & $\oplus$ & $\oplus$ & $\oplus$ \\
$s_0[i-7] \oplus s_1[i-5]$ & $s_0[i-6] \oplus s_1[i-4]$ & $s_0[i-5] \oplus s_1[i-3]$ & $s_0[i-4] \oplus s_1[i-2]$ & $s_0[i-3] \oplus s_1[i-1]$ & $s_0[i-2] \oplus s_1[i]$\\
\hline
\end{tabular}}
\subfigure[DE-SCo Code Construction for $(B_1,T_1) = (1,2)$ and $(B_2,T_2) = (2,5)$]{
\begin{tabular}{x{65pt}|x{65pt}|x{65pt}|x{65pt}|x{65pt}|x{65pt}}
$\displaystyle [i-1]$ & $\displaystyle [i]$ & $\displaystyle [i+1]$ & $\displaystyle [i+2]$ & $\displaystyle [i+3]$ & $\displaystyle [i+4]$ \\
\hline
\fbox{$s_0[i-1]$} & \fboxrule=1pt \fbox{$s_0[i]$} & $s_0[i+1]$ & $s_0[i+2]$ & $s_0[i+3]$ & $s_0[i+4]$ \\
\fboxrule=1pt \fbox{$s_1[i-1]$} & \fbox{$s_1[i]$} & $s_1[i+1]$ & $s_1[i+2]$ & $s_1[i+3]$ & $s_1[i+4]$ \\
\hline
$s_0[i-3] \oplus s_1[i-2]$ & $s_0[i-2] \oplus s_1[i-1]$ & \fbox{$s_0[i-1] \oplus s_1[i]$} & $s_0[i] \oplus s_1[i+1]$ & $s_0[i+1] \oplus s_1[i+2]$ & $s_0[i+2] \oplus s_1[i+3]$ \\
$\oplus$ & $\oplus$ & $\oplus$ & $\oplus$ & $\oplus$ & $\oplus$ \\
$s_1[i-6] \oplus s_0[i-5]$ & $s_1[i-5] \oplus s_0[i-4]$ & $s_1[i-4] \oplus s_0[i-3]$ & $s_1[i-3] \oplus s_0[i-2]$ & $s_1[i-2] \oplus s_0[i-1]$ & \fboxrule=1pt \fbox{$s_1[i-1] \oplus s_0[i]$}\\
\hline
\end{tabular}}
\everymath{\displaystyle}
\caption{Rate $2/3$ code constructions that satisfy user 1 with $(B_1,T_1) = (1,2)$ and user $2$ with $B_2 = 2$. The two points $\{(1,2)-(2,6)\}$ and $\{(1,2)-(2,5)\}$ lies in region (a).}
\label{table:Code1225}
\end{table}
\endgroup

\subsection{Example}
Consider an example with the first and second users experiencing burst erasures of length $B_1 = 1$ and $B_2 = 2$ symbols respectively (i.e., $\al = 2$) and the corresponding delay for the first user is $T_1 = 2$.  From Prop.~\ref{prop:IASCo} we have that $T_2 = 6$. Table~\ref{table:Code1225}(a) illustrates the IA-SCo construction. For comparison the optimal DE-SCo construction achieving $T_2 = 5$, proposed in~\cite{Our_JSAC} is provided in Table~\ref{table:Code1225}(b).

The construction of the IA-SCo code is as follows. We split each source symbol $\bs[i]$ into two sub-symbols $s_0[i]$ and $s_1[i]$ of equal size.  Let $p_1[i] = s_0[i-2] \oplus s_1[i-1] $ be the parity check associated with the $(1,2)$ SCo-code~\cite{Martinian_Thesis} and let $p_2[i] = s_0[i-4]\oplus s_1[i-2]$ be the parity check for the $(2,4)$ SCo-code~\cite{Martinian_Thesis}.  The parity check row is obtained by combining $q[i] = p_1[i] \oplus p_2[i-T_1]$ i.e., by shifting $p_2[i]$ by $T_1$ units and then combining.  The parity check stream $q[\cdot]$ are then concatenated with the source symbols as shown in Table~\ref{table:Code1225}(a).

When an erasure of one symbol occurs say at $t= i-1$ for user $1$, it needs to recover $\bs[i-1]$ at time $t=i+1$. Note that  user $1$ can cancel the  second row of parity checks $p_2[\cdot],$ which combines unerased sub-symbols. For user $2$ suppose that a burst erasure of length $B_2=2$ symbols occurs at times ${t =i-2, i-1}$.  User $2$ simply ignores the parity checks $q[i]$ and $q[i+1]$. Starting from ${t=i+2}$, the parity checks $p_1[\cdot]$ are functions of symbols $\bs[i], \bs[i+1],\ldots$ and do not involve the erased symbols $\bs[i-1]$ and $\bs[i-2]$. Therefore we can subtract $p_1[\cdot]$ from $q[i+2],\ldots, q[i+6]$ and recover $p_2[i], \ldots, p_2[i+4],$ which suffice to recover the missing symbols.

\subsection{General Construction}
The main idea behind the general construction is to start with two single-user codes for the two users, $(B_1,T_1)$ and $(\al B_1,\al T_1)$ and delay the parity checks of the second by $T_1$ so that they can be combined with the parity checks of user $1$ without causing any interference to the two users.

Throughout our discussion we let $T_1 = T$ and $B_1 = B$ and $B_2 = \al B$ and $T_2= \al T + T$.

\subsubsection{Code Construction}
\begin{itemize}
\item Let ${\cal C}_1$ be the single-user code of user 1~\cite{Martinian_Thesis, Our_JSAC}. Assume that the source symbols $\bs[i]$ are divided into $T$ sub-symbols $(s_0[i],\dots,s_{T-1}[i])$ and combined to produce $B$ parity check sub-symbols $\bp^{\rm{I}}[i] = (p^{\rm{I}}_0[i],\dots,p^{\rm{I}}_{B-1}[i])$ according to
\begin{align}
p_j^{\rm{I}}[i] = s_j[i- T] + h_j(s_B[i-(j+T-B)],\ldots,s_{T-1}[i-(j+1)]), \quad j=0,\ldots, B-1.
\label{eq:ScoParityI}
\end{align}

\item Let $\cC_2$ be a rate $C_1$ SCo with parameters $(\al B, \al T)$ also obtained by splitting the source symbols into $T$ sub-symbols $(s_0[i],\ldots,s_{T-1}[i])$ combined to produce $B$ parity checks $\bp^{\rm{II}}[i] = (p^{\rm{II}}_0[i],\ldots, p^{\rm{II}}_{B-1}[i])$ according to the vertical interleaving property in~\cite{Our_JSAC}, i.e.,
\begin{align}
p_j^{\rm{II}}[i] = s_j[i-\al T] + h_j(s_B[i-\al(j+T-B)],\ldots, s_{T-1}[i-(j+1)\al]), j=0,\ldots, B-1.
\label{eq:ScoParityII}
\end{align}

\item Construct $\cC_M$ whose symbols have the form
$\left( \bs[i], \bq[i]\right)$, where $ \bq[i]= \bp^{\rm{I}}[i]+ \bp^{\rm{II}}[i-T]$. Intuitively the stream of parity checks
$ \bp^{\rm{II}}[\cdot]$ is delayed by $T$ symbols and then the resulting non-interfering streams are combined.
\end{itemize}

Clearly the rate of $\cC_M$ equals $C_1$. We need to show that user 1 and user 2 can recover from erasure bursts of $B$ and $B_2=\al B$ within delays of $T$ and $T_2=\al T + T$ respectively.

\subsubsection{User 1 Decoding}
Assume that the symbols at time $i,\ldots, i+B-1$ are erased on user 1's channel. By virtue of $\cC_1$, symbol $ \bs[i+k]$ (for $k=0,1,\ldots, B-1$) can be recovered by time $i+k+T$ using parity checks $ \bp^{\rm{I}}[i+B],\ldots, \bp^{\rm{I}}[i+k+T]$. Thus it suffices to show that we can recover $ \bp^{\rm{I}}[i+k]$
from $ \bq[i+k]$ for $k=B,\ldots,B+T-1$.

First note that the $\bp^{\rm{I}}[i+k]$ for $k=B,\ldots, T$ can be directly recovered from $\bq[i+k]$ since the interfering parity checks $\bp^{\rm{II}}[\cdot]$ only consist of source symbols before time $i$. Indeed the parity check at $k=T$ is$$ \bq[i+T] = \bp^{\rm{I}}[i+T]+ \bp^{\rm{II}}[i]$$
and from~\eqref{eq:ScoParityII} the sub-symbols in $\bp^{\rm{II}}[i]$ only depend on the source symbols before time $i$. Thus upon receiving $\bq[i+T]$ user 1 can recover the erased symbol $\bs[i]$. Furthermore, we can also compute $\bp^{\rm{II}}[i+1]$ which only consists of source symbols up to time $i$ and upon receiving $\bq[i+T+1]$ can compute $\bp^{\rm{I}}[i+T+1]$ from
$$\bp^{\rm{I}}[i+T+1]=\bq[i+T+1]-\bp^{\rm{II}}[i+1].$$In turn it recovers $\bs[i+1]$. Continuing this process it can recover all the erased symbols $\bs[i+k]$ by time $i+T+k$.

\subsubsection{User 2 Decoding}
Suppose that the symbols at time $i,\ldots, i+B_2-1$ are erased on user 2's channel. By virtue of $\cC_2$, symbol $ \bs[i+k]$ (for $k=0,1,\ldots, B_2-1$) can be recovered by time $i+k+\al T$ using parity checks $ \bp^{\rm{II}}[i+B_2],\ldots, \bp^{\rm{II}}[i+k+\al T]$. To establish~\eqref{eq:necc} it suffices to show that symbols $ \bp^{\rm{II}}[i+B_2],\ldots, \bp^{\rm{II}}[i+k+\al T]$ can be recovered from
symbols $ \bq[i+T+B_2],\ldots, \bq[i+k+\al T+T]$. Indeed since
$$ \bq[i+B_2+T] = \bp^{\rm{I}}[i+B_2+T] + \bp^{\rm{II}}[i+B_2],$$
it suffices to observe that user 2 can cancel $\bp^{\rm{I}}[i+B_2+T]$ upon receiving $\bq[i+B_2+T]$. It however immediately follows from \eqref{eq:ScoParityI} that $\bp^{\rm{I}}[i+B_2+T]$ involves source symbols at time $i+B_2$ or later (the construction limits the memory in the channel input stream to previous $T$ symbols). The symbols $ \bp^{\rm{I}}[\cdot]$ after this time also depend on $ \bs[\cdot]$ at time $i+B_2$ or later.


\section{Proof Of Theorem~\ref{thm:Multicast_UB}}
\label{sec:Multicast_UB}

\begin{figure}
\vspace{-2em}
  \centering
    \subfigure[Step (1)]
    {
      \includegraphics[scale = 0.3]{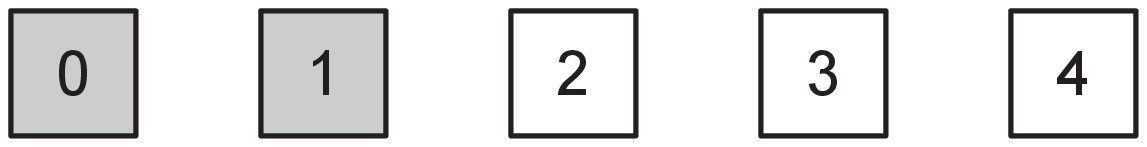}
			\label{fig:Converse1224_a}
		}
		\subfigure[Step (2)]
    {
      \includegraphics[scale = 0.3]{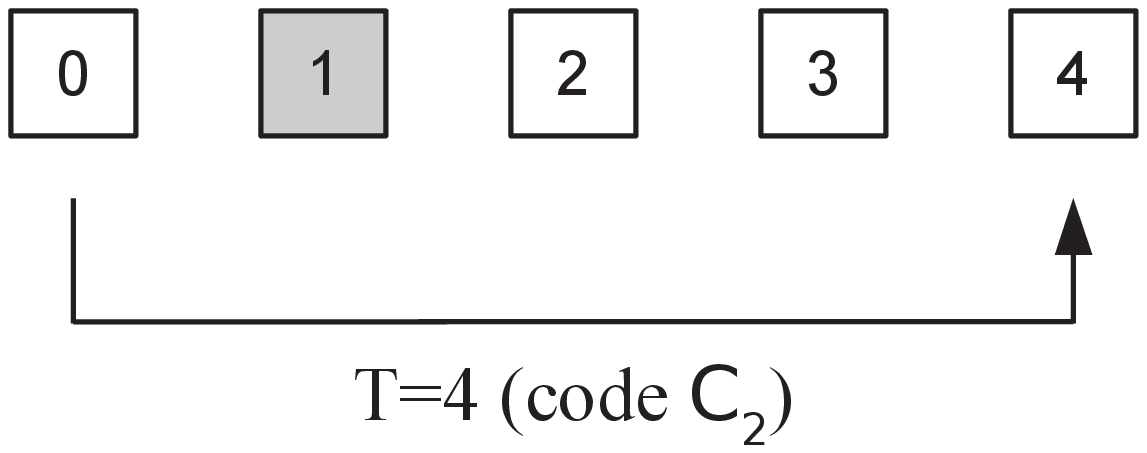}
			\label{fig:Converse1224_b}
		}
		\subfigure[Step (3)]
		{
      \includegraphics[scale = 0.3]{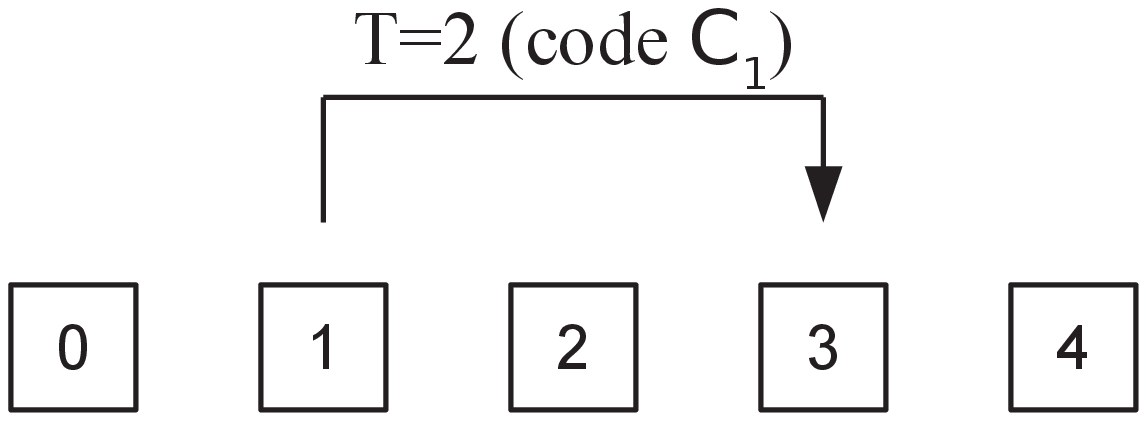}
			\label{fig:Converse1224_c}
		}
  \caption{Main steps of finding the upper-bound for the $\{(1,2)-(2,4)\}$ point through one period illustration of the Periodic Erasure Channel. Grey and white squares resemble erased and unerased symbols respectively.} \label{fig:Converse1224}
\end{figure}

We first provide an example to illustrate the upper bound using periodic erasure channel. Then we outline the general periodic erasure channel (PEC) based argument. Finally we provide a rigorous information theoretic converse.

\subsection{Example}
The main steps of this proof can be illustrated by first considering a specific example, $\{(1,2)-(2,4)\}$ which is shown in Fig.~\ref{fig:Converse1224}. We consider a periodic erasure channel with each period having two consecutive erasures followed by three unerased symbols. Thus, one can start by using code $\mathcal{C}_2 = (2,4)$ to recover the erasure at time $0,$ wtih a delay of $4$, leaving only one erasure at time $1$ (c.f. Fig.~\ref{fig:Converse1224_b}). Now, code $\mathcal{C}_1 = (1,2)$ can be used as it is capable of recovering this one erasure within a delay of $2$ (i.e., by time $3$) (c.f. Fig.~\ref{fig:Converse1224_c}). Let us assume that the code is systematic and thus one can recover the symbols at time $2$, $3$ and $4$ from their corresponding unerased channel symbols. Thus, one can recover a total of $5$ source symbols from $3$ unerased channel symbols which implies that $3/5$ is an upper-bound of this channel.

\subsection{PEC based Converse}
For the general case of Theorem~\ref{thm:Multicast_UB}, we start by the case, $T_2 > T_1 + B_1$ and then consider $T_2 \leq T_1 + B_1$.
\begin{lem}
When $T_2 > T_1 + B_1$, suppose there exists a sequence of feasible encoding functions $\{f_t(\cdot)\}$ and decoding functions $\{\g_{1t}(\cdot)\}$ and $\{\g_{2t}(\cdot)\}$. Then
there also exist  decoding functions $\g_t(\cdot)$ that can reproduce the source symbols $ \bs[t]$, over a channel with periodic bursts as stated below
\begin{equation}
\by[t] = \begin{cases}
\star, & t \in \left[T^k, T^k + B_2 - 1\right] \\
\bx[t],& t \in \left[T^k + B_2, T^{k+1}-1 \right]
\end{cases}
\end{equation}
where $T^k = k T_2 + k(B_2 - B_1)$, $k=0,1,\ldots$
\label{lem:Lemma_Periodic}
\end{lem}

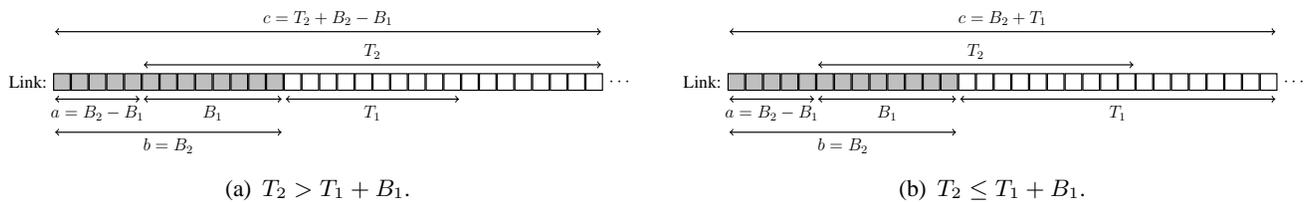
\begin{figure}[htbp]
	\centering
	\subfigure[$T_2 > T_1 + B_1$.]
	{
	\resizebox{0.47\columnwidth}{!}{
	\begin{tikzpicture}[node distance=0mm]
		\node                       (x1start) {Link:};
		\node[esym, right = of x1start]  (x100) {};
		\node[esym, right = of x100]     (x101) {};
		\node[esym, right = of x101]     (x102) {};
		\node[esym, right = of x102]     (x103) {};
		\node[esym, right = of x103]     (x104) {};
		\node[esym, right = of x104]     (x105) {};
		\node[esym, right = of x105]     (x106) {};
		\node[esym, right = of x106]     (x107) {};
		\node[esym, right = of x107]     (x108) {};
		\node[esym, right = of x108]     (x109) {};
		\node[esym, right = of x109]     (x110) {};
		\node[esym, right = of x110]     (x111) {};
		\node[esym, right = of x111]     (x112) {};
		\node[usym, right = of x112]     (x113) {};
		\node[usym, right = of x113]     (x114) {};
		\node[usym, right = of x114]     (x115) {};
		\node[usym, right = of x115]     (x116) {};
		\node[usym, right = of x116]     (x117) {};
		\node[usym, right = of x117]     (x118) {};
		\node[usym, right = of x118]     (x119) {};
		\node[usym, right = of x119]     (x120) {};
		\node[usym, right = of x120]     (x121) {};
		\node[usym, right = of x121]     (x122) {};
		\node[usym, right = of x122]     (x123) {};
		\node[usym, right = of x123]     (x124) {};
		\node[usym, right = of x124]     (x125) {};
		\node[usym, right = of x125]     (x126) {};
		\node[usym, right = of x126]     (x127) {};
		\node[usym, right = of x127]     (x128) {};
		\node[usym, right = of x128]     (x129) {};
		\node[usym, right = of x129]     (x130) {};
		\node      [right = of x130]     (x1end) {$\cdots$};

		\dimdn{x100}{x112}{-10mm}{$b = B_2$};

		\dimdn{x100}{x104}{-2mm}{$a = B_2 - B_1$};
		\dimdn{x105}{x112}{-2mm}{$B_1$};
		\dimdn{x113}{x122}{-2mm}{$T_1$};
		\dimup{x105}{x130}{2mm}{$T_2$};
		\dimup{x100}{x130}{10mm}{$c = T_2+B_2-B_1$};
	\end{tikzpicture}}
	\label{fig:Periodic_Erasure_Channel_1}
	}
	\subfigure[$T_2 \leq T_1 + B_1$.]
	{
	\resizebox{0.47\columnwidth}{!}{
	\begin{tikzpicture}[node distance=0mm]
		\node                       (x1start) {Link:};
		\node[esym, right = of x1start]  (x100) {};
		\node[esym, right = of x100]     (x101) {};
		\node[esym, right = of x101]     (x102) {};
		\node[esym, right = of x102]     (x103) {};
		\node[esym, right = of x103]     (x104) {};
		\node[esym, right = of x104]     (x105) {};
		\node[esym, right = of x105]     (x106) {};
		\node[esym, right = of x106]     (x107) {};
		\node[esym, right = of x107]     (x108) {};
		\node[esym, right = of x108]     (x109) {};
		\node[esym, right = of x109]     (x110) {};
		\node[esym, right = of x110]     (x111) {};
		\node[esym, right = of x111]     (x112) {};
		\node[usym, right = of x112]     (x113) {};
		\node[usym, right = of x113]     (x114) {};
		\node[usym, right = of x114]     (x115) {};
		\node[usym, right = of x115]     (x116) {};
		\node[usym, right = of x116]     (x117) {};
		\node[usym, right = of x117]     (x118) {};
		\node[usym, right = of x118]     (x119) {};
		\node[usym, right = of x119]     (x120) {};
		\node[usym, right = of x120]     (x121) {};
		\node[usym, right = of x121]     (x122) {};
		\node[usym, right = of x122]     (x123) {};
		\node[usym, right = of x123]     (x124) {};
		\node[usym, right = of x124]     (x125) {};
		\node[usym, right = of x125]     (x126) {};
		\node[usym, right = of x126]     (x127) {};
		\node[usym, right = of x127]     (x128) {};
		\node[usym, right = of x128]     (x129) {};
		\node[usym, right = of x129]     (x130) {};
		\node      [right = of x130]     (x1end) {$\cdots$};

		\dimdn{x100}{x112}{-10mm}{$b = B_2$};

		\dimdn{x100}{x104}{-2mm}{$a = B_2 - B_1$};
		\dimdn{x105}{x112}{-2mm}{$B_1$};
		\dimdn{x113}{x130}{-2mm}{$T_1$};
		\dimup{x105}{x122}{2mm}{$T_2$};
		\dimup{x100}{x130}{10mm}{$c = B_2+T_1$};
	\end{tikzpicture}}
	\label{fig:Periodic_Erasure_Channel_2}
	}
	\caption{One period illustration of the Periodic Erasure Channel in Fig.~\ref{fig:PEC_SCo} to be used for proving the multicast upper-bound provided in Theorem~\ref{thm:Multicast_UB}.}
	\label{fig:Periodic_Erasure_Channel}
\end{figure}

An illustration of one period of the proposed periodic-erasure channel (from $T^0$ to $T^1$) in the case $T_2 > T_1 + B_1$ is shown in Fig.~\ref{fig:Periodic_Erasure_Channel}(a). The capacity of the periodic erasure channel in Lemma~\ref{lem:Lemma_Periodic} is
\begin{equation}
C \leq \frac{T_2 - B_1}{T_2 - B_1 + B_2}.
\end{equation}

To establish Lemma~\ref{lem:Lemma_Periodic}, it suffices to show that by time $T^k-1$, the receiver is able to recover symbols $ \bx[0],\ldots, \bx[T^k-1]$. We first show that by time $T^1-1$, the receiver is able to recover symbols $ \bx[0],\ldots, \bx[T^1-1]$. Since only symbols $ \bx[0],\ldots, \bx[B_2-1]$ are erased by time $T^1-1$ we focus on these symbols.

Consider a single-burst channel that introduces a burst of length $B_2$ from times $t=0,1,\ldots,B_2 -1$. Note that this channel behaves identically to the periodic burst channel up to time $T^1-1$. Applying the decoder $\g_{2t}(\cdot)$ for $t=0,1,\ldots, (B_2 - B_1)-1$, the receiver recovers symbols $ \bs[0],\ldots, \bs[t]$ with a delay of $T_2$ i.e., by time $T^1 -1$ and hence it also recovers the channel packets $ \bx[0],\ldots, \bx[t]$ via~\eqref{eq:Code_Function}. It remains to show that the symbols at time $t= (B_2-B_1),\ldots,B_2-1$ are also recovered by time $T^1-1$. One cannot apply the decoder $\g_{2t}(\cdot)$ to recover these symbols since the decoding will require symbols beyond time $T^1$, which are available on the single-burst channel but not on the periodic burst channel. However, to recover these symbols we use the multicast property of the code as follows. Consider a channel that introduces a single erasure burst of length $B_1$ between times $t=(B_2 - B_1),\ldots, B_2 -1$. Note that up to time $T^1-1$, this channel is identical to our periodic burst-erasure channel (which has recovered $ \bx[0],\ldots, \bx[B_2-B_1-1]$). For this channel, and hence the periodic erasure channel, using the decoder $\g_{1t}(\cdot)$ the source symbols are recovered by time $B_2 + T_1 -1 \le T^1 -1$. Furthermore via~\eqref{eq:Code_Function}, the erased channel symbols $ \bx[B_2 - B_1],\ldots, \bx[B_2 -1]$ are also recovered by time $T^1 -1$. Since the channel introduces periodic bursts, the same argument can be repeated to recover all symbols up to time $T^k-1$ for each $k$.

The same argument applies in the case when $T_2 \leq T_1 + B_1$ (in Fig.~\ref{fig:Periodic_Erasure_Channel}.(b)) except that the periodic bursts are stated as,
\begin{equation}
 \by[t] = \begin{cases}
\star, & t \in \left[T^k, T^k + B_2 - 1\right] \\
 \bx[t],& t \in \left[T^k + B_2, T^{k+1}-1 \right]
\end{cases}
\end{equation}
where $T^k = k (T_1 + B_2)$ and the theorem follows.

\subsection{Information Theoretic Converse}

Recall that our PEC argument assumed that (1) the channel packets $\bx[t]$ are deterministic functions of the source packets up to time $t$, (2) the channel code is systematic and (3) the recovery must happen with zero error. All of these assumptions can be removed by resorting to the information theoretic converse discussed next.

We start by proving the first case $T_2 > T_1 + B_1$. We use the periodic erasure channel shown in Fig.~\ref{fig:Periodic_Erasure_Channel}.(a), where each period has $B_2$ erasures followed by $T_2-B_1$ non-erasures. We can assign
\begin{align*}
	&a = B_2 - B_1, \quad
	b = B_2, \quad
	c = B_2 + T_2 - B_1 \quad \text{(period length)}, \\
	&W_i = \bx\sqstack{ic+b}{(i+1)c-1}, \quad
	V_i = \bs\sqstack{ic}{(i+1)c-1}.
\end{align*}

We define the capability of the $\mathcal{C}_1 = (B_1,T_1)$ and $\mathcal{C}_2 = (B_2,T_2)$ codes by,
\begin{align}
H\Bigl(\bs[i] \Big| \bx\sqstack{i+B_1}{i+T_1} \bx\sqstack{0}{i-1}\Bigr) = 0 \label{eq:Code1_Def_thm1_caseA} \\
H\Bigl(\bs[i] \Big| \bx\sqstack{i+B_2}{i+T_2} \bx\sqstack{0}{i-1}\Bigr) = 0 \label{eq:Code2_Def_thm1_caseA},
\end{align}

We  use mathematical induction to prove that for $n \geq 0$
\begin{equation}
	H(W_0^{n}) \geq H(V_0^{n-1}) + H\Bigl(W_n  \Big|  V_0^{n-1} \bx\sqstack{0}{nc-1}\Bigr). \label{eq:induction_formula_thm1_caseA}
\end{equation}
The base case for \eqref{eq:induction_formula_thm1_caseA} is given by substituting $n=0$ into it:
\begin{align}
	H(W_0) \geq H(V_0^{-1}) + H\Bigl(W_0  \Big|  V_0^{-1} \bx\sqstack{0}{-1}\Bigr) \geq H(W_0)
\end{align}
which is obviously true. For the induction step, let us start by assuming that \eqref{eq:induction_formula_thm1_caseA} is true for $n=k$,
\begin{equation}
\label{eq:MuSCo_nk}
	H(W_0^{k}) \geq H(V_0^{k-1}) + H\Bigl(W_k  \Big|  V_0^{k-1} \bx\sqstack{0}{kc-1}\Bigr).
\end{equation}

In the first part of the induction step, some entropy manipulations are applied (c.f. Appendix.~\ref{app:MuSCo}), to show that:
\begin{align}
\label{eq:MuSCo_a}
H(W_0^{k}) \geq H\Bigl(V_0^{k-1} \bs\sqstack{kc}{kc+b-1}\Bigr) + H\Bigl(W_{k}  \Big|  V_0^{k-1} \bs\sqstack{kc}{kc+b-1} \bx\sqstack{0}{kc+b-1}\Bigr)
\end{align}
These entropy manipulations can be summarized in two main steps, the first of which is recovering the first $a=B_2-B_1$ source symbols, $\bs\sqstack{kc}{kc+a-1}$ using code $\cC_2$ defined in~\eqref{eq:Code2_Def_thm1_caseA} due to the availability of $W_k$, while the second step is recovering the next $b-a=B_1$ source symbols, $\bs\sqstack{kc+a}{kc+b-1}$ using $\cC_1$ defined in~\eqref{eq:Code1_Def_thm1_caseA}.

In the second part, we add $H(W_{k+1} | W_0^k)$ to both sides of the inequality. Because the channel code is not necessarily systematic, we will use the additional channel packets in $W_{k+1}$ to help decode the source packets $\bs\sqstack{kc+b}{(k+1)c-1}$ (detailed steps are shown in Appendix.~\ref{app:MuSCo}).
\begin{align}
\label{eq:MuSCo_b}
	H&(W_0^{k+1}) \geq H(V_0^k) + H\Bigl(W_{k+1}  \Big|  V_0^k \bx\sqstack{0}{(k+1)c-1}\Bigr).
\end{align}

The working in~\eqref{eq:MuSCo_b} shows that if~\eqref{eq:induction_formula_thm1_caseA} is true for $n=k$, then it is also true for $n=k+1$. By induction,~\eqref{eq:induction_formula_thm1_caseA} is true for $n \geq 0$. Finally,
\begin{align*}
	H(W_0^{n}) &\geq H(V_0^{n-1}) + H\Bigl(W_n  \Big|  V_0^{n-1} \bx\sqstack{0}{nc-1}\Bigr) \geq H(V_0^{n-1}).
\end{align*}
Using the fact that all of the channel packets have the same entropy, and all of the source packets have the same entropy, we can continue to get
\begin{align}
	H(W_0^{n}) &\geq H(V_0^{n-1}) \nonumber \\
	(n+1) \cdot (T_2-B_1) \cdot H(\bx) &\geq n \cdot (T_2+B_2-B_1) \cdot H(\bs) \nonumber \\
	\frac{n+1}{n} \cdot \frac{T_2-B_1}{T_2 + B_2 - B_1} &\geq \frac{H(\bs)}{H(\bx)}.
\end{align}
Finally, we get
\begin{equation}
	R = \frac{H(\bs)}{H(\bx)} \leq \frac{T_2-B_1}{T_2 - B_1 + B_2}. \; (\mathrm{as} \; n \rightarrow \infty) \label{eq:upper_bound_CaseA}
\end{equation}
Therefore, any $\{(B_1,T_1),(B_2,T_2)\}$ code with $T_2 > T_1 + B_1$ must satisfy \eqref{eq:upper_bound_CaseA}.

For the case with $T_2 \leq T_1 + B_1$, the same proof applies except that the values of $a$, $b$ and $c$ are updated as follows,
\begin{align*}
	a = B_2 - B_1, \quad
	b = B_2, \quad
	c = B_2 + T_1 \quad \text{(period length)},
\end{align*}
and again we end up having,
\begin{align}
	H(W_0^{n}) &\geq H(V_0^{n-1}) \nonumber \\
	(n+1) \cdot T_1 \cdot H(\bx) &\geq n \cdot (T_1 + B_2) \cdot H(\bs) \nonumber \\
	\frac{n+1}{n} \cdot \frac{T_1}{T_1 + B_2} &\geq \frac{H(\bs)}{H(\bx)}.
\end{align}

In other words,
\begin{equation}
	R = \frac{H(\bs)}{H(\bx)} \leq \frac{T_1}{T_1 + B_2}. \; (\mathrm{as} \; n \rightarrow \infty) \label{eq:upper_bound_CaseB}
\end{equation}
Therefore, any $(B_1,T_1),(B_2,T_2)$ code with $T_2 \leq T_1 + B_1$ must satisfy \eqref{eq:upper_bound_CaseB}.


\section{Code Construction in Region (e) (Theorem~\ref{thm:Region_e})}
\label{sec:Region_e_CC}

Recall that region (e) in Fig.~\ref{fig:Regions} and Fig.~\ref{fig:Regions2} is contained within $T_2 \ge B_1 + T_1$, $T_2 \ge B_2$ and $T_2 \le B_1 + B_2$. Since the capacity $C_e$  given by
\begin{equation}\label{eq:Ce2}C_e = \frac{T_1}{2T_1 + B_1 + B_2 - T_2}\end{equation} is constant along each
45-degree line starting from the line $T_2 = B_1 + T_1$ in the $(B_2, T_2)$ plane in Fig.~\ref{fig:Regions2},  we can parameterize the $T_2$ and $B_2$ as
\begin{align*}
T_2 = T_1 + B_1 + m, \quad \quad B_2 = T_1 + k + m,
\end{align*}
where $m \geq 0$ is the number of steps upwards on the 45 line in Fig.~\ref{fig:Regions2} starting from $T_2 = T_1 + B_1$ line dividing regions (e) and (f), and where $k$ is an integer taking values from $0$ to $B_1$, which horizontally spans region (e) from $T_2 = B_1 + B_2$ to $T_2 = B_2$. Substituting into~\eqref{eq:Ce2} we have
\begin{align}
C_e = \frac{T_1}{2T_1+k},\label{eq:Ce3}
\end{align}
which we will show is achievable.

In Appendix~\ref{sec:e-example} we provide two examples of the code constructions with parameters $\{(4,5), (7,10)\}$  and $\{(3,5), (7,9)\}$.  These examples compliment the general description below and might be worthwhile reading in parallel with this section.

The construction generates three layers of parity checks and carefully combines them to satisfy the decoding constraints of both the receivers. The main construction steps are described below.

\begin{itemize}
\item Split each source symbols $\bs[i]$ in $T_1$ sub-symbols $$\bs[i] = (s_0[i],\dots,s_{T_1-1}[i])$$

\item Apply a $\mathcal{C}_1 = (B_1,T_1)$ single user SCo code to the source symbols $\bs[i]$ producing $B_1$ parity check sub-symbols $$\bp^{\rm{I}}[i] = (p^{\rm{I}}_0[i],\dots,p^{\rm{I}}_{B_1-1}[i])$$ at each time by combining the source sub-symbols along the main diagonal,
\begin{align}
p_j^{\rm{I}}[i]	= s_j[i-T_1] + h_j^{\rm{II}}(s_{B_1}[i-j-T_1+B_1],\dots,s_{T_1 - 1}[i-j-1]), \quad j = \{0,1,\dots,B_1-1\}. \nonumber
\end{align}

\item Apply a $\mathcal{C}_2$ repetition code to the source symbols $\bs[i]$ with a delay of $T_2$, i.e., the corresponding parity check symbols are,
\begin{align}
\bp^{\rm{II}}[i] = (p^{\rm{II}}_0[i],\dots,p^{\rm{II}}_{T_1-1}[i]) = (s_0[i-T_2],\dots,s_{T_1-1}[i-T_2]) = \bs[i-T_2].
\end{align}

\item Concatenate the two streams $\bp^{\rm{I}}[\cdot]$ and $\bp^{\rm{II}}[\cdot]$ with a \textit{partial} overlap as illustrated
in~\eqref{eq:comb-1}. In particular, the two streams of parity checks $\bp^{\rm{I}}[\cdot]$ and $\bp^{\rm{II}}[\cdot]$
are concatenated with the last $B_1 - k$ rows of the first added to upper most $B_1-k$ rows of the second.

\begingroup
\singlespacing
\begin{align}
\tilde{\bx}[i] =
\left[
\begin{array}{rl}
& s_0[i] \\
& \vdots\\
& s_{T-1}[i] \\
& p^{\rm{I}}_0[i] \\
& \vdots \\
& p^{\rm{I}}_{k-1}[i] \\
p^{\rm{I}}_{k}[i] &+ s_0[i-T_2] \\
& \vdots \\
p^{\rm{I}}_{B_1-1}[i] &+ s_{B_1-k-1}[i-T_2] \\
& s_{B_1-k}[i-T_2] \\
& \vdots \\
& s_{T_1-1}[i-T_2]
\end{array}
\right]
\label{eq:comb-1}
\end{align}
\endgroup

We further combine the last $T_1-({B_1-k})$  parity checks of $\bp^{\rm{II}}[\cdot]$ with additional parity checks of code $\cC_3$ as explained below.

\item Consider the two cases:
\begin{enumerate}[(A)]
\item $T_1 \leq 2(B_1 - k)$\\
\begin{itemize}
\item Apply a $\mathcal{C}_3=(B_3,T_3) = (T_1-(B_1-k),B_1-k)$ single user SCo code on the last $B_1 - k$ \emph{parity check} sub-symbols of $\cC_1$,  $(p^{\rm{I}}_{k}[.],\dots,p^{\rm{I}}_{B_1-1}[.])$ constructing $T_1-(B_1-k)$ parity checks $$\bp^{3}[i] = (p^{3}_0[i],\dots,p^{3}_{T_1-(B_1-k)-1}[i])$$ at each time by combining the last $B_1 - k$ \emph{parity check} sub-symbols, $(p^{\rm{I}}_{k}[.],\dots,p^{\rm{I}}_{B_1-1}[.])$, along the main diagonal, i.e.,
\begin{align}
&p_j^{3}[i]	= p_{k+j}^{\rm{I}}[i-T_3] + h_j^3(p_{k+B_3}^{\rm{I}}[i-j-T_3+B_3],\dots,p_{k+T_3-1}^{\rm{I}}[i-j-1]), \nonumber
\end{align}
for $j = \{0,1,\dots,T_1-(B_1-k)-1\}$.
\item Combine a $\Delta_{3} = -T_1$ shifted version of the produced stream of parity checks $\bp^{3}[.]$ to the last $T_1-(B_1-k)$ rows of $\bx[.]$, thus,
\begingroup
\singlespacing
\begin{align}
\bx[i] =
\left[
\begin{array}{rl}
& \bs[i] \\
& p^{\rm{I}}_0[i] \\
& \vdots \\
& p^{\rm{I}}_{k-1}[i] \\
p^{\rm{I}}_{k}[i] &+s_0[i-T_2] \\
& \vdots \\
p^{\rm{I}}_{B_1-1}[i] &+s_{B_1-k-1}[i-T_2] \\
s_{B_1-k}[i-T_2] &+p^3_{0}[i+T_1]\big{|}_i \\
& \vdots \\
s_{T_1-1}[i-T_2] &+p^3_{T_1-(B_1-k)-1}[i+T_1]\big{|}_i
\end{array}
\right]
\label{eq:bx-nc}
\end{align}
\endgroup


where $$\bp^{3}[t_1]\big{|}_{t_2} = (p^{3}_0[t_1]\big{|}_{t_2},\dots,p^{3}_{T_1-(B_1-k)-1}[t_1]\big{|}_{t_2})$$ is the parity-check $\bp^{3}[t_1]$ shifted to time $t_2$.
\end{itemize}

We note that the construction of $\bx[i]$ in~\eqref{eq:bx-nc} requires us to have access to source symbols after time $i$ as the parity checks $\bp^3[i+T_1]$
may include  source symbols after time $i$. Since our encoder is causal we cannot have access to these source symbols. Instead we transmit only the causal part of the
underlying parity checks.  In particular, we decompose each parity check into two parts as follows. For any $t_2 \le t_1$ we have,
\begin{align}
 p^3_j[t_1]\big{|}_{t_2}= \tilde{ p}^3_j[t_1]\big{|}_{t_2} +\hat{ p}^3_j[t_1]\big{|}_{t_2}
\end{align}
where $\tilde{p}^3_j[t_1]\big{|}_{t_2}$ denotes the causal part of the parity check with respect to $t_2$
whereas $\hat{p}^3_j[t_1]\big{|}_{t_2}$ denotes the non-causal part of the parity check with respect to $t_2$ i.e.,
\begin{align}
\tilde{p}^3_j[t_1]\big{|}_{t_2} &= f_j(\bs[t_2], \bs[t_2-1],\bs[t_2-2]\ldots) \label{eq:p-causal}\\
\hat{p}^3_j[t_1]\big{|}_{t_2}&= g_j( \bs[t_2+1],\bs[t_2+2]\ldots). \label{eq:p-nc}
\end{align}

The resulting input symbol at time $i$ is given by
\begingroup
\singlespacing
\begin{align}
\bx[i] =
\left[
\begin{array}{rl}
& \bs[i] \\
& p^{\rm{I}}_0[i] \\
& \vdots \\
& p^{\rm{I}}_{k-1}[i] \\
p^{\rm{I}}_{k}[i] &+ s_0[i-T_2] \\
& \vdots \\
p^{\rm{I}}_{B_1-1}[i] &+ s_{B_1-k-1}[i-T_2] \\
s_{B_1-k}[i-T_2] &+ \tilde{p}^3_{0}[i+T_1]\big{|}_i \\
& \vdots \\
s_{T_1-1}[i-T_2] &+ \tilde{p}^3_{T_1-(B_1-k)-1}[i+T_1]\big{|}_i
\end{array}
\right]
\label{eq:bx-c}
\end{align}
\endgroup

 The symbol $\bx[i]$ in~\eqref{eq:bx-c} is the transmitted symbol at time $i$.

\item $T_1 > 2(B_1 - k)$

Since, $B_1-k > T_1 - (B_1-k)$, a SCo of parameters $(T_1 - (B_1-k),B_1-k)$ constructed in case (A) is not feasible and is thus replaced by a set of SCo codes.
For the associated values of $T_1,$ $B_1$ and $k$ we let
\begin{align}T_1-(B_1-k) = r(B_1-k) + q, \quad q < (B_1-k).\end{align}

\begin{itemize}
\item Let $$\mathcal{C}_{3,n} = (B_{3,n},T_{3,n}) = (B_1-k,B_1-k),~n=1,\dots,r,$$ be a set of $r$ SCo \emph{repetition} codes applied on the last $B_1 - k$ \emph{parity check} sub-symbols $(p^{\rm{I}}_{k}[i],\dots,p^{\rm{I}}_{B_1-1}[i])$ and repeating them to construct $r$ sets of parity check vectors each of size $B_1-k$ as follows,
\begin{align}
\label{eq:C_B}
\bp^{3,n}[i] = (p^{3,n}_{0}[i],\dots,p^{3,n}_{B_1-k-1}[i]) = (p^{\rm{I}}_{k}[i+n(B_1-k)],\dots,p^{\rm{I}}_{B_1-1}[i+n(B_1-k)]),
\end{align}
at each time, i.e., a $\cC_{3,n}$ code is a $(B_1-k,B_1-k)$ SCo repetition code shifted back by $(n+1)(B_1-k)$.
\item Let $\mathcal{C}_{3,{r+1}}$ be a $(B_{3,{r+1}},T_{3,{r+1}}) = (q,B_1-k)$ SCo again applied on the last $B_1 - k$ \emph{parity check} sub-symbols $(p^{\rm{I}}_{k}[i],\dots,p^{\rm{I}}_{B_1-1}[i])$ and then constructing $q$ parity checks $\bp^{3,{r+1}}[i] = (p^{3,{r+1}}_{0}[i],\dots,p^{3,{r+1}}_{q-1}[i])$ at each time by combining the last $B_1 - k$ \emph{parity check} sub-symbols, $(p^{\rm{I}}_{k}[.],\dots,p^{\rm{I}}_{B_1-1}[.])$, along the main diagonal.
\item Concatenate the set of streams $\bp^{3,n}[.]$ for $n=1,\dots,r$ and $\bp^{3,{r+1}}[.]$ after introducing a shift of $\Delta_{3,r+1} = -T_1$ in the later. The output symbol at time $i$ is,
\begingroup
\singlespacing
\begin{align}
\bx[i] =
\left[
\begin{array}{rl}
& \bs[i] \\
& p^{\rm{I}}_0[i] \\
& \vdots \\
& p^{\rm{I}}_{k-1}[i] \\
p^{\rm{I}}_{k}[i] &+ s_0[i-T_2] \\
& \vdots \\
p^{\rm{I}}_{B_1-1}[i] &+ s_{B_1-k-1}[i-T_2] \\
s_{B_1-k}[i-T_2] &+ \tilde{p}^3_{0}[i] \\
& \vdots \\
s_{T_1-1}[i-T_2] &+ \tilde{p}^3_{T_1-(B_1-k)-1}[i]
\end{array}
\right]
\end{align}
\endgroup
where
\begin{align}
(\tilde{p}^3_{0}[i],\dots,\tilde{p}^3_{T_1-(B_1-k)-1}[i]) = (\tilde{\bp}^{3,1}[i]\big{|}_i,\dots,\tilde{\bp}^{3,r}[i]\big{|}_i,\tilde{\bp}^{3,{r+1}}[i+T_1]\big{|}_i)
\end{align}
is the concatenation of the $r+1$ parity check sub-streams for the codes $\mathcal{C}_{3,n}$ for $n=1,\dots,r+1$, respectively. Since each of the first $r$ of these sub-streams is composed of $B_1-k$ parity check sub-symbols while the last of which is composed of $q$ parity check sub-symbols, then the $\bp^{3}[i]$ has a sum of $r(B_1-k) + q = T_1 - (B_1-k)$ parity check sub-symbols which will be denoted by the parity check sub-symbols of code $\cC_3$ (the set of codes $\{ \cC_{3,1},\cC_{3,2}, \dots, \cC_{3,{r+1}} \}$), and hence can be combined with the last $T_1 - (B_1-k)$ parity check sub-symbols of code $\cC_2$, $(p^{\rm{II}}_{B_1-k}[i],\dots,p^{\rm{II}}_{T_1-1}[i])$.
\end{itemize}
\end{enumerate}
\end{itemize}

\begin{figure}
\centering
\resizebox{\columnwidth}{!}{\includegraphics{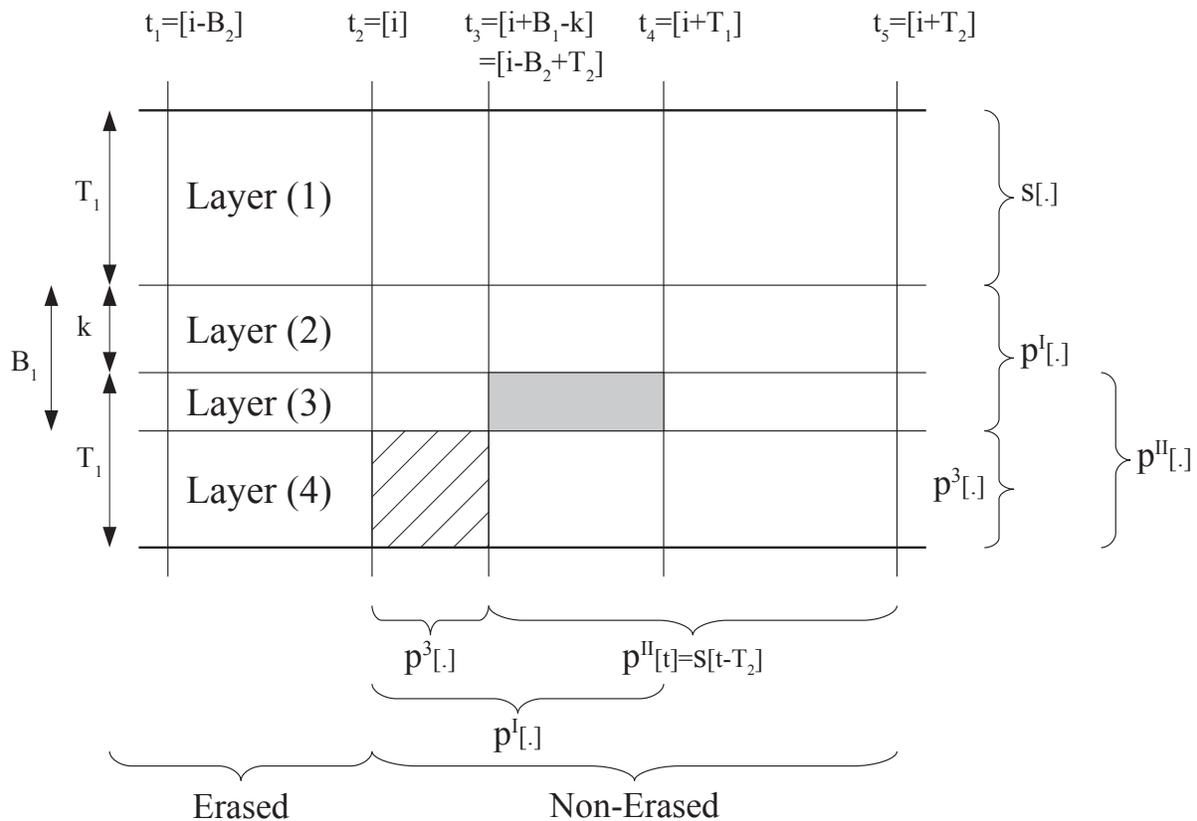}}
\caption{A graphical illustration of both the Encoding and Decoding Steps for a general point lying in the (e) Region. The labels on the right show the layers spanned by each set of parity check sub-symbols. The labels at the bottom show the intervals in which each set of parity check sub-symbols combine erased source sub-symbols.}
\label{fig:eRegion_CC}
\end{figure}

Since there are $T_1$ source sub-symbols and two streams of parity checks one with $B_1$ and the other with $T_1$ parity check sub-symbols for every $T_1$ source sub-symbols but partially overlapping in $B_1 - k$ rows, it follows that the rate of the code is $\frac{T_1}{2T_1 +B_1 - (B_1 - k)} = \frac{T_1}{2T_1 + k} = C_e$ (c.f. Fig.~\ref{fig:eRegion_CC}).\\

A graphical representation of such coding scheme is illustrated in Fig.~\ref{fig:eRegion_CC}.  The horizontal axis represents time while the vertical axis represents the index of sub-symbols in the channel symbol at each time instant. We divide each channel symbol into four layers,
\begin{itemize}
\item Layer (1) contains the $T_1$ source sub-symbols
\item Layer (2) contains the first $k$ of the $B_1$ parity check sub-symbols, $\bp^{\rm{I}}[\cdot]$ produced by code $\cC_1$
\item Layer (3) has the remaining $B_1-k$ parity check sub-symbols $\bp^{\rm{I}}[\cdot]$ of $\cC_1$
combined with the first $B_1-k$ of the the parity check sub-symbols, $\bp^{\rm{II}}[\cdot]$ produced by the repetition code, $\cC_2$.
\item Layer (4) has the remaining $T_1 - (B_1-k)$ parity check sub-symbols of $\cC_2$ combined with the parity checks of $\cC_3$.
\end{itemize}
Note that two overlaps between codes exist in this coding scheme. The first is between codes $\cC_1$ and $\cC_2$ and takes place in layer (3), while the second is between $\cC_2$ and $\cC_3$ and takes place in layer (4).


\subsection{Decoding of User 1}
A burst erasure of length $B_1$ in the interval $\mathcal{I}_1 = [i-B_1,i-1]$ can be directly recovered using the stream of parity checks $\bp^{\rm{I}}[\cdot]$ in the interval $[i,i+T_1-1]=[t_2,t_4)$ (c.f. Fig.~\ref{fig:eRegion_CC}) produced by code $\mathcal{C}_1$ within a delay of $T_1$. The  overlapping parity checks $\bp^{\rm{II}}[t] = \bs[t-T_2]$ in this interval consist of source symbols from the interval $\mathcal{I}_2 = [i-T_2,i+T_1-T_2-1] = [i-T_1-B_1-m,i-B_1-m-1]$ which are unerased (i.e., $\mathcal{I}_2 \cap \mathcal{I}_1 = \Phi$ since $m \geq 0$).

\subsection{Decoding of User 2}
Suppose that the symbols in the interval $\mathcal{I}_2 = [i-B_2,i-1]$ are erased by the channel of user 2.
We start by summarizing the main  decoding steps. Thereafter we describe each step in detail. 
\begin{itemize}
\item \textbf{Step (1) (Recovery of $\bp^{\rm{I}}[\cdot]$):} The parity checks of code ${\mathcal C}_3$, $\bp^{3}[\cdot]$ in the interval $[i,\dots,i+T_2-B_2-1]$ (in layer (4)) are capable of recovering the last $B_1-k$ sub-symbols of $\bp^{\rm{I}}[t]$ for $t \in \{ i+T_2-B_2, \dots, i+T_1-1 \}$ by time $t$.
\item \textbf{Step (2) (Removal of $\bp^{\rm{I}}[\cdot]$):} Subtract $\bp^{\rm{I}}[\cdot]$ in layer (3) starting at $i-B_2+T_2$.
\item \textbf{Step (3) (Removal of $\bp^{3}[\cdot]$):} Compute and subtract $\bp^{3}[\cdot]$ in layer (4) starting at $i-B_2+T_2$.
\item \textbf{Step (4) (Recovery using $\bp^{\rm{II}}[\cdot]$):} Use $\bp^{\rm{II}}[t]$ for $t \in \{ i+T_2-B_2,\dots, i+T_2-1 \}$ to recover the erased source symbols, $(\bs[i-B_2],\dots,\bs[i-1])$.
\end{itemize}

\subsubsection*{{\bf Step (1) (Recovery of $\bp^{\rm{I}}[\cdot]$)}}

Step (1) involves applying code $\cC_3$ in computing some missing parity checks $\bp^{\rm{I}}[t]$. This is the most elaborate step and is established in the following lemma.
\begin{lem}
\label{lem:step1}
The parity check sub-symbols $p^{\rm{I}}_j[t]$ for $t \in \{ i+T_2-B_2, \dots, i+T_1-1 \}$ and $j \in \{k,\dots,B_1-1 \}$ can be recovered using the parity check symbols $\bp^{3}[\cdot]$ in the interval $[i,\dots,i+T_2-B_2-1]$ (in layer (4)) by time $t$, i.e., with a zero delay.
\end{lem}
Since the proof  of Lemma~\ref{lem:step1} is rather long it is deferred to Appendix.~\ref{app:step1}.

\subsubsection*{{\bf Step (2) (Removal of $\bp^{\rm{I}}[\cdot]$)}}

Next we show that the parity check sub-symbols of $\cC_2$ in layer (3) are free of interference starting at $t_3 = i+B_1-k$. This is because the parity check sub-symbols of $\cC_1$ in the interval $[i+B_1-k,i+T_1-1] = [t_3,t_4-1]$ are recovered in Step (1) and those appearing at time $i+T_1$ and later are functions of unerased source symbols at times $i$ and later (this follows from the fact that a $(B,T)$ SCo code has a memory of $T$).

\subsubsection*{{\bf Step (3) (Removal of $\bp^{3}[\cdot]$)}}

We next claim that the rest of parity check sub-symbols of $\cC_2$ in layer (4) are also free of interference. In case (A) considered before, this follows immediately from the memory of the SCo code. The parity check sub-symbols of $\cC_3$ has a memory of $T_3 = B_1-k$ and thus these parity checks at time $i+B_1-k$ combine parity check sub-symbols of $\cC_1$ of time $i+B_1-k - T_3 + T_1 = i + T_1$ and later (where the addition of $T_1$ is due to the shift back applied on these parity checks). Moreover, we have shown in Step (2) that the parity check sub-symbols of $\cC_1$ at time $i+T_1$ and later combines only unerased source symbols from time $i$ and the claim follows. While for case (B), the same argument follows in the last $q$ rows. But for the first $r(B_1-k)$ rows of layer (4), the parity checks of $\cC_3$ are  repetition codes. These are  either recovered in Step (2) or contain only unerased source symbols.

\subsubsection*{{\bf Step (4) (Recovery using $\bp^{\rm{II}}[\cdot]$)}}

Step (4) uses the previous two steps to recover the parity check sub-symbols of $\cC_2$ in layers (3) and (4) starting at $\bp^{\rm{II}}[i+B_1-k] = \bs[i+B_1-k-T_2] = \bs[i-B_2]$ and thus the erased source-symbols can be recovered.

\section{The Converse For Region (e) (Theorem~\ref{thm:Region_e})}
\label{sec:Region_e_UB}
We want to prove that the capacity is at most $\frac{T_1}{2T_1+B_1+B_2-T_2}$ in the (e)-region defined by the inequalities $ B_2 \leq T_2 < B_2+B_1$ and $T_2 \geq T_1 + B_1$.

\begin{figure}
  \centering
    \subfigure[Step (1)]
    {
      \includegraphics[scale = 0.35]{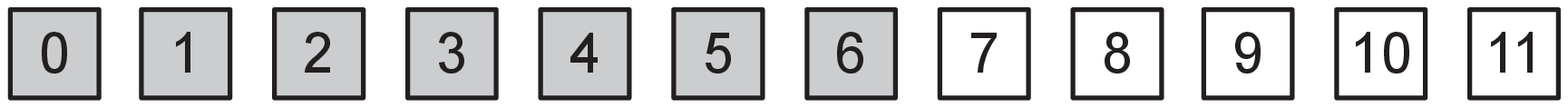}
			\label{fig:Converse_45710_a}
		}
		\subfigure[Step (2)]
    {
      \includegraphics[scale = 0.35]{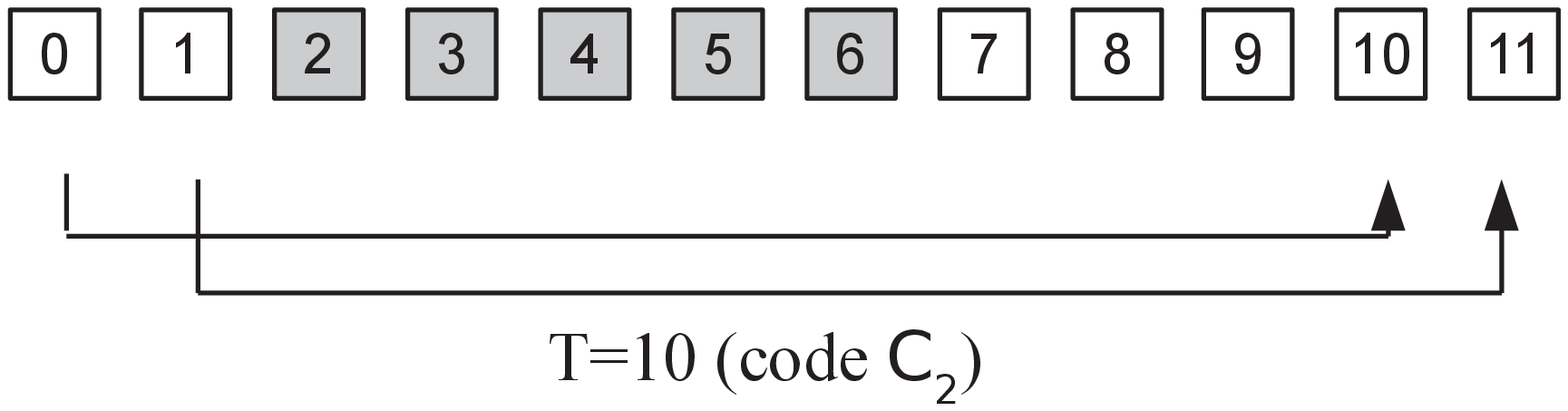}
			\label{fig:Converse_45710_b}
		}
		\subfigure[Step (3)]
		{
      \includegraphics[scale = 0.35]{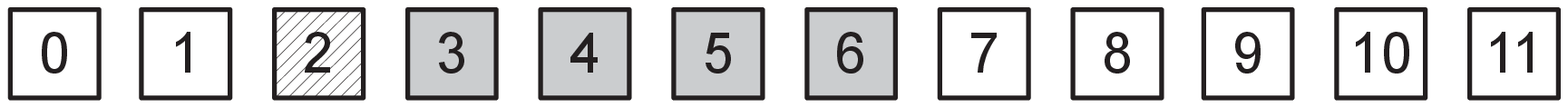}
			\label{fig:Converse_45710_c}
		}
		\subfigure[Step (4)]
		{
      \includegraphics[scale = 0.35]{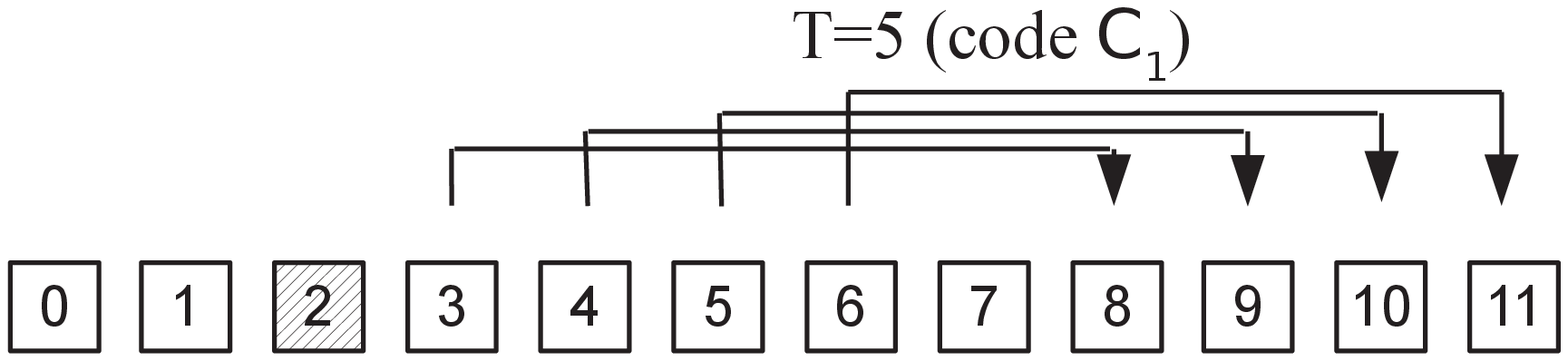}
			\label{fig:Converse_45710_d}
		}
  \caption{Main steps of finding the upper-bound for the $\{(4,5)-(7,10)\}$ point lying in Region (e) through one period illustration of the Periodic Erasure Channel. Grey and white squares resemble erased and unerased symbols respectively while hatched squares resemble symbols revealed to the receiver.} \label{fig:Converse_45710}
\end{figure}

We start by considering the example $\{ (4,5)-(7,10) \}$ illustrating the steps of the converse proof. We again use the periodic erasure channel strategy with a period of length $12$ and the first $7$ of which are erased. With $7$ erasures, code $\mathcal{C}_2 = (7,10)$ can recover the first two symbols at time $0$ and $1$ by time $10$ and $11$, respectively (c.f. Fig.~\ref{fig:Converse_45710_b}). Since code $\mathcal{C}_1 = (4,5)$ is not capable of recovering the remaining $5$ erasures, we reveal the first of which to the decoder. Now, $\mathcal{C}_1$ can recover the source symbols at times $3$ to $6$ by times $8$ to $11$, respectively (i.e., incurring a delay of $5$ symbols). Again with the assumption of systematic encoding, one can see that a rate of $5/11$ upper-bounds the capacity of this channel as $5$ channel symbols where able to decode $6$ of the erased source symbols.

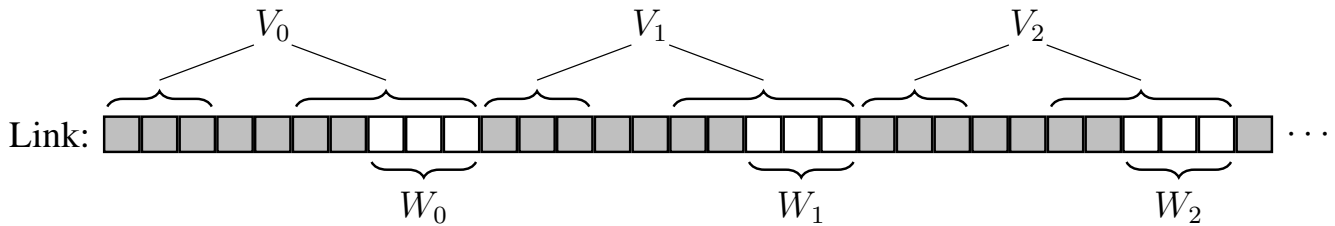
\begin{figure}[htbp]
	\centering
	\resizebox{\columnwidth}{!}{
	\begin{tikzpicture}[node distance=0mm]
		\node                       (x1start) {Link:};
		\node[esym, right = of x1start]  (x100) {};
		\node[esym, right = of x100]     (x101) {};
		\node[esym, right = of x101]     (x102) {};
		\node[esym, right = of x102]     (x103) {};
		\node[esym, right = of x103]     (x104) {};
		\node[esym, right = of x104]     (x105) {};
		\node[esym, right = of x105]     (x106) {};
		\node[usym, right = of x106]     (x107) {};
		\node[usym, right = of x107]     (x108) {};
		\node[usym, right = of x108]     (x109) {};
		\node[esym, right = of x109]     (x110) {};
		\node[esym, right = of x110]     (x111) {};
		\node[esym, right = of x111]     (x112) {};
		\node[esym, right = of x112]     (x113) {};
		\node[esym, right = of x113]     (x114) {};
		\node[esym, right = of x114]     (x115) {};
		\node[esym, right = of x115]     (x116) {};
		\node[usym, right = of x116]     (x117) {};
		\node[usym, right = of x117]     (x118) {};
		\node[usym, right = of x118]     (x119) {};
		\node[esym, right = of x119]     (x120) {};
		\node[esym, right = of x120]     (x121) {};
		\node[esym, right = of x121]     (x122) {};
		\node[esym, right = of x122]     (x123) {};
		\node[esym, right = of x123]     (x124) {};
		\node[esym, right = of x124]     (x125) {};
		\node[esym, right = of x125]     (x126) {};
		\node[usym, right = of x126]     (x127) {};
		\node[usym, right = of x127]     (x128) {};
		\node[usym, right = of x128]     (x129) {};
		\node[esym, right = of x129]     (x130) {};
		\node      [right = of x130]     (x1end) {$\cdots$};

		\braceup{x100}{x102}{1mm}{};
		\braceup{x105}{x109}{1mm}{};
		\braceup{x110}{x112}{1mm}{};
		\braceup{x115}{x119}{1mm}{};
		\braceup{x120}{x122}{1mm}{};
		\braceup{x125}{x129}{1mm}{};

		\draw([yshift=4mm]x101.north) -- ([yshift=8mm]x104.north west);
		\draw([yshift=4mm]x107.north) -- ([yshift=8mm]x104.north east);
		\draw([yshift=7mm]x104.north) node[anchor=south] {$V_0$};
		\draw([yshift=4mm]x111.north) -- ([yshift=8mm]x114.north west);
		\draw([yshift=4mm]x117.north) -- ([yshift=8mm]x114.north east);
		\draw([yshift=7mm]x114.north) node[anchor=south] {$V_1$};
		\draw([yshift=4mm]x121.north) -- ([yshift=8mm]x124.north west);
		\draw([yshift=4mm]x127.north) -- ([yshift=8mm]x124.north east);
		\draw([yshift=7mm]x124.north) node[anchor=south] {$V_2$};

		\bracedn{x107}{x109}{-1mm}{$W_0$};
		\bracedn{x117}{x119}{-1mm}{$W_1$};
		\bracedn{x127}{x129}{-1mm}{$W_2$};
		

	\end{tikzpicture}}
	\caption{The periodic erasure channel used to prove an upper bound on capacity in region (e) indicating which symbols are in groups $W_i$ and $V_i$. Grey and white squares resemble erased and unerased symbols respectively.}
	\label{fig:PEC_eRegion}
\end{figure}

For the general case, the periodic erasure channel to be used is shown in Fig. \ref{fig:PEC_eRegion}, where each period has $B_2$ erasures followed by $T_1$ non-erasures. We can assign
\begin{align*}
	&a = T_1 + B_2 - T_2, \quad
	b = B_2 - B_1, \quad
	c = B_2, \quad
	d = B_2 + T_1 \quad \text{(period length)}, \\
	&W_i = \bx\sqstack{id+c}{(i+1)d-1}, \quad
	V_i = \bs\sqstack{id}{id+a-1} \bs\sqstack{id+b}{(i+1)d-1}.
\end{align*}
The idea behind the converse proof is similar to before, but instead we have two decoding functions to use.

We use the decoder of receiver 2 to recover $\bs\sqstack{0}{a-1}$ within a delay of $T_2$ using the channel packets $\bx\sqstack{c}{d-1}$. We then reveal the channel symbols $\bx\sqstack{a}{b-1}$. The decoder of receiver 1 can now be used to recover the next $B_1$ source packets, which are the packets $\bs\sqstack{b}{c-1}$, using $\bx\sqstack{c}{d-1}$ again. In general, we may not have a systematic code, so even if $\bx\sqstack{c}{d-1}$ is received, we may not be able to recover the corresponding source packet $\bs\sqstack{c}{d-1}$. Instead, $\bs\sqstack{c}{d-1}$ can be recovered using the second decoder and the first and second sets of channel packets that are not erased, i.e. $\bx\sqstack{c}{d-1}$ and $\bx\sqstack{d+c}{2d-1}$.

So far, we have recovered $(T_1+B_2-T_2)+B_1+T_1 = 2T_1+B_1+B_2-T_2$ source packets, using $2T_1$ channel packets. We do not include the source packets $\bs\sqstack{a}{b-1}$, because it cannot be decoded from the information in the unerased channel packets. The channel has a period of $B_2+T_1$ packets, and if we had $n$ periods, then we would be able to recover $n(2T_1+B_1+B_2-T_2)$ source packets using $(n+1)T_1$ channel packets. Therefore, we can suppose that the upper bound on the multicast streaming capacity is given by
\begin{align}
	n \cdot (2T_1+B_1+B_2-T_2) \cdot H(\bs) &\leq (n+1)\cdot T_1 \cdot H(\bx) \nonumber \\
	R = \frac{H(\bs)}{H(\bx)} &\leq \frac{n+1}{n} \cdot \frac{T_1}{2T_1+B_1+B_2-T_2} \nonumber \\
	&\xrightarrow{n\rightarrow \infty} \frac{T_1}{2T_1+B_1+B_2-T_2}
\end{align}

The more formal proof is given below.

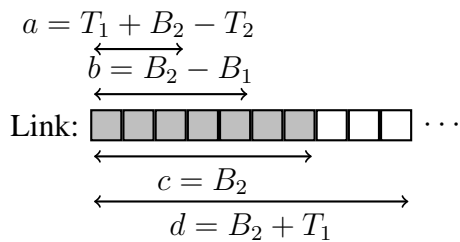
\begin{figure}[htbp]
	\centering
	\begin{tikzpicture}[node distance=0mm]
		\node                       (x1start) {Link:};
		\node[esym, right = of x1start]  (x100) {};
		\node[esym, right = of x100]     (x101) {};
		\node[esym, right = of x101]     (x102) {};
		\node[esym, right = of x102]     (x103) {};
		\node[esym, right = of x103]     (x104) {};
		\node[esym, right = of x104]     (x105) {};
		\node[esym, right = of x105]     (x106) {};
		\node[usym, right = of x106]     (x107) {};
		\node[usym, right = of x107]     (x108) {};
		\node[usym, right = of x108]     (x109) {};
		\node      [right = of x109]     (x1end) {$\cdots$};

		\dimup{x100}{x102}{8mm}{$a = T_1+B_2-T_2$};
		\dimup{x100}{x104}{2mm}{$b = B_2 - B_1$};
		\dimdn{x100}{x106}{-2mm}{$c = B_2$};
		\dimdn{x100}{x109}{-8mm}{$d = B_2 + T_1$};

	\end{tikzpicture}
	\caption{One period of the periodic erasure channel in Fig.~\ref{fig:PEC_eRegion}, with labels.}
	\label{fig:OnePeriod_eRegion}
\end{figure}

\begin{proof}
From the $(B_1, T_1)$ decoder, we have for $i \geq 0$:
\begin{align}
	H\Bigl(\bs\sqstack{id+b}{id+c-1} \Big| \bx\sqstack{0}{id+b-1} W_i\Bigr) = 0 \label{eq:dldr_e_region_rule2}
\end{align}
From the $(B_2, T_2)$ decoder, we have for $i \geq 0$:
\begin{align}
	H\Bigl(\bs\sqstack{id}{id+a-1} \Big| \bx\sqstack{0}{id-1} W_i\Bigr) &= 0 \label{eq:dldr_e_region_rule1} \\
	H\Bigl(\bs\sqstack{id+c}{(i+1)d-1} \Big| \bx\sqstack{0}{id+c-1} W_i^{i+1}\Bigr) &= 0. \label{eq:dldr_e_region_rule3}
\end{align}
We want to use mathematical induction to prove that for $n \geq 0$
\begin{equation}
	H(W_0^{n}) \geq H(V_0^{n-1}) + H\Bigl(W_n \Big| V_0^{n-1} \bx\sqstack{0}{nd-1}\Bigr). \label{eq:dldr_e_region_target_ineq}
\end{equation}
The base case for \eqref{eq:dldr_e_region_target_ineq} is given by substituting $n=0$ into it:
\begin{align}
	H(W_0) \geq H(V_0^{-1}) + H\Bigl(W_0 \Big| V_0^{-1} \bx\sqstack{0}{-1}\Bigr) \geq H(W_0)
\end{align}
which is obviously true. Let us assume that \eqref{eq:dldr_e_region_target_ineq} is true for $n=k$. This gives:
\begin{equation}
\label{eq:eRegion_nk}
	H(W_0^{k}) \geq H(V_0^{k-1}) + H\Bigl(W_k \Big| V_0^{k-1} \bx\sqstack{0}{kd-1}\Bigr).
\end{equation}
We can manipulate the expression in two parts. In the first part, we use $W_k$ to recover the source packets $\bs\sqstack{kd}{kd+a-1}$ and $\bs\sqstack{kd+b}{kd+c-1}$ and one can write,
\begin{align}
\label{eq:eRegion_a}
	H(W_0^{k}) \geq H\Bigl(V_0^{k-1} \bs\sqstack{kd}{kd+a-1} \bs\sqstack{kd+b}{kd+c-1}\Bigr) + H\Bigl(W_{k} \Big| V_0^{k-1} \bs\sqstack{kd}{kd+a-1} \bs\sqstack{kd+b}{kd+c-1} \bx\sqstack{0}{kd+c-1}\Bigr),
\end{align}
where the first term on the R.H.S. gives the entropy of the source symbols recovered in previous periods $V_0^{k-1}$ as well as the source symbols recovered in this step due to the availability of $W_k$. The second term gives the remaining ambiguity in $W_k$ to be used in the next step. The detailed steps from~\eqref{eq:eRegion_nk} to~\eqref{eq:eRegion_a} is shown in Appendix.~\ref{app:eRegion}.

In the second part, we add $H(W_{k+1} | W_0^k)$ to both sides of the inequality. Because the channel code is not necessarily systematic, we will use the additional channel packets in $W_{k+1}$ to help decode the source packets $\bs\sqstack{kd+c}{(k+1)d-1}$. The corresponding steps provided in Appendix.~\ref{app:eRegion} shows that,
\begin{align}
\label{eq:eRegion_b}
	H(W_0^{k+1}) \geq H(V_0^{k}) + H\Bigl(W_{k+1} \Big| V_0^{k} \bx\sqstack{0}{(k+1)d-1}\Bigr)
\end{align}

The working in \eqref{eq:dldr_e_region_ind_step} shows that if \eqref{eq:dldr_e_region_target_ineq} is true for $n=k$, then it is also true for $n=k+1$. By induction, \eqref{eq:dldr_e_region_target_ineq} is true for $n \geq 0$. Finally,
\begin{align*}
	H(W_0^{n}) \geq H(V_0^{n-1}) + H\Bigl(W_{n} \Big| V_0^{n-1} \bx\sqstack{0}{nd-1}\Bigr) \geq H(V_0^{n-1}).
\end{align*}
Using the fact that all of the channel packets have the same entropy, and all of the source packets have the same entropy, we can continue to get
\begin{align}
	H(W_0^{n}) &\geq H(V_0^{n-1}) \nonumber \\
	(n+1) \cdot T_1 \cdot H(\bx) &\geq n \cdot (2T_1 + B_1 + B_2 - T_2) \cdot H(\bs) \nonumber \\
	\frac{n+1}{n} \cdot \frac{T_1}{2T_1+B_1 + B_2 - T_2} &\geq \frac{H(\bs)}{H(\bx)}.
\end{align}
Finally, we get
\begin{equation}
	R = \frac{H(\bs)}{H(\bx)} \leq \frac{T_1}{2T_1+B_1 + B_2 - T_2}. \; (\mathrm{as} \; n \rightarrow \infty) \label{eq:dldr_e_region_final}
\end{equation}
Therefore, any $(B_1,T_1),(B_2,T_2)$ code in the (e)-region must satisfy \eqref{eq:dldr_e_region_final}.
\end{proof}



\section{Achievability Scheme In Region (f) At $T_1 = B_1$ (Theorem~\ref{thm:Region_f1})}
\label{sec:Region_f_CC}

\begingroup
\singlespacing
\everymath{\scriptstyle}
\begin{table}[t]
\begin{tabular}{x{65pt}|x{65pt}|x{65pt}|x{65pt}|x{65pt}|x{65pt}}
$\displaystyle [i]$ & $\displaystyle [i+1]$ & $\displaystyle [i+2]$ & $\displaystyle [i+3]$ & $\displaystyle [i+4]$ & $\displaystyle [i+5]$\\
\hline
\fboxrule=1pt \fbox{$s_0[i]$} & $s_0[i+1]$ & $s_0[i+2]$ & $s_0[i+3]$ & $s_0[i+4]$ & $s_0[i+5]$ \\
\fboxrule=1pt \fbox{$s_1[i]$} & $s_1[i+1]$ & $s_1[i+2]$ & $s_1[i+3]$ & $s_1[i+4]$ & $s_1[i+5]$ \\
\hline
$s_0[i-4]$ & $s_0[i-3]$ & $s_0[i-2]$ & $s_0[i-1]$ & \fboxrule=1pt \fbox{$s_0[i]$} & $s_0[i+1]$ \\
$s_1[i-4]$ & $s_1[i-3]$ & $s_1[i-2]$ & $s_1[i-1]$ & \fboxrule=1pt \fbox{$s_1[i]$} & $s_1[i+1]$ \\
$s_0[i-6] + s_1[i-5]$ & $s_0[i-5] + s_1[i-4]$ & $s_0[i-4] + s_1[i-3]$ & $s_0[i-3] + s_1[i-2]$ & $s_0[i-2] + s_1[i-1]$ & $s_0[i-1] + s_1[i]$ \\
\hline\end{tabular}
\everymath{\displaystyle}
\caption{Mu-SCo Code Construction for $(B_1,T_1) = (4,4)$ and $(B_2,T_2) = (5,6)$. This point achieves the upper-bound given in Theorem~\ref{thm:Region_f1} as $T_1=B_1=4$.}
\label{table:Code4456}
\end{table}
\endgroup

We begin with an example of  $\{ (4,4)-(5,6) \}$ Mu-SCo construction of rate $2/5$, as shown in Table~\ref{table:Code4456}. A $(4,4)$ SCo repetition code is then applied resulting in the first two rows of parity checks and then a $(B_2-B_1,T_2-T_1)=(1,2)$ SCo is applied and the resulting parity checks are shifted by $T_1=4$ forming the last row. 
Note that the first user can recover from any burst erasure of length $4$ within a delay of $4$ symbols using the first two rows of parity check sub-symbols. For the second user, assume a burst erasure of length $5$ takes place from time $i-5$ to $i-1$. Notice that  user $2$ recovers $s_1[i-5]$
and $s_0[i-5]$ respectively from the last two parity checks at time $t=i+1$ i.e., with a delay of $T_2=6$.  The rest of the erased source symbols are recovered with a delay of $T_1=4$  using the repetition code.

\subsection{Code Construction}
Our proposed code construction, which achieves the minimum delay for user $1$ i.e., $T_1=B_1$ is as folows
\begin{itemize}
\item Let $\mathcal{C}_1$ be the single user ${(B_1,T_1) = (T_1,T_1)}$ SCo obtained by splitting each source symbol $\bs[i]$ into ${(T_2 - B_1) = (T_2 - T_1)}$ sub-symbols $$\bs[i] = (s_0[i],\dots,s_{T_2-T_1-1}[i])$$ and repeating them to produce $(T_2 - T_1)$ parity check sub-symbols.
\begin{align}
\bp^{\rm{I}}[i] = (p^{\rm{I}}_0[i],\dots,p^{\rm{I}}_{T_2-T_1-1}[i]) = (s_0[i-T_1],\dots,s_{T_2-T_1-1}[i-T_1]) = \bs[i-T_1].
\end{align}
\item Let $\mathcal{C}_2$ be a $(B_2-B_1,T_2-T_1)$ SCo also obtained by splitting each source symbol $\bs[i]$ into $(T_2 - B_1) = (T_2 - T_1)$ sub-symbols $(s_0[i],\dots,s_{T_2-T_1-1}[i])$ and then constructing  $(B_2 - B_1)$ parity checks $\bp^{\rm{II}}[i] = (p^{\rm{II}}_0[i],\dots,p^{\rm{II}}_{B_2-B_1-1}[i])$ at each time by combining the source sub-symbols along the main diagonal.
\item Concatenate the two streams $p^{\rm{I}}[\cdot]$ and $p^{\rm{II}}[\cdot]$ after introducing a shift of $T_1$ in the second stream. The output symbol at time $i$ is $\bx[i] = (\bs[i],\bp^{\rm{I}}[i],\bp^{\rm{II}}[i-T_1])$.
\end{itemize}

Since there are $T_2 - T_1$ and $B_2-B_1$ parity check sub-symbols for every $T_2-T_1$ source sub-symbols, it follows that the rate of the code is $\frac{T_2-T_1}{2(T_2-T_1) + (B_2-B_1)} = C_f^+$.

\subsection{Decoding at User 1}
A burst erasure of length $B_1$ can be directly recovered using the stream of parity checks $p^{\rm{I}}[\cdot]$ produced by code $\mathcal{C}_1$ within a delay of $T_1$. Recall that this immediately follows since the parity checks of the two codes are concatenated and not added.

\subsection{Decoding at User 2}
Suppose that the symbols at time $i-B_2,\dots,i-1$ are erased by the channel of user 2. We first show how the receiver can recover $\bs[t]$ for $t \in [i-B_2,i-B_1-1]$ at time  ${t+T_2}$.  To recover $\bs[t],$ the code $\cC_2$ which is a $(T_2-T_1, B_2-B_1)$ code, can be used provided  that the corresponding parity checks starting at time $i-B_1$ are available. Due to the forward shift of $T_1=B_1$ applied in our construction, these parity checks appear starting at time ${t=i}$ and are clearly not erased. Secondly for the recovery of $\bs[t]$ we also need the source symbols in the interval $[i-B_1, t+T_2-T_1]$. The $\cC_1$ repetition code guarantees that these are in fact available by time $t+T_2$. This shows that all the erased symbols in the interval $[i-B_2, i-B_1-1]$ can be recovered. The remaining symbols in the interval $[i-B_1, i-1]$ are recovered using the $\cC_1$ repetition code.

\section{ Upper-Bound For Region (f) (Theorem~\ref{thm:Region_f1})}
\label{sec:Region_f_UB1}

\begin{figure}[htbp]
	\centering
	\resizebox{\columnwidth}{!}{
	\begin{tikzpicture}[node distance=0mm]
		\node                       (x1start) {Link:};
		\node[esym, right = of x1start]  (x100) {};
		\node[esym, right = of x100]     (x101) {};
		\node[esym, right = of x101]     (x102) {};
		\node[esym, right = of x102]     (x103) {};
		\node[esym, right = of x103]     (x104) {};
		\node[esym, right = of x104]     (x105) {};
		\node[esym, right = of x105]     (x106) {};
		\node[usym, right = of x106]     (x107) {};
		\node[usym, right = of x107]     (x108) {};
		\node[usym, right = of x108]     (x109) {};
		\node[esym, right = of x109]     (x110) {};
		\node[esym, right = of x110]     (x111) {};
		\node[esym, right = of x111]     (x112) {};
		\node[esym, right = of x112]     (x113) {};
		\node[esym, right = of x113]     (x114) {};
		\node[esym, right = of x114]     (x115) {};
		\node[esym, right = of x115]     (x116) {};
		\node[usym, right = of x116]     (x117) {};
		\node[usym, right = of x117]     (x118) {};
		\node[usym, right = of x118]     (x119) {};
		\node[esym, right = of x119]     (x120) {};
		\node[esym, right = of x120]     (x121) {};
		\node[esym, right = of x121]     (x122) {};
		\node[esym, right = of x122]     (x123) {};
		\node[esym, right = of x123]     (x124) {};
		\node[esym, right = of x124]     (x125) {};
		\node[esym, right = of x125]     (x126) {};
		\node[usym, right = of x126]     (x127) {};
		\node[usym, right = of x127]     (x128) {};
		\node[usym, right = of x128]     (x129) {};
		\node[esym, right = of x129]     (x130) {};
		\node      [right = of x130]     (x1end) {$\cdots$};

		\braceup{x100}{x104}{1mm}{};
		\braceup{x107}{x109}{1mm}{};
		\braceup{x110}{x114}{1mm}{};
		\braceup{x117}{x119}{1mm}{};
		\braceup{x120}{x124}{1mm}{};
		\braceup{x127}{x129}{1mm}{};
		
		\draw([yshift=4mm]x102.north) -- ([yshift=8mm]x105.north west);
		\draw([yshift=4mm]x108.north) -- ([yshift=8mm]x105.north east);
		\draw([yshift=7mm]x105.north) node[anchor=south] {$V_0$};
		\draw([yshift=4mm]x112.north) -- ([yshift=8mm]x115.north west);
		\draw([yshift=4mm]x118.north) -- ([yshift=8mm]x115.north east);
		\draw([yshift=7mm]x115.north) node[anchor=south] {$V_1$};
		\draw([yshift=4mm]x122.north) -- ([yshift=8mm]x125.north west);
		\draw([yshift=4mm]x128.north) -- ([yshift=8mm]x125.north east);
		\draw([yshift=7mm]x125.north) node[anchor=south] {$V_2$};
		
		\bracedn{x107}{x109}{-1mm}{$W_0$};
		\bracedn{x117}{x119}{-1mm}{$W_1$};
		\bracedn{x127}{x129}{-1mm}{$W_2$};


	\end{tikzpicture}}
	\caption{The periodic erasure channel used to prove the first upper bound in region (f) showing the locations of the symbols in groups $V_i$ and $W_i$. Grey and white squares resemble erased and unerased symbols respectively.}
	\label{fig:PEC_f1Region}
\end{figure}
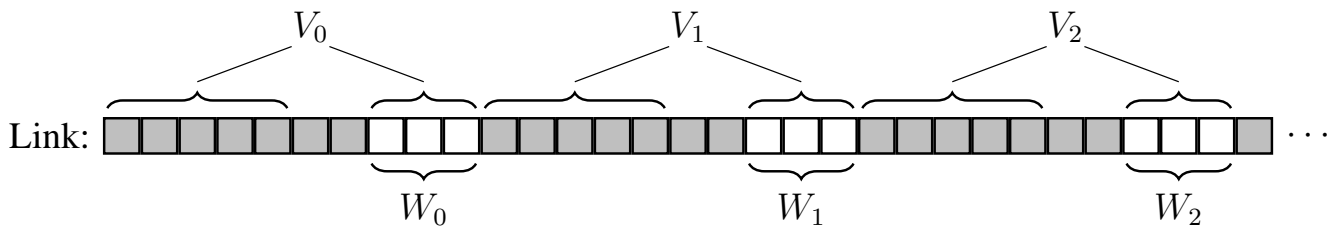

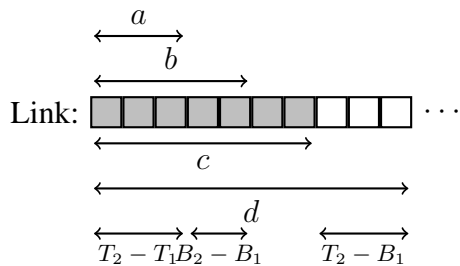
\begin{figure}[htbp]
	\centering
	\begin{tikzpicture}[node distance=0mm]
		\node                       (x1start) {Link:};
		\node[esym, right = of x1start]  (x100) {};
		\node[esym, right = of x100]     (x101) {};
		\node[esym, right = of x101]     (x102) {};
		\node[esym, right = of x102]     (x103) {};
		\node[esym, right = of x103]     (x104) {};
		\node[esym, right = of x104]     (x105) {};
		\node[esym, right = of x105]     (x106) {};
		\node[usym, right = of x106]     (x107) {};
		\node[usym, right = of x107]     (x108) {};
		\node[usym, right = of x108]     (x109) {};
		\node      [right = of x109]     (x1end) {$\cdots$};

		\dimup{x100}{x102}{8mm}{$a$};
		\dimup{x100}{x104}{2mm}{$b$};
		\dimdn{x100}{x106}{-2mm}{$c$};
		\dimdn{x100}{x109}{-8mm}{$d$};

		\dimdn{x107}{x109}{-14mm}{\footnotesize$T_2-B_1$\normalsize};

		\dimdn{x100}{x102}{-14mm}{\footnotesize$T_2-T_1$\normalsize};
		\dimdn{x103}{x104}{-14mm}{\footnotesize$B_2-B_1$\normalsize};
	\end{tikzpicture}
	\caption{One period of the periodic erasure channel in Fig.~\ref{fig:PEC_f1Region}, with labels}
	\label{fig:OnePeriod_f1Region}
\end{figure}

The converse proof for region (f) is similar to the proof for region (e). We shall use Fig.~\ref{fig:PEC_f1Region} and~\ref{fig:OnePeriod_f1Region} to illustrate the periodic erasure channel used in this proof. Each period, in this case, contains $B_2$ erasures followed by $T_2-B_1$ non-erasures, for a total of $B_2+T_2-B_1$ symbols.

The first $B_2-B_1$ source symbols can be recovered with code $\mathcal{C}_2$, from $\bx\sqstack{B_2}{B_2+T_2-B_1}$, which are the $T_2-B_1$ unerased channel symbols. We can see that $s_0$ is recovered at time $T_2$, while $s_{B_2-B_1-1}$ is recovered at time $B_2+T_2-B_1-1$. Code $\mathcal{C}_1$ recovers the next $T_2-T_1$ source symbols, which is $\bs\sqstack{B_2-B_1}{B_2-B_1+T_2-T_1-1}$. We then reveal the remaining channel symbols in the block of $B_2$ erased symbols, which are the symbols $\bx\sqstack{B_2-B_1+T_2-T_1}{B_2-1}$. Finally, code $\mathcal{C}_2$ is used to recover $\bs\sqstack{B_2}{B_2+T_2-B_1-1}$, using two sets of $T_2-B_1$ unerased channel symbols, which are $\bx\sqstack{B_2}{B_2+T_2-B_1-1}$ and $\bx\sqstack{2B_2+T_2-B_1}{2B_2+2T_2-2B_1-1}$.

In this one period of $B_2+T_2-B_1$ symbols, we have recovered $\bs\sqstack0{B_2-B_1-1}$, $\bs\sqstack{B_2-B_1}{B_2-B_1+T_2-T_1-1}$ and $\bs\sqstack{B_2}{B_2+T_2-B_1}$. This is a total of $2(T_2-B_1) + (B_2-T_1)$ source symbols recovered by $2(T_2-B_1)$ channel symbols. We can extrapolate that $n(2(T_2-B_1)+(B_2-T_1))$ source symbols can be recovered by $(n+1)(T_2-B_1)$ channel symbols. As in region (d) proof, we can suppose that the upper bound on the capacity is:
\begin{align}
	n (2(T_2-B_1)+(B_2&-T_1) H(\bs) \leq (n+1)(T_2-B_1) H(\bx) \nonumber \\
	\frac{H(\bs)}{H(\bx)} &\leq \frac{(n+1)(T_2-B_1)}{n(2(T_2-B_1)+(B_2-T_1))} \nonumber \\
	&\xrightarrow{n\rightarrow \infty} \frac{T_2-B_1}{2(T_2-B_1)+(B_2-T_1)} \nonumber \\
	\therefore C_f^+ &= \frac{T_2-B_1}{2(T_2-B_1)+(B_2-T_1)} \label{eq:e_conv_ub}
\end{align}

For the formal proof, we assign the following:


\begin{align*}
	&a = B_2 - B_1, \quad
	b = B_2 - B_1 + T_2 - T_1, \quad
	c = B_2, \quad
	d = B_2 + T_2 - B_1 \quad \text{(period length)}, \\
	&W_i = \bx\sqstack{id+c}{(i+1)d-1}, \quad
	V_i = \bs\sqstack{id}{id+b-1} \bs\sqstack{id+c}{(i+1)d-1}.
\end{align*}
From code $\mathcal{C}_1$, we have for $i\geq0$:
\begin{align}
	&H\Bigl(\bs\sqstack{id+a}{id+b-1}  \Big|  \bx\sqstack{0}{id+a-1} W_i\Bigr) = 0 \label{eq:e_rule2}
\end{align}
From code $\mathcal{C}_2$, we have for $i \geq 0$:
\begin{align}
	&H\Bigl(\bs\sqstack{id}{id+a-1}  \Big|  \bx\sqstack{0}{id-1} W_i\Bigr) = 0 \label{eq:e_rule1} \\
	&H\Bigl(\bs\sqstack{id+c}{(i+1)d-1}  \Big|  \bx\sqstack{0}{id+c-1} W_i^{i+1}\Bigr) = 0. \label{eq:e_rule3}
\end{align}
We want to show, using mathematical induction, that for $n \geq 0$
\begin{align}
	H(W_0^{n}) \geq H(V_0^{n-1}) + H\Bigl(W_{n}  \Big|  V_0^{n-1} \bx\sqstack{0}{nd-1}\Bigr). \label{eq:e_conv_induction}
\end{align}
The base case for \eqref{eq:e_conv_induction} is given by:
\begin{align}
H(W_0) = H(V_0^{-1}) + H\Bigl(W_0  \Big|  V_0^{-1} \bx\sqstack{0}{-1} \Bigr) \geq H(W_0)
\end{align}
which is true. For the induction step, we assume~\eqref{eq:e_conv_induction} is true for $n=k$,
\begin{align}
\label{eq:f1Region_nk}
	H(W_0^{n}) \geq H(V_0^{n-1}) + H\Bigl(W_{n}  \Big|  V_0^{n-1} \bx\sqstack{0}{nd-1}\Bigr).
\end{align}

The second term of the R.H.S. can be used to recover $\bs\sqstack{kd}{kd+a-1}$ and $\bs\sqstack{kd+a}{kd+b-1}$ through codes $\cC_2$ and $\cC_1$, respectively. The corresponding entropy manipulations are provided in Appendix.~\ref{app:f1Region} and the following is deduced,
\begin{align}
\label{eq:f1Region_a}
	H(W_0^{k}) &\geq H\Bigl(V_0^{k-1} \bs\sqstack{kd}{kd+b-1}\Bigr) + H\Bigl(W_{k}  \Big|  V_0^{k-1} \bs\sqstack{kd}{kd+b-1} \bx\sqstack{0}{kd+c-1}\Bigr)
\end{align}

Next, we add $H(W_{k+1}|W_0^k)$ to both sides and show that the newly added $W_{k+1}$ is capable of recovering the source symbols $\bs\sqstack{kd+c}{(k+1)d-1}$ corresponding to $W_k$,
\begin{align}
\label{eq:f1Region_b}
H(W_0^{k+1}) \geq H(V_0^{k}) + H\Bigl(W_{k+1}  \Big|  V_0^{k} \bx\sqstack{0}{(k+1)d-1}\Bigr)
\end{align}
The working out of~\eqref{eq:f1Region_b} is provided in Appendix.~\ref{app:f1Region}.

The working in~\eqref{eq:f1Region_a} and~\eqref{eq:f1Region_b} shows that if~\eqref{eq:e_conv_induction} is true for $n=k$, then it is true for $n=k+1$. By induction,~\eqref{eq:e_conv_induction} is true for $n \geq 0$. Therefore,
\begin{align*}
	H(W_0^{n}) \geq H(V_0^{n-1}) + H\Bigl(W_{n}  \Big|  V_0^{n-1} \bx\sqstack{0}{nd-1}\Bigr) \geq H(V_0^{n-1}).
\end{align*}
We can use the fact that the source symbols have the same entropy and the same for channel symbols to obtain:
\begin{align}
H(W_0^{n}) &\geq H(V_0^{n-1}) \nonumber \\
	(n+1) \cdot (T_2-B_1) \cdot H(\bx) &\geq n \cdot (2(T_2-B_1)+(B_2-T_1)) \cdot H(\bs) \nonumber \\
\end{align}
In other words,
\begin{equation}
	R = \frac{H(\bs)}{H(\bx)} \leq \frac{T_2-B_1}{2(T_2-B_1)+(B_2-T_1)}. \; (\mathrm{as} \; n \rightarrow \infty) \label{eq:upper_bound_f1}
\end{equation}
Therefore,~\eqref{eq:upper_bound_f1} governs any $\{ (B_1,T_1),(B_2,T_2) \}$ code in the (f)-region.



\section{Code Construction For Region (f) At $T_2 = B_2$ (Theorem~\ref{thm:Region_f_T2B2})}
\label{sec:Region_f_CC2}

\begingroup
\singlespacing
\everymath{\scriptstyle}
\begin{table}[t]
\begin{tabular}{x{65pt}|x{65pt}|x{65pt}|x{65pt}|x{65pt}|x{65pt}}
$\displaystyle [i]$ & $\displaystyle [i+1]$ & $\displaystyle [i+2]$ & $\displaystyle [i+3]$ & $\displaystyle [i+4]$ & $\displaystyle [i+5]$\\
\hline
\fboxrule=1pt \fbox{$s_0[i]$} & $s_0[i+1]$ & $s_0[i+2]$ & $s_0[i+3]$ & $s_0[i+4]$ & $s_0[i+5]$ \\
\fboxrule=1pt \fbox{$s_1[i]$} & $s_1[i+1]$ & $s_1[i+2]$ & $s_1[i+3]$ & $s_1[i+4]$ & $s_1[i+5]$ \\
\fboxrule=1pt \fbox{$s_2[i]$} & $s_2[i+1]$ & $s_2[i+2]$ & $s_2[i+3]$ & $s_2[i+4]$ & $s_2[i+5]$ \\
\hline
$s_0[i-3] + s_2[i-1]$ & $s_0[i-2] + s_2[i]$ & $s_0[i-1] + s_2[i+1]$ & $s_0[i] + s_2[i+2]$ & $s_0[i+1] + s_2[i+3]$ & $s_0[i+2] + s_2[i+4]$ \\
$s_1[i-3] + s_2[i-2]$ & $s_1[i-2] + s_2[i-1]$ & $s_1[i-1] + s_2[i]$ & $s_1[i] + s_2[i+1]$ & $s_1[i+1] + s_2[i+2]$ & $s_1[i+2] + s_2[i+3]$ \\
$s_0[i-4]$ & $s_0[i-3]$ & $s_0[i-2]$ & $s_0[i-1]$ & \fboxrule=1pt \fbox{$s_0[i]$} & $s_0[i+1]$ \\
$s_1[i-4]$ & $s_1[i-3]$ & $s_1[i-2]$ & $s_1[i-1]$ & \fboxrule=1pt \fbox{$s_1[i]$} & $s_1[i+1]$ \\
$s_2[i-4]$ & $s_2[i-3]$ & $s_2[i-2]$ & $s_2[i-1]$ & \fboxrule=1pt \fbox{$s_2[i]$} & $s_2[i+1]$ \\
\hline\end{tabular}
\everymath{\displaystyle}
\caption{Mu-SCo Code Construction for $(B_1,T_1) = (2,3)$ and $(B_2,T_2) = (4,4)$. The rate of $3/8$ of such Mu-SCo is the capacity given in Theorem~\ref{thm:Region_f_T2B2} for $T_2=B_2$ case in region (f).}
\label{table:Code2344}
\end{table}
\endgroup

We simply use a concatenation of two codes --- one for user $1$ and one for user $2$.
In particular, we divide each source symbol into $T_1$ sub-symbols, apply a $(B_1,T_1)$ SCo to get $B_1$ parity check sub-symbols, apply the $(T_2,T_2)$ SCo which is just a repetition code resulting in $T_1$ parity check sub-symbols and finally concatenate them to have $B_1+T_1$ parity check sub-symbols for each $T_1$ source sub-symbol (i.e., a rate of $\frac{T_1}{2T_1 + B_1} = C_{f(T_2 = B_2)}$).

Consider the  example  of $\{ (2,3)-(4,4) \}$ code in Table~\ref{table:Code2344}. Each source symbols is divided into $T_1=3$ sub-symbols. A $(B_1,T_1)=(2,3)$ SCo is applied to generate the first two rows of parity check sub-symbols which are concatenated to three more rows of parity check sub-symbols generated by the $(B_2,T_2) = (4,4)$ repetition code. User $1$ and $2$ can recover from bursts of length $2$ and $4$ within delays of $3$ and $4$ respectively by considering the corresponding rows of parity checks while neglecting the other rows.



\section{The Converse For Region (f) At $T_2 = B_2$ (Theorem~\ref{thm:Region_f_T2B2})}
\label{sec:Region_f_UB}

\begin{figure}
  \centering
    \subfigure[Step (1)]
    {
      \includegraphics[scale = 0.49]{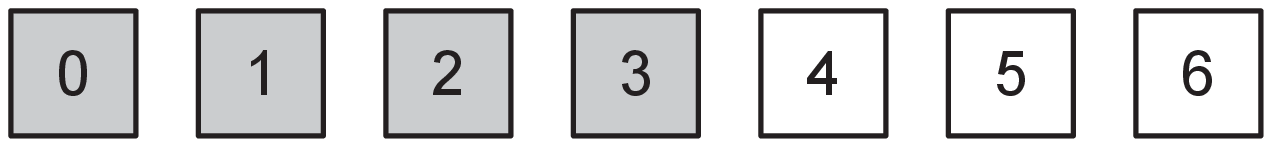}
			\label{fig:Converse_2344_a}
		}
		\subfigure[Step (2)]
    {
      \includegraphics[scale = 0.49]{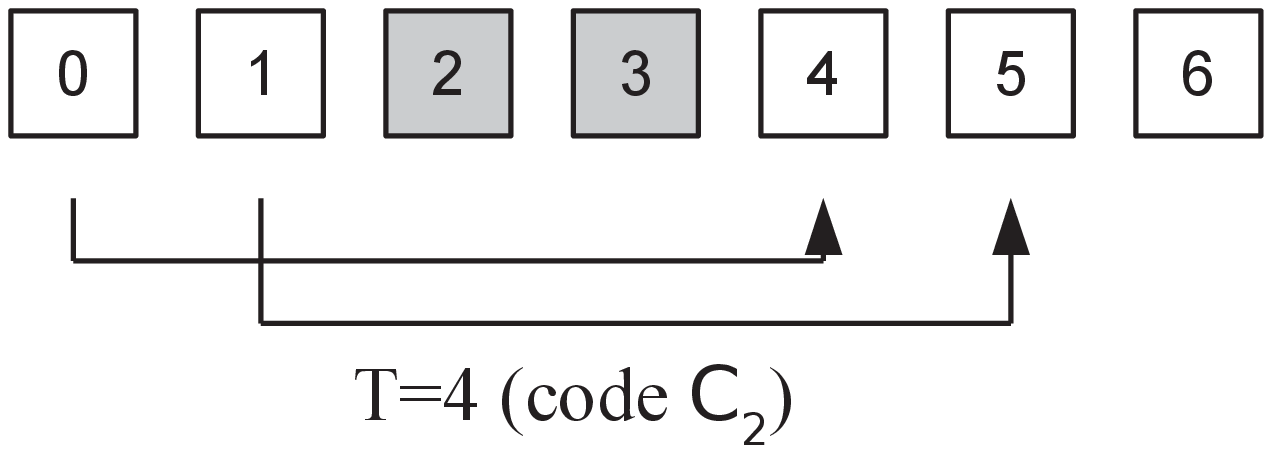}
			\label{fig:Converse_2344_b}
		}
		\subfigure[Step (3)]
		{
      \includegraphics[scale = 0.49]{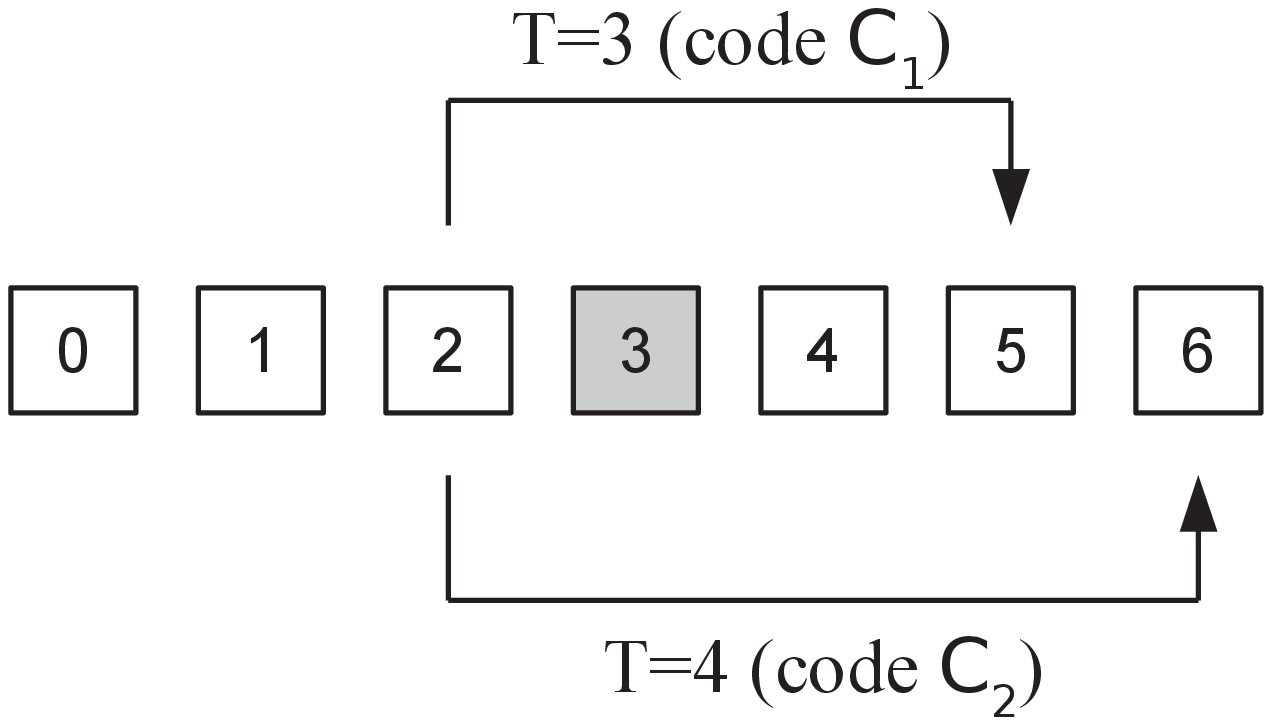}
			\label{fig:Converse_2344_c}
		}
		\subfigure[Step (4)]
		{
      \includegraphics[scale = 0.49]{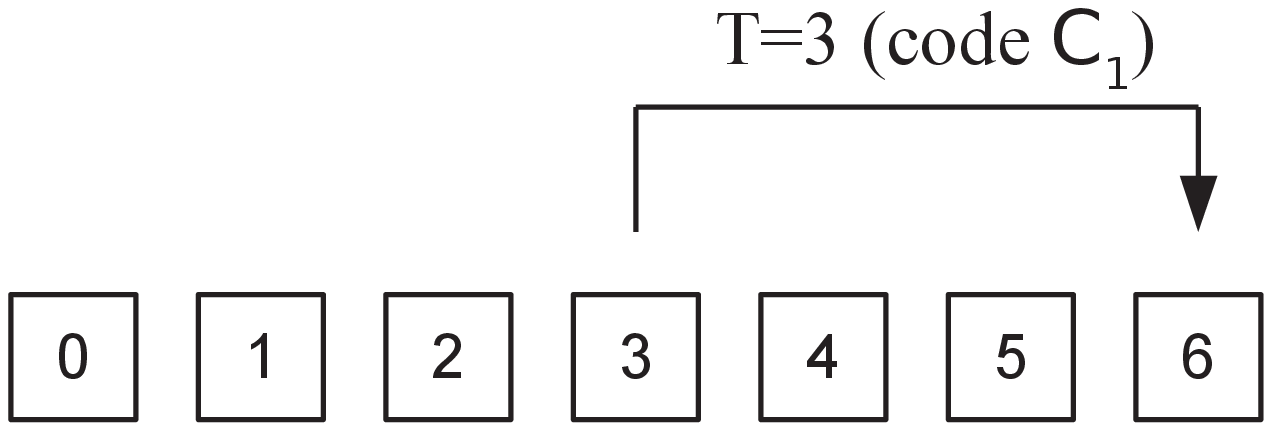}
			\label{fig:Converse_2344_d}
		}
  \caption{Main steps of finding the upper-bound for the $\{(2,3)-(4,4)\}$ point lying in Region (f) through one period illustration of the Periodic Erasure Channel. Grey and white squares resemble erased and unerased symbols respectively.} \label{fig:Converse_2344}
\end{figure}

The converse for Theorem~\ref{thm:Region_f_T2B2} depends on double recovery of some source symbols, once using code $\mathcal{C}_1$ and another using $\mathcal{C}_2$. We illustrate the main idea of such converse through considering the specific point $\{ (2,3)-(4,4) \}$ shown in  Fig.~\ref{fig:Converse_2344}. We start by considering a periodic erasure channel with period length $7$. The first $4$ symbols are erased while the rest are unerased. With $4$ erasures, code $\mathcal{C}_2 = (4,4)$ can recover the first two symbols at time $0$ and $1$ by time $4$ and $5$, respectively. We note that the channel symbol at time $i$ is sufficient to recover the source symbol at time $i-4$ (i.e., no more channel symbols are required). In step (3) in Fig.~\ref{fig:Converse_2344} gives the main idea of this converse. Since, there are two remaining erasures, the source symbol at time $2$ can be recovered using $\mathcal{C}_1=(2,3)$ within a delay of $3$ (i.e., by time $5$). Also, the same  source symbol can be decoded using $\mathcal{C}_2$ by time $6$ (double recovery). The remaining erasure can be recovered using $\mathcal{C}_1$ by time $6$. Moreover, the repetition code $\mathcal{C}_2 = (4,4)$ can recover the source symbols at time $4$, $5$ and $6$ from their corresponding channel symbols. Therefore, the three channel symbols are capable of recovering a total of $8$ source symbols (symbol at time $2$ is recovered twice) which implies that a rate of $3/8$ is an upper-bound.

For the general case, the corresponding periodic erasure channel to be used for proving the upper-bound is given in Figure \ref{fig:PEC_fRegion_T2B2}. Each period has $B_2$ erasures followed by $T_1$ non-erasures.

\begin{figure}[htbp]
	\centering
	\resizebox{\columnwidth}{!}{
	\begin{tikzpicture}[node distance=0mm]
		\node                       (x1start) {Link:};
		\node[esym, right = of x1start]  (x100) {};
		\node[esym, right = of x100]     (x101) {};
		\node[esym, right = of x101]     (x102) {};
		\node[esym, right = of x102]     (x103) {};
		\node[esym, right = of x103]     (x104) {};
		\node[usym, right = of x104]     (x105) {};
		\node[usym, right = of x105]     (x106) {};
		\node[usym, right = of x106]     (x107) {};
		\node[usym, right = of x107]     (x108) {};
		\node[esym, right = of x108]     (x109) {};
		\node[esym, right = of x109]     (x110) {};
		\node[esym, right = of x110]     (x111) {};
		\node[esym, right = of x111]     (x112) {};
		\node[esym, right = of x112]     (x113) {};
		\node[usym, right = of x113]     (x114) {};
		\node[usym, right = of x114]     (x115) {};
		\node[usym, right = of x115]     (x116) {};
		\node[usym, right = of x116]     (x117) {};
		\node[esym, right = of x117]     (x118) {};
		\node[esym, right = of x118]     (x119) {};
		\node[esym, right = of x119]     (x120) {};
		\node[esym, right = of x120]     (x121) {};
		\node[esym, right = of x121]     (x122) {};
		\node[usym, right = of x122]     (x123) {};
		\node[usym, right = of x123]     (x124) {};
		\node[usym, right = of x124]     (x125) {};
		\node[usym, right = of x125]     (x126) {};
		\node[esym, right = of x126]     (x127) {};
		\node[esym, right = of x127]     (x128) {};
		\node[esym, right = of x128]     (x129) {};
		\node[esym, right = of x129]     (x130) {};
		\node      [right = of x130]     (x1end) {$\cdots$};

		\dimup{x100}{x104}{2mm}{$B_2$};
		\dimup{x105}{x108}{2mm}{$T_1$};
		\dimup{x109}{x113}{2mm}{$B_2$};
		\dimup{x114}{x117}{2mm}{$T_1$};
		\dimup{x118}{x122}{2mm}{$B_2$};
		\dimup{x123}{x126}{2mm}{$T_1$};
	\end{tikzpicture}}
	\caption{The periodic erasure channel used to prove an upper bound on capacity in region (f) for the special case $T_2=B_2$.}
	\label{fig:PEC_fRegion_T2B2}
\end{figure}

It so happens that the $B_2=T_2$ restriction means that we can prove the converse by only analyzing one period. The reason will be made clear later. But this simplifies the proof and allows us to study the technique of double counting source packets more easily.

\begin{figure}[htbp]
	\centering
	\begin{tikzpicture}[node distance=0mm]
		\node                       (x1start) {Link:};
		\node[esym, right = of x1start]  (x100) {};
		\node[esym, right = of x100]     (x101) {};
		\node[esym, right = of x101]     (x102) {};
		\node[esym, right = of x102]     (x103) {};
		\node[esym, right = of x103]     (x104) {};
		\node[esym, right = of x104]     (x105) {};
		\node[esym, right = of x105]     (x106) {};
		\node[esym, right = of x106]     (x107) {};
		\node[usym, right = of x107]     (x108) {};
		\node[usym, right = of x108]     (x109) {};
		\node[usym, right = of x109]     (x110) {};
		\node[usym, right = of x110]     (x111) {};
		\node[usym, right = of x111]     (x112) {};
		\node[usym, right = of x112]     (x113) {};
		\node      [right = of x113]     (x1end) {$\cdots$};

		\dimdn{x100}{x105}{-2mm}{$T_1$};
		\dimup{x104}{x107}{2mm}{$B_1$};
		\dimup{x108}{x113}{2mm}{$T_1$};
		\dimdn{x100}{x107}{-10mm}{$B_2$};
	\end{tikzpicture}
	\caption{One period of the periodic erasure channel in Fig.~\ref{fig:PEC_fRegion_T2B2}, with labels.}
	\label{fig:dldr_one_per_f_region}
\end{figure}

In Figure \ref{fig:dldr_one_per_f_region}, we have the first period of the erasure channel. The key is to show that the received channel packets $\bx\sqstack{B_2}{B_2+T_1-1}$ alone can recover all of the source packets in the period, but there is enough information in the channel packets to recover some of the source packets twice. The fact that we have two decoders allows some of the source packets to be decoded by mutually exclusive groups of channel packets, but when we put all of the channel packets together, the redundant information in the channel packets does affect the maximum achievable rate of the code.

The source packets that can be recovered by $\bx\sqstack{B_2}{B_2+T_1-1}$ are $\bs\sqstack{0}{T_1-1}$, $\bs\sqstack{B_2-B_1}{B_2-1}$ and $\bs\sqstack{B_2}{B_2+T_1-1}$. As Figure \ref{fig:dldr_one_per_f_region} shows, the first two groups of source packets overlap. The overlap consists of the packets $\bs\sqstack{B_2-B_1}{T_1-1}$. The reason why we can use a single period in the proof is because the $B_2=T_2$ constraint allows us to decode the final group of source packets $\bs\sqstack{B_2}{B_2+T_1-1}$ using only the packets $\bx\sqstack{B_2}{B_2+T_1-1}$ and does not require any future channel packets.

Assuming that what we have just described is possible, then we have $T_1$ channel packets recovered $2T_1+B_1$ source packets. We should be able to write the relation:
\begin{align}
	(2T_1+B_1) \cdot H(\bs) &\leq T_1 \cdot H(\bx) \nonumber \\
	R = \frac{H(\bs)}{H(\bx)} &\leq \frac{T_1}{2T_1+B_1}
\end{align}
The formal proof shows that this is indeed possible.

\begin{proof}
We can split the proof into three major parts.

1. The source packets $\bs\sqstack{0}{T_1-B_1-1}$ can be recovered from the channel packets $\bx\sqstack{B_2}{B_2+T_1-B_1-1}$ using the $(B_2,B_2)$ decoder, so we can write
\begin{equation}
	H\Bigl(\bs\sqstack{0}{T_1-B_1-1} \Big| \bx\sqstack{B_2}{B_2+T_1-B_1-1}\Bigr) = 0. \label{eq:dldr_f_region_step1}
\end{equation}
Next, we can write
{\allowdisplaybreaks
\begin{align}
	H\Bigl(\bx\sqstack{B_2}{B_2+T_1-B_1-1}\Bigr) &= H\Bigl(\bs\sqstack{0}{T_1-B_1-1} \bx\sqstack{B_2}{B_2+T_1-B_1-1}\Bigr) - H\Bigl(\bs\sqstack{0}{T_1-B_1-1} \Big| \bx\sqstack{B_2}{B_2+T_1-B_1-1}\Bigr) \nonumber \\
	&\overset{(a)}{=} H\Bigl(\bs\sqstack{0}{T_1-B_1-1} \bx\sqstack{B_2}{B_2+T_1-B_1-1}\Bigr) \nonumber \\
	&= H\Bigl(\bs\sqstack{0}{T_1-B_1-1}\Bigr) + H\Bigl(\bx\sqstack{B_2}{B_2+T_1-B_1-1} \Big| \bs\sqstack{0}{T_1-B_1-1}\Bigr) \nonumber \\
	&\geq H\Bigl(\bs\sqstack{0}{T_1-B_1-1}\Bigr) + H\Bigl(\bx\sqstack{B_2}{B_2+T_1-B_1-1} \Big| \bs\sqstack{0}{T_1-B_1-1} \bx\sqstack{0}{T_1-B_1-1}\Bigr). \label{eq:dldr_f_region_step1_result}
\end{align}}%
We used \eqref{eq:dldr_f_region_step1} to remove the negative term before step (a).

2. In this step, we want to prove the following inequality for $m \geq B_2+T_1-B_1-1$:
\begin{align}
	\sum_{i=B_2}^{m} H(\bx[i]) \geq H\Bigl(\bs\sqstack{0}{m-B_2}\Bigr) + H\Bigl(\bs\sqstack{B_2-B_1}{m-T_1}\Bigr) + H\Bigl(\bx\sqstack{B_2}{m} \Big| \bs\sqstack{0}{m-B_2} \bs\sqstack{B_2-B_1}{m-T_1} \bx\sqstack{0}{m-B_2}\Bigr) \label{eq:dldr_f_region_target_ineq}
\end{align}

Using the first decoder with a $(B_1,T_1)$ property, we can write the following relation:
\begin{equation}
	H\Bigl(\bs[i-T_1] \Big| \bx\sqstack{i-T_1+B_1}{i} \bx\sqstack{0}{i-T_1-1}\Bigr) = 0. \label{eq:dldr_f_region_relation1}
\end{equation}
Using the $(B_2, B_2)$ decoder, we can write the following relation:
\begin{equation}
	H\Bigl(\bs[i-B_2] \Big| \bx[i] \bx\sqstack{0}{i-B_2-1}\Bigr) = 0 \label{eq:dldr_f_region_relation2}
\end{equation}
which can be used in the following steps
\begin{align}
	H\Bigl(\bx[i] \Big| \bx\sqstack{0}{i-B_2-1}\Bigr) &= H\Bigl(\bs[i-B_2] \bx[i] \Big| \bx\sqstack{0}{i-B_2-1}\Bigr) - H\Bigl(\bs[i-B_2] \Big| \bx[i] \bx\sqstack{0}{i-B_2-1}\Bigr) \nonumber \\
	&\overset{(a)}{=} H\Bigl(\bs[i-B_2] \bx[i] \Big| \bx\sqstack{0}{i-B_2-1}\Bigr) \nonumber \\
	&= H(\bs[i-B_2]) + H\Bigl(\bx[i] \Big| \bs[i-B_2] \bx\sqstack{0}{i-B_2-1}\Bigr). \nonumber
\end{align}
Therefore,
\begin{align}
	H(\bx[i]) &\geq H(\bs[i-B_2]) + H\Bigl(\bx[i] \Big| \bs[i-B_2] \bx\sqstack{0}{i-B_2}\Bigr). \label{eq:dldr_f_region_relation2b}
\end{align}
The second decoder \eqref{eq:dldr_f_region_relation2} was used to remove the negative term before step (a).

Now we can use mathematical induction to prove \eqref{eq:dldr_f_region_target_ineq}. For the base case, we substitute $m=B_2+T_1-B_1-1$
\begin{align}
	\sum_{i=B_2}^{B_2+T_1-B_1-1} H(\bx[i]) &\geq H\Bigl(\bs\sqstack{0}{T_1-B_1-1}\Bigr) + H\Bigl(\bs\sqstack{B_2-B_1}{B_2-B_1-1}\Bigr) \nonumber \\
	&\quad + H\Bigl(\bx\sqstack{B_2}{B_2+T_1-B_1-1} \Big| \bs\sqstack{0}{T_1-B_1-1} \bs\sqstack{B_2-B_1}{B_2-B_1-1} \bx\sqstack{0}{T_1-B_1-1}\Bigr) \nonumber \\
	&= H\Bigl(\bs\sqstack{0}{T_1-B_1-1}\Bigr) + H\Bigl(\bx\sqstack{B_2}{B_2+T_1-B_1-1} \Big| \bs\sqstack{0}{T_1-B_1-1} \bx\sqstack{0}{T_1-B_1-1}\Bigr).
\end{align}
This is proved by the result of~\eqref{eq:dldr_f_region_step1_result}.

Assume that \eqref{eq:dldr_f_region_target_ineq} is true for $m=j$, which gives us
\begin{align}
	\sum_{i=B_2}^{j} H(\bx[i]) \geq H\Bigl(\bs\sqstack{0}{j-B_2}\Bigr) + H\Bigl(\bs\sqstack{B_2-B_1}{j-T_1}\Bigr) + H\Bigl(\bx\sqstack{B_2}{j} \Big| \bs\sqstack{0}{j-B_2} \bs\sqstack{B_2-B_1}{j-T_1} \bx\sqstack{0}{j-B_2}\Bigr). \label{eq:dldr_f_region_ind_form}
\end{align}
We add $H(\bx[j+1])$ to both sides, and use~\eqref{eq:dldr_f_region_relation1} and~\eqref{eq:dldr_f_region_relation2} to recover the source symbols $\bs[j+1-B_2]$ and $\bs[j+1-T_1]$ respectively giving:
\begin{align}
\label{eq:fRegion_a}
	\sum_{i=B_2}^{j+1} H(\bx[i]) \geq H\Bigl(\bs\sqstack{0}{j+1-B_2}\Bigr) + H\Bigl(\bs\sqstack{B_2-B_1}{j+1-T_1}\Bigr) + H\Bigl(\bx\sqstack{B_2}{j+1} \Big| \bs\sqstack{0}{j+1-B_2} \bs\sqstack{B_2-B_1}{j+1-T_1} \bx\sqstack{0}{j+1-B_2}\Bigr).
\end{align}
The corresponding detailed steps are provided in Appendix.~\ref{app:fRegion}. By induction, we have proved \eqref{eq:dldr_f_region_target_ineq} for $m \geq B_2+T_1-B_1-1$.

3. We substitute $m = B_2+T_1-1$ into \eqref{eq:dldr_f_region_target_ineq}
\begin{align}
	\sum_{i=B_2}^{B_2+T_1-1} H(\bx[i]) \geq H\Bigl(\bs\sqstack{0}{T_1-1}\Bigr) + H\Bigl(\bs\sqstack{B_2-B_1}{B_2-1}\Bigr) + H\Bigl(\bx\sqstack{B_2}{B_2+T_1-1} \Big| \bs\sqstack{0}{T_1-1} \bs\sqstack{B_2-B_1}{B_2-1} \bx\sqstack{0}{T_1-1}\Bigr). \label{eq:dldr_f_region_step3a}
\end{align}
We can recover $\bs\sqstack{B_2}{B_2+T_1-1}$ from $\bx\sqstack{B_2}{B_2+T_1-1}$ given the previous channel symbols $\bx\sqstack{0}{B_2-1}$ using decoder 2, so we can write
\begin{equation}
	H\Bigl(\bs\sqstack{B_2}{B_2+T_1-1} \Big| \bx\sqstack{0}{B_2+T_1-1}\Bigr) = 0. \label{eq:dldr_f_region_step3b}
\end{equation}

Using~\eqref{eq:dldr_f_region_step3b}, we continue with \eqref{eq:dldr_f_region_step3a} to get (c.f. Appendix.~\ref{app:fRegion}):
\begin{align}
\label{eq:fRegion_b}
	\sum_{i=B_2}^{B_2+T_1-1} H(\bx[i]) \geq H\Bigl(\bs\sqstack{0}{T_1-1}\Bigr) + H\Bigl(\bs\sqstack{B_2-B_1}{B_2+T_1-1}\Bigr)
\end{align}

Finally, we use the fact that all source symbols have the same entropy and all channel symbols have the same entropy to write,
\begin{align}
	\sum_{i=B_2}^{B_2+T_1-1} H(\bx[i]) &\geq H\Bigl(\bs\sqstack{0}{T_1-1}\Bigr) + H\Bigl(\bs\sqstack{B_2-B_1}{B_2+T_1-1}\Bigr) \nonumber \\
	T_1 \cdot H(\bx) &\geq (2T_1 + B_1)\cdot H(\bs) \nonumber \\
	R = \frac{H(\bs)}{H(\bx)} &\leq \frac{T_1}{2T_1+B_1}
\end{align}
which is the proper upper bound.
\end{proof}
\section{Conclusion}
We study a multicast extension of the low-delay codes for streaming  over burst erasure channels. The proposed setup has several interesting implications.
From a capacity point of view, we observe an interesting interplay between the delay of the two receivers. In particular, in the large delay regime we characterize the capacity and observe a surprising delay-slackness property i.e., for most parameters, the delay of one of the receivers can be reduced up to a certain critical value without reducing the capacity. In the low-delay regime the capacity has only been partially characterized. New code constructions are developed for various regimes.  Our constructions generate parity checks in multiple layers and carefully combine them to meet the require decoding constraints.

Our ongoing work involves further investigating the capacity  in the low-delay regime. Furthermore the results in this work are a step towards developing
robust streaming code constructions, which can be used in time-varying channel conditions where the burst-length cannot be determined apriori.

\appendices

\section{Proof Of~\eqref{eq:SCo_a} and~\eqref{eq:SCo_b}}
\label{app:SCo}

The steps to get the result in~\eqref{eq:SCo_a} is as follows: Using~\eqref{eq:SCo_a} we have that
\begin{align}
	H(W_0^k) \geq H(V_0^{k-1}) + H\Bigl(W_k \Big| V_0^{k-1} \bx\sqstack{0}{k(B+T)-1}\Bigr).
\end{align}
This can be further simplified as follows.
{\allowdisplaybreaks
\begin{align}
	H&(W_0^k) \geq H(V_0^{k-1}) + H\Bigl(W_k \Big| V_0^{k-1} \bx\sqstack{0}{k(B+T)-1}\Bigr) \nonumber \\
	&\overset{(a)}{=} H(V_0^{k-1}) + H\Bigl(\bs\sqstack{k(B+T)}{k(B+T)+B-1} W_k \Big| V_0^{k-1} \bx\sqstack{0}{k(B+T)-1}\Bigr) \nonumber \\
	&\quad - H\Bigl(\bs\sqstack{k(B+T)}{k(B+T)+B-1} \Big| V_0^{k-1} \bx\sqstack{0}{k(B+T)-1} W_k\Bigr) \nonumber \\
	&\overset{(b)}{=} H(V_0^{k-1}) + H\Bigl(\bs\sqstack{k(B+T)}{k(B+T)+B-1} W_k \Big| V_0^{k-1} \bx\sqstack{0}{k(B+T)-1}\Bigr) \nonumber \\
	&\overset{(c)}{=} H(V_0^{k-1}) + H\Bigl(\bs\sqstack{k(B+T)}{k(B+T)+B-1} \Big| V_0^{k-1} \bx\sqstack{0}{k(B+T)-1}\Bigr) \nonumber \\
	&\quad + H\Bigl(W_k \Big| V_0^{k-1} \bs\sqstack{k(B+T)}{k(B+T)+B-1} \bx\sqstack{0}{k(B+T)-1}\Bigr) \nonumber \\
	&\overset{(d)}{=} H(V_0^{k-1}) + H\Bigl(\bs\sqstack{k(B+T)}{k(B+T)+B-1} \Big| V_0^{k-1} \Bigr) + H\Bigl(W_k \Big| V_0^{k-1} \bs\sqstack{k(B+T)}{k(B+T)+B-1} \bx\sqstack{0}{k(B+T)-1}\Bigr) \nonumber \\
	&\overset{(e)}{\geq} H\Bigl(V_0^{k-1} \bs\sqstack{k(B+T)}{k(B+T)+B-1} \Bigr) + H\Bigl(W_k \Big| V_0^{k-1} \bs\sqstack{k(B+T)}{k(B+T)+B-1} \bx\sqstack{0}{k(B+T)+B-1}\Bigr).
\end{align}
}%
Step (a) uses the joint entropy expansion formula, step (b) uses \eqref{eq:slsr_gen_first_relation} to remove the negative term and step (c) is a joint entropy expansion. Step (d) uses the fact that source packets are independent of each other, so therefore the source packets $\bs\sqstack{k(B+T)}{k(B+T)+B-1}$ must be independent of the past channel packets $\bx\sqstack{0}{k(B+T)-1}$. Step (e) joins the first two terms from (d), and also uses the fact that conditioning reduces entropy in the last term and the result in~\eqref{eq:SCo_a} follows.

To get the result in~\eqref{eq:SCo_b}, we start by adding $H(W_{k+1} | W_0^k)$ to both sides of~\eqref{eq:SCo_a} to get,
{\allowdisplaybreaks
\begin{align}
	H&(W_0^{k+1}) \nonumber \\
	&\geq H\Bigl(V_0^{k-1} \bs\sqstack{k(B+T)}{k(B+T)+B-1} \Bigr) + H\Bigl(W_k \Big| V_0^{k-1} \bs\sqstack{k(B+T)}{k(B+T)+B-1} \bx\sqstack{0}{k(B+T)+B-1}\Bigr) + H(W_{k+1} | W_0^k) \nonumber \\
	&\overset{(a)}{\geq} H\Bigl(V_0^{k-1} \bs\sqstack{k(B+T)}{k(B+T)+B-1} \Bigr) + H\Bigl(W_k \Big| R\Bigr) + H\Bigl(W_{k+1} \Big| R \; W_k\Bigr) \nonumber \\
	&\overset{(b)}{\geq} H\Bigl(V_0^{k-1} \bs\sqstack{k(B+T)}{k(B+T)+B-1} \Bigr) + H\Bigl(W_k^{k+1} \Big| R\Bigr) \nonumber \\
	&\overset{(c)}{=} H\Bigl(V_0^{k-1} \bs\sqstack{k(B+T)}{k(B+T)+B-1} \Bigr) + H\Bigl(\bs\sqstack{k(B+T)+B}{(k+1)(B+T)-1} W_k^{k+1} \Big| R\Bigr) - H\Bigl(\bs\sqstack{k(B+T)+B}{(k+1)(B+T)-1} \Big| R \; W_k^{k+1}\Bigr) \nonumber \\
	&\overset{(d)}{=} H\Bigl(V_0^{k-1} \bs\sqstack{k(B+T)}{k(B+T)+B-1} \Bigr) + H\Bigl(\bs\sqstack{k(B+T)+B}{(k+1)(B+T)-1} W_k^{k+1} \Big| R\Bigr) \nonumber \\
	&\overset{(e)}{=} H\Bigl(V_0^{k-1} \bs\sqstack{k(B+T)}{k(B+T)+B-1} \Bigr) + H\Bigl(\bs\sqstack{k(B+T)+B}{(k+1)(B+T)-1} \Big| R\Bigr) \nonumber \\
	&\quad + H\Bigl(W_k^{k+1} \Big| V_0^{k-1} \bs\sqstack{k(B+T)}{(k+1)(B+T)-1} \bx\sqstack{0}{k(B+T)+B-1}\Bigr) \nonumber \\
	&\overset{(f)}{=} H\Bigl(V_0^{k-1} \bs\sqstack{k(B+T)}{k(B+T)+B-1} \Bigr) + H\Bigl(\bs\sqstack{k(B+T)+B}{(k+1)(B+T)-1} \Big| V_0^{k-1} \bs\sqstack{k(B+T)}{k(B+T)+B-1} \Bigr) \nonumber \\
		&\quad + H\Bigl(W_k^{k+1} \Big| V_0^{k} \bx\sqstack{0}{k(B+T)+B-1}\Bigr) \nonumber \\
	&\overset{(g)}{=} H\Bigl(V_0^{k-1} \bs\sqstack{k(B+T)}{(k+1)(B+T)-1} \Bigr) + H\Bigl(W_k \Big| V_0^{k} \bx\sqstack{0}{k(B+T)+B-1}\Bigr) + H\Bigl(W_{k+1} \Big| V_0^{k} \bx\sqstack{0}{k(B+T)+B-1} W_k\Bigr) \nonumber \\
	&\geq H(V_0^{k}) + H\Bigl(W_{k+1} \Big| V_0^{k} \bx\sqstack{0}{(k+1)(B+T)-1} \Bigr),
\end{align}
}%
where $R = V_0^{k-1} \bs\sqstack{k(B+T)}{k(B+T)+B-1} \bx\sqstack{0}{k(B+T)+B-1}$. Step (a) introduces extra conditions in the final term, so entropy is reduced, step (b) uses the joint entropy formula, step (c) uses the joint entropy expansion formula and step (d) uses \eqref{eq:slsr_gen_second_relation} to remove the negative term in (c). Step (e) uses the joint entropy formula again to expand the second term of (d) and step (f) uses the fact that source packets are independent of previous channel packets. Step (g) once again uses the joint entropy formula and~\eqref{eq:SCo_b} follows.

\section{Proof of~\eqref{eq:MuSCo_a} and~\eqref{eq:MuSCo_b}}
\label{app:MuSCo}

The working out of~\eqref{eq:MuSCo_a} is as follows:
{\allowdisplaybreaks
\begin{align}
	H&(W_0^{k}) \geq H(V_0^{k-1}) + H\Bigl(W_k  \Big|  V_0^{k-1} \bx\sqstack{0}{kc-1}\Bigr) \nonumber \\
	&= H(V_0^{k-1}) + H\Bigl(\bs\sqstack{kc}{kc+a-1} W_{k}  \Big|  V_0^{k-1} \bx\sqstack{0}{kc-1}\Bigr) - H\Bigl(\bs\sqstack{kc}{kc+a-1}  \Big|  V_0^{k-1} \bx\sqstack{0}{kc-1} W_{k}\Bigr) \nonumber \\
	&\overset{(a)}{=} H(V_0^{k-1}) + H\Bigl(\bs\sqstack{kc}{kc+a-1} W_{k}  \Big|  V_0^{k-1} \bx\sqstack{0}{kc-1}\Bigr) \nonumber \\
	&= H(V_0^{k-1}) + H\Bigl(\bs\sqstack{kc}{kc+a-1}  \Big|  V_0^{k-1} \bx\sqstack{0}{kc-1}\Bigr) + H\Bigl(W_{k}  \Big|  V_0^{k-1} \bs\sqstack{kc}{kc+a-1} \bx\sqstack{0}{kc-1}\Bigr) \nonumber \\
	&\overset{(b)}{=} H\Bigl(V_0^{k-1} \bs\sqstack{kc}{kc+a-1}\Bigr) + H\Bigl(W_{k}  \Big|  V_0^{k-1} \bs\sqstack{kc}{kc+a-1} \bx\sqstack{0}{kc-1}\Bigr) \nonumber \\
	&\geq H\Bigl(V_0^{k-1} \bs\sqstack{kc}{kc+a-1}\Bigr) + H\Bigl(W_{k}  \Big|  V_0^{k-1} \bs\sqstack{kc}{kc+a-1} \bx\sqstack{0}{kc+a-1}\Bigr) \nonumber \\
	&= H\Bigl(V_0^{k-1} \bs\sqstack{kc}{kc+a-1}\Bigr) + H\Bigl(\bs\sqstack{kc+a}{kc+b-1} W_{k}  \Big|  V_0^{k-1} \bs\sqstack{kc}{kc+a-1} \bs\sqstack{0}{kc+a-1}\Bigr) \nonumber \\
		&\quad - H\Bigl(\bs\sqstack{kc+a}{kc+b-1}  \Big|  V_0^{k-1} \bs\sqstack{kc}{kc+a-1} \bx\sqstack{0}{kc+a-1} W_{k}\Bigr) \nonumber \\
	&\overset{(c)}{=} H\Bigl(V_0^{k-1} \bs\sqstack{kc}{kc+a-1}\Bigr) + H\Bigl(\bs\sqstack{kc+a}{kc+b-1} W_{k}  \Big|  V_0^{k-1} \bs\sqstack{kc}{kc+a-1} \bx\sqstack{0}{kc+a-1}\Bigr) \nonumber \\
	&= H\Bigl(V_0^{k-1} \bs\sqstack{kc}{kc+a-1}\Bigr) + H\Bigl(\bs\sqstack{kc+a}{kc+b-1}  \Big|  V_0^{k-1} \bs\sqstack{kc}{kc+a-1} \bx\sqstack{0}{kc+a-1}\Bigr) \nonumber \\
		&\quad + H\Bigl(W_{k}  \Big|  V_0^{k-1} \bs\sqstack{kc}{kc+a-1} \bs\sqstack{kc+a}{kc+b-1} \bx\sqstack{0}{kc+a-1}\Bigr) \nonumber \\
	&\overset{(d)}{=} H\Bigl(V_0^{k-1} \bs\sqstack{kc}{kc+b-1}\Bigr) + H\Bigl(W_{k}  \Big|  V_0^{k-1} \bs\sqstack{kc}{kc+b-1} \bx\sqstack{0}{kc+a-1}\Bigr) \nonumber \\
	&\geq H\Bigl(V_0^{k-1} \bs\sqstack{kc}{kc+b-1}\Bigr) + H\Bigl(W_{k}  \Big|  V_0^{k-1} \bs\sqstack{kc}{kc+b-1} \bx\sqstack{0}{kc+b-1}\Bigr)
\end{align}
}
We use \eqref{eq:Code1_Def_thm1_caseA} to remove the negative term before step (a). Similarly, we remove the negative term before step (c) using \eqref{eq:Code2_Def_thm1_caseA}. Steps (b) and (d) use the fact that source packets are independent of each other and of previous channel packets.

While for~\eqref{eq:MuSCo_b}, we start by adding $H(W_{k+1} | W_0^k)$ to both sides of~\eqref{eq:MuSCo_a} to get,
{\allowdisplaybreaks
\begin{align}
	H&(W_0^{k+1}) \geq H\Bigl(V_0^{k-1} \bs\sqstack{kc}{kc+b-1}\Bigr) + H\Bigl(W_{k}  \Big|  V_0^{k-1} \bs\sqstack{kc}{kc+b-1} \bx\sqstack{0}{kc+b-1}\Bigr) + H(W_{k+1} | W_0^k) \nonumber \\
	&\overset{(e)}{\geq} H\Bigl(V_0^{k-1} \bs\sqstack{kc}{kc+b-1}\Bigr) + H\Bigl(W_{k}^{k+1}  \Big|  V_0^{k-1} \bs\sqstack{kc}{kc+b-1} \bx\sqstack{0}{kc+b-1}\Bigr) \nonumber \\
	&= H\Bigl(V_0^{k-1} \bs\sqstack{kc}{kc+b-1}\Bigr) + H\Bigl(\bs\sqstack{kc+b}{(k+1)c-1} W_{k}^{k+1}  \Big|  V_0^{k-1} \bs\sqstack{kc}{kc+b-1} \bx\sqstack{0}{kc+b-1}\Bigr) \nonumber \\
		&\quad - H\Bigl(\bs\sqstack{kc+b}{(k+1)c-1}  \Big|  V_0^{k-1} \bs\sqstack{kc}{kc+b-1} \bx\sqstack{0}{kc+b-1} W_{k}^{k+1}\Bigr) \nonumber \\
	&\overset{(f)}{=} H\Bigl(V_0^{k-1} \bs\sqstack{kc}{kc+b-1}\Bigr) + H\Bigl(\bs\sqstack{kc+b}{(k+1)c-1} W_{k}^{k+1}  \Big|  V_0^{k-1} \bs\sqstack{kc}{kc+b-1} \bx\sqstack{0}{kc+b-1}\Bigr) \nonumber \\
	&= H\Bigl(V_0^{k-1} \bs\sqstack{kc}{kc+b-1}\Bigr) + H\Bigl(\bs\sqstack{kc+b}{(k+1)c-1}  \Big|  V_0^{k-1} \bs\sqstack{kc}{kc+b-1} \bx\sqstack{0}{kc+b-1}\Bigr) \nonumber \\
		&\quad + H\Bigl(W_{k}^{k+1}  \Big|  V_0^{k-1} \bs\sqstack{kc}{(k+1)c-1} \bx\sqstack{0}{kc+b-1}\Bigr) \nonumber \\
	&\overset{(g)}{=} H\Bigl(V_0^{k-1} \bs\sqstack{kc}{(k+1)c-1}\Bigr) + H\Bigl(W_{k}^{k+1}  \Big|  V_0^{k-1} \bs\sqstack{kc}{(k+1)c-1} \bx\sqstack{0}{kc+b-1}\Bigr) \nonumber \\
	&\geq H(V_0^k) + H\Bigl(W_{k}^{k+1}  \Big|  V_0^k \bx\sqstack{0}{(k+1)c-1}\Bigr) \nonumber \\
	&= H(V_0^k) + H\Bigl(W_{k}  \Big|  V_0^k \bx\sqstack{0}{(k+1)c-1}\Bigr) + H\Bigl(W_{k+1}  \Big|  V_0^k \bx\sqstack{0}{(k+1)c-1} W_{k}\Bigr) \nonumber \\
	&\geq H(V_0^k) + H\Bigl(W_{k+1}  \Big|  V_0^k \bx\sqstack{0}{(k+1)c-1}\Bigr) \label{eq:inductive_step_thm1_caseA}
\end{align}
}%
Step (e) follows by the fact that conditioning reduces entropy knowing that $W_{0}^{k-1} \subset \bx\sqstack{0}{kc+b-1}$ and thus $H\Bigl(W_{k+1} \Big| V_0^{k-1} \bs\sqstack{kc}{kc+b-1} \bx\sqstack{0}{kc+b-1}\Bigr) \leq H(W_{k+1} | W_{k})$, and again we remove the negative term before step (f) using \eqref{eq:Code2_Def_thm1_caseA}. Step (g) uses the fact that source packets are independent of each other and~\eqref{eq:MuSCo_b} follows.

\section{Examples of Code Construction in the (e) Region}
\label{sec:e-example}
We give the construction for two specific points in this region,  Table~\ref{table:Code45710} shows the code construction for the point $\{(4,5)-(7,10)\}$
whereas Table~\ref{table:Code3579} shows the code construction for the point $\{(3,5)-(7,9)\}|$. In both cases $k=1$ and $m=1$. The former satisfies
$T_1 < 2(B_1-k)$ whereas the latter satisfies $T_1 > 2(B_1-k)$.

\subsection{Example (1): $\{ (4,5)-(7,10) \} \Rightarrow k=1, m=1$}
\label{subsec:45710}

\begingroup
\singlespacing
\everymath{\scriptstyle}
\begin{table}[t]
\begin{tabular}{c|x{61pt}|x{61pt}|x{61pt}|x{61pt}|x{61pt}|x{61pt}}
& $\displaystyle [i]$ & $\displaystyle [i+1]$ & $\displaystyle [i+2]$ & $\displaystyle [i+3]$ & $\displaystyle [i+4]$ & $\displaystyle [i+5]$\\
\hline
(1) 	& $s_0[i]$ & $s_0[i+1]$ & $s_0[i+2]$ & $s_0[i+3]$ & $s_0[i+4]$ & $s_0[i+5]$ \\
	& $s_1[i]$ & $s_1[i+1]$ & $s_1[i+2]$ & $s_1[i+3]$ & $s_1[i+4]$ & $s_1[i+5]$ \\
	& $s_2[i]$ & $s_2[i+1]$ & $s_2[i+2]$ & $s_2[i+3]$ & $s_2[i+4]$ & $s_2[i+5]$ \\
	& $s_3[i]$ & $s_3[i+1]$ & $s_3[i+2]$ & $s_3[i+3]$ & $s_3[i+4]$ & $s_3[i+5]$ \\
	& $s_4[i]$ & $s_4[i+1]$ & $s_4[i+2]$ & $s_4[i+3]$ & $s_4[i+4]$ & $s_4[i+5]$ \\
\hline
(2)	& $p_0[i]$ & $p_0[i+1]$ & $p_0[i+2]$ & $p_0[i+3]$ & $p_0[i+4]$ & $p_0[i+5]$ \\
\hline
(3)	& $s_0[i-10] + p_1[i]$ 	& $s_0[i-9] + p_1[i+1]$ & $s_0[i-8] + p_1[i+2]$ & $s_0[i-7] + p_1[i+3]$ & $s_0[i-6] + p_1[i+4]$ & $s_0[i-5] + p_1[i+5]$ \\
	& $s_1[i-10] + p_2[i]$ 	& $s_1[i-9] + p_2[i+1]$ & $s_1[i-8] + p_2[i+2]$ & $s_1[i-7] + p_2[i+3]$ & $s_1[i-6] + p_2[i+4]$ & $s_1[i-5] + p_2[i+5]$ \\
	& $s_2[i-10] + p_3[i]$ 	& $s_2[i-9] + p_3[i+1]$ & $s_2[i-8] + p_3[i+2]$ & $s_2[i-7] + p_3[i+3]$ & $s_2[i-6] + p_3[i+4]$ & $s_2[i-5] + p_3[i+5]$ \\
\hline
(4)	& $s_3[i-10]$ & $s_3[i-9]$ & $s_3[i-8]$ & $s_3[i-7]$ & $s_3[i-6]$ & $s_3[i-5]$\\
	& $+$ & $+$ & $+$ & $+$ & $+$ & $+$ \\
	& $\tl{p}_1[i+2] + \tl{p}_3[i+4]$	& $\tl{p}_1[i+3] + \tl{p}_3[i+5]$	& $\tl{p}_1[i+4] + \tl{p}_3[i+6]$	& $\tl{p}_1[i+5] + \tl{p}_3[i+7]$	& $\tl{p}_1[i+6] + \tl{p}_3[i+8]$	& $\tl{p}_1[i+7] + \tl{p}_3[i+9]$	\\
\cline{2-7}
	& $s_4[i-10]$ & $s_4[i-9]$ & $s_4[i-8]$ & $s_4[i-7]$ & $s_4[i-6]$ & $s_4[i-5]$\\
	& $+$ & $+$ & $+$ & $+$ & $+$ & $+$ \\
	& $\tl{p}_2[i+2] + \tl{p}_3[i+3]$ & $\tl{p}_2[i+3] + \tl{p}_3[i+4]$ & $\tl{p}_2[i+4] + \tl{p}_3[i+5]$ & $\tl{p}_2[i+5] + \tl{p}_3[i+6]$ & $\tl{p}_2[i+6] + \tl{p}_3[i+7]$ & $\tl{p}_2[i+7] + \tl{p}_3[i+8]$ \\
\hline\end{tabular}
\everymath{\displaystyle}
\caption{Rate $5/11$ Mu-SCo Code Construction for the point, $(B_1,T_1) = (4,5)$ and $(B_2,T_2) = (7,10)$ lying in region (e). This point is also illustrating case (A) defined by $T_1 \leq 2(B_1-k)$. For the causal part of parity check sub-symbols of $\cC_1$ shifted back to time $i-t$, we write $\tl{p}_j[i]$ instead of $\tl{p}_j[i]\big{|}_{i-t}$ for simplicity.}
\label{table:Code45710}
\end{table}
\endgroup

The code construction achieving the optimal rate of $5/11$ is illustrated in Table~\ref{table:Code45710}. In this example, we walk through the steps of both the encoder and the decoder. We note that this point resembles case (A) defined by $T_1 \leq 2(B_1-k)$ in the general code construction given in Section.~\ref{sec:Region_e_CC}.

\begin{itemize}
\item Encoder
\begin{itemize}
\item Each source symbol is divided into $T_1 = 5$ sub-symbols $(s_0[.],\dots,s_4[.])$. A $\cC_1 = (4,5)$ is applied along the diagonal of such source sub-symbols producing $B_1 = 4$ parity check sub-symbols $(p_0[.],\dots,p_3[.])$ defined as follows,
\begin{align}
\label{eq:Code45710}
p_0[i] &= s_0[i-5] + s_4[i-1] \nonumber \\
p_1[i] &= s_1[i-5] + s_4[i-2] \nonumber \\
p_2[i] &= s_2[i-5] + s_4[i-3] \nonumber \\
p_3[i] &= s_3[i-5] + s_4[i-4]
\end{align}
\item Then, the $T_1 = 5$ parity check-symbols of code $\cC_2 = (10,10)$ which are repetitions of the source sub-symbols such that $p_j^{\rm{II}}[i] = s_j[i-10]$ for $j \in \{0,\dots,4\}$ are concatenated to the parity checks of $\cC_1$ with partial overlap of $B_1-k = 3$ rows as shown in Table~\ref{table:Code45710}.
\item A $\cC_3 = (T_1-(B_1-k),B_1-k) = (2,3)$ SCo code is applied on the last $B_1 - k = 3$ rows of parity check sub-symbols of $\cC_1$, $(p_1[.],p_2[.],p_3[.])$ producing $T_1 - (B_1-k) = 2$ parity check sub-symbols, $(p_0^3[.],p_1^3[.])$. The produced parity checks is shifted back by $T_1 = 5$ and combined with the last two rows of parity check sub-symbols of $\cC_2$.
\end{itemize}
We note that applying a shift back of $T_1 = 5$ on the parity check sub-symbols of $\cC_3$ explains why $p_0^{3}[i] = p_1[i+2] + p_3[i+4]$ appears at time $i$ and not $i+5$. Moreover, since $p_1[i+2]+p_3[i+4]$ in general combines source sub-symbols at time $i+3$ and earlier, they can not appear at time $i$ as this violates the causality of the code construction. Thus, the causal part of such parity checks shifted to any time instant $t$ (denoted by $\tl{p}_j[.]\big{|}_t$) is to be sent instead. For example, the first parity check sub-symbol of $\cC_3$ at time $i$ is $p_0^3[i+5] = p_1[i+2] + p_3[i+4] = s_1[i-3] + s_4[i+1] + s_3[i-1] + s_4[i]$. The causal part of this parity check is sent instead, i.e., $\tl{p}_0^3[i+5]\big{|}_i = \tl{p}_1[i+2]\big{|}_i + \tl{p}_3[i+4]\big{|}_i = s_1[i-3] + s_3[i-1]$.\\
According to Fig.~\ref{fig:eRegion_CC}, we divide each channel packet into four layers,
\begin{itemize}
\item Layer (1) contains the first five rows which are the source sub-symbols.
\item Layer (2) contains the next row.
\item Layer (3) contains the next three rows where overlap between the parity checks of codes $\cC_1$ and $\cC_2$ takes place.
\item Layer (4) contains the last two rows. The overlap between the parity checks of codes $\cC_2$ and $\cC_3$ takes place.
\end{itemize}

\item Decoder \\
With a burst erasure of length $B_1 = 4$ taking place at times $[i-4,i-1]$, the decoder at user 1 simply uses the first four rows of parity checks at times $[i,i+4]$ after subtracting the unerased source sub-symbols $s_0[t],s_1[t],s_2[t]$ for $t \in \{ i-10, \dots, i-6 \}$. For user 2, we assume a burst erasure of length $B_2 = 7$ at times $[i-7,i-1]$. The decoding steps are as follows.
\begin{itemize}
\item \textbf{Step (1):} Recover $p_j[i+3]$ and $p_j[i+4]$ for $j = \{1,2,3\}$.

\begin{enumerate}[(a)]
\item In layer (3), spanning the second, third and fourth rows of parity checks, one can see that the parity check sub-symbols of $\cC_2$ in the interval $[i,i+2]$ are unerased source sub-symbols. Thus, the corresponding combined parity check sub-symbols of $\cC_1$ can be computed in this interval. 
\item In the same layer but in the interval $[i+5,\infty)$, the parity check sub-symbols of $\cC_1$ are of indices $i+5$ and later. Using the fact that $(B_1,T_1)$ SCo code has a memory of $T_1$ symbols, it can be easily shown that these parity check sub-symbols combine only source sub-symbols of time $i$ and later which are not erased and thus can be computed as well (c.f. \eqref{eq:Code45710}). 
\item Steps (a) and (b) show that all the parity check sub-symbols of $\cC_1$ in layer (3) can be computed except for the interval $[i+3,i+4]$. 
\item The parity check sub-symbols of $\cC_2$ in layer (4) spanning the last two rows of parity check sub-symbols in the interval $[i,i+2]$ are again unerased source sub-symbols and thus can be cancelled and the corresponding parity check sub-symbols of $\cC_3$ can be computed in this interval.
\item The parity-check sub-symbols of $\cC_3$ in the interval $[i,i+2]$,
\begin{align}
\left(
\begin{array}{ccc}
p_0^3[i+5]\big{|}_i & p_0^3[i+6]\big{|}_{i+1} & p_0^3[i+7]\big{|}_{i+2} \\
p_1^3[i+5]\big{|}_i & p_1^3[i+6]\big{|}_{i+1} & p_1^3[i+7]\big{|}_{i+2}
 \end{array}
\right),\label{eq:p-arr}
\end{align}
can recover the remaining two columns of parity-check sub-symbols of $\cC_1$ in the interval $[i+3,i+4]$ lying in layer (3),
\begin{align*}
\left(
\begin{array}{cc}
p_1[i+3] & p_1[i+4] \\
p_2[i+3] & p_2[i+4] \\
p_3[i+3] & p_3[i+4]
 \end{array}
\right),
\end{align*}
since $\cC_3$ is a $(2,3)$ SCo code whose parity-check sub-symbols are shifted back by $T_1 = 5$.

However, only the causal part of the parity checks of $\cC_3$ are available. Thus, the non-causal part is to be computed and added to the causal-part to recover the original parity checks of the SCo code. Using~\eqref{eq:Code45710}, it can be seen that the recovery of the non-causal part does not require the availability of source sub-symbols after time\footnote{A proof of this in the general case is provided in the proof of Lemma~\ref{lem:step1} in Appendix~\ref{app:step1}.} $i+3$. For example, $p_0^3[i+5]\big{|}_{i} = p_1[i+2] + p_3[i+4] = s_1[i-3]+s_4[i]+s_3[i-1]+s_4[i]$, while $\tilde{p}_0^3[i+5]\big{|}_{i} = \tilde{p}_1[i+2]\big{|}_{i} + \tilde{p}_3[i+4]\big{|}_{i} = s_1[i-3]+s_3[i-1]$, i.e., the non-causal part of $p_0^3[i+5]\big{|}_{i}$ is $\bar{p}_0^3[i+5]\big{|}_{i} = 2s_4[i]$ which is clearly available before time $i+3$. Thus the non-causal portions of all the parity checks are computed and then~\eqref{eq:p-arr} is applied.


\end{enumerate}

\item  \textbf{Step (2):}
After recovering these parity check sub-symbols, the decoder can cancel their effect in the second, third and fourth rows of parity checks (layer (3)) at times $i+3$ and $i+4$. Moreover, in the same rows and starting at time $i+5$ all parity checks of code $\cC_1$ combine only unerased source symbols (c.f. \eqref{eq:Code45710}) and thus can be cancelled as well.

\item \textbf{Step (3):}
Furthermore, one can see that the parity check sub-symbols of $\cC_3$ interfering in the last two rows (layer (4)) starting at time $i+3$ combine parity check sub-symbols of $\cC_1$ of indices $i+5$ and later which was shown before to combine unerased source sub-symbols (c.f.~\eqref{eq:Code45710}). \\

According to Step (2) and (3), the parity checks of $\cC_2 = (10,10)$ repetition code in layers (3) and (4) are now free of any interference from $i+3$ and later. Thus, the decoder of user 2 is capable of recovering the erased source sub-symbols in the interval ${[i-7,i-1]}$.
\end{itemize}

\end{itemize}

\subsection{Example (2): $\{ (3,5)-(7,9) \} \Rightarrow k=1, m=1$}
\label{subsec:3579}

\begingroup
\singlespacing
\everymath{\scriptstyle}
\begin{table}[t]
\begin{tabular}{c|x{62pt}|x{62pt}|x{62pt}|x{62pt}|x{62pt}|x{62pt}}
& $\displaystyle [i]$ & $\displaystyle [i+1]$ & $\displaystyle [i+2]$ & $\displaystyle [i+3]$ & $\displaystyle [i+4]$ & $\displaystyle [i+5]$\\
\hline
1 & $s_0[i]$ & $s_0[i+1]$ & $s_0[i+2]$ & $s_0[i+3]$ & $s_0[i+4]$ & $s_0[i+5]$ \\
 	& $s_1[i]$ & $s_1[i+1]$ & $s_1[i+2]$ & $s_1[i+3]$ & $s_1[i+4]$ & $s_1[i+5]$ \\
 	& $s_2[i]$ & $s_2[i+1]$ & $s_2[i+2]$ & $s_2[i+3]$ & $s_2[i+4]$ & $s_2[i+5]$ \\
 	& $s_3[i]$ & $s_3[i+1]$ & $s_3[i+2]$ & $s_3[i+3]$ & $s_3[i+4]$ & $s_3[i+5]$ \\
 	& $s_4[i]$ & $s_4[i+1]$ & $s_4[i+2]$ & $s_4[i+3]$ & $s_4[i+4]$ & $s_4[i+5]$ \\
\hline
2 & $p_0[i]$ & $p_0[i+1]$ & $p_0[i+2]$ & $p_0[i+3]$ & $p_0[i+4]$ & $p_0[i+5]$ \\
\hline
3 & $s_0[i-9] + p_1[i]$ 	& $s_0[i-8] + p_1[i+1]$ & $s_0[i-7] + p_1[i+2]$ & $s_0[i-6] + p_1[i+3]$ & $s_0[i-5] + p_1[i+4]$ & $s_0[i-4] + p_1[i+5]$ \\
  & $s_1[i-9] + p_2[i]$ 	& $s_1[i-8] + p_2[i+1]$ & $s_1[i-7] + p_2[i+2]$ & $s_1[i-6] + p_2[i+3]$ & $s_1[i-5] + p_2[i+4]$ & $s_1[i-4] + p_2[i+5]$ \\
\hline
4 & $s_2[i-9] + \tl{p}_1[i+2]$ 	& $s_2[i-8] + \tl{p}_1[i+3]$ & $s_2[i-7] + \tl{p}_1[i+4]$ & $s_2[i-6] + \tl{p}_1[i+5]$ & $s_2[i-5] + \tl{p}_1[i+6]$ & $s_2[i-4] + \tl{p}_1[i+7]$ \\
  & $s_3[i-9] + \tl{p}_2[i+2]$ 	& $s_3[i-8] + \tl{p}_2[i+3]$ & $s_3[i-7] + \tl{p}_2[i+4]$ & $s_3[i-6] + \tl{p}_2[i+5]$ & $s_3[i-5] + \tl{p}_2[i+6]$ & $s_3[i-4] + \tl{p}_2[i+7]$ \\\cline{2-7}
  & $s_4[i-9]$ & $s_4[i-8]$ & $s_4[i-7]$ & $s_4[i-6]$ & $s_4[i-5]$ & $s_4[i-4]$\\
	& $+$ & $+$ & $+$ & $+$ & $+$ & $+$ \\
  & $\tl{p}_1[i+3] + \tl{p}_2[i+4]$ & $\tl{p}_1[i+4] + \tl{p}_2[i+5]$ & $\tl{p}_1[i+5] + \tl{p}_2[i+6]$ & $\tl{p}_1[i+6] + \tl{p}_2[i+7]$ & $\tl{p}_1[i+7] + \tl{p}_2[i+8]$ & $\tl{p}_1[i+8] + \tl{p}_2[i+9]$ \\
\hline\end{tabular}
\everymath{\displaystyle}
\caption{Rate $5/11$ Mu-SCo Code Construction for the point, $(B_1,T_1) = (3,5)$ and $(B_2,T_2) = (7,9)$ lying in region (e). This point is also illustrating case (B) defined by $T_1 > 2(B_1-k)$. For the causal part of parity check sub-symbols of $\cC_1$ shifted back to time $i-t$, we write $\tl{p}_j[i]$ instead of $\tl{p}_j[i]\big{|}_{i-t}$ for simplicity.}
\label{table:Code3579}
\end{table}
\endgroup

Again the capacity equals $5/11$. The code construction achieving such rate is illustrated in Table~\ref{table:Code3579}. The reason we give the detailed encoding and decoding steps for one more example is to show the main differences between case (A): $T_1 \leq 2(B_1-k)$  illustrated by the previous example $\{ (4,5)-(7,10) \}$ and case (B): $T_1 > 2(B_1-k)$ illustrated by this example, $\{ (3,5)-(7,9) \}$.

\begin{itemize}
\item Encoder
\begin{itemize}
\item Each source symbol is divided into $T_1 = 5$ sub-symbols $(s_0[.],\dots,s_4[.])$ (layer (1)). A $\cC_1 = (3,5)$ is applied along the diagonal of such source sub-symbols producing $B_1 = 3$ parity check sub-symbols $(p_0[.],p_1[.],p_2[.])$ defined as follows,
\begin{align}
\label{eq:Code3579}
p_0[i] &= s_0[i-5] + s_3[i-2] \nonumber \\
p_1[i] &= s_1[i-5] + s_4[i-2] \nonumber \\
p_2[i] &= s_2[i-5] + s_3[i-4] + s_4[i-3] 
\end{align}
\item Then, the $T_1 = 5$ parity check-symbols of code $\cC_2 = (9,9)$ which are repetitions of the corresponding source sub-symbols are concatenated to the parity checks of $\cC_1$ with partial overlap of $B_1-k = 2$ rows as shown in Table~\ref{table:Code3579}.
\item Since $T_1 = 5 > 4 = 2(B_1-k)$, this point falls in case (B), one can write $T_1 - (B_1-k) = r(B_1-k) + q$ as $3 = 1(2) + 1$, i.e., $r=1$ and $q=1$. Thus, $r+1 = 2$ SCo codes are to be constructed. The first is a repetition code of parameters $\cC_{3,1} = (B_1-k,B_1-k) = (2,2)$ is applied on the last $B_1 - k = 2$ rows of parity check sub-symbols of $\cC_1$, $(p_1[.],p_2[.])$ producing $(B_1-k) = 2$ parity check sub-symbols, $(p_0^3[\cdot],p_1^3[\cdot])$ which are then shifted back by $2(B_1-k) = 4$ symbols, while the second is a $\cC_{3,2} = (q,B_1-k) = (1,2)$ SCo code applied again on the last two rows of parity check sub-symbols of $\cC_1$ along the main diagonal producing one row of parity check sub-symbols, $p_2^3[\cdot]$ which is shifted back by $T_1 = 5$ symbols. The parity check sub-symbols of $\cC_{3,1}$ and $\cC_{3,2}$ (denoted by $\cC_3$) are then concatenated forming $T_1 - (B_1-k) = 3$ rows of parity check sub-symbols and then combined with the last three rows of parity check sub-symbols of $\cC_2$ (layer (3)).
\end{itemize}
The same causality argument stated in the previous example applies and the causal parts of the corresponding parity check sub-symbols shifted to any time instant $t$ denoted by $\tl{p}_j[.]\big{|}_t$ are sent instead (c.f. Table~\ref{table:Code3579}). \\
Similar to the previous example, we divide each channel packet into four layers (c.f. Fig.~\ref{fig:eRegion_CC}),
\begin{itemize}
\item Layer (1) contains the first five rows which are the source sub-symbols.
\item Layer (2) contains the next row.
\item Layer (3) contains the next two rows where overlap between the parity checks of codes $\cC_1$ and $\cC_2$ takes place.
\item Layer (4) contains the last three rows. The overlap between the parity checks of codes $\cC_2$ and $\cC_3$ takes place.
\end{itemize}

\item Decoding: \newline
For user $1$, the decoding is similar to the previous example. We assume a burst erasure of length $B_1 = 3$ taking place at times $[i-3,i-1]$. One can recover the parity checks of code $\cC_1$ in the first three rows of parity checks at times $[i,i+4]$ after subtracting the unerased combined source sub-symbols $s_0[t],s_1[t],s_2[t]$ for $t \in \{ i-9, \dots, i-5 \}$. For user 2, we assume a burst erasure of length $B_2 = 7$ in the interval $[i-7,i-1]$. The decoding steps are as follows.

\begin{itemize}

\item

Recover $p_j[i+2]$, $p_j[i+3]$ and $p_j[i+4]$ for $j = \{1,2\}$.

\begin{enumerate}[(a)]
\item In layer (3), spanning the second and third rows of parity checks, one can see that the parity check sub-symbols of $\cC_2$ in the interval $[i,i+1]$ are unerased source sub-symbols. Thus, the overlapping parity check sub-symbols of $\cC_1$ can be computed in this interval. 
\item In the same layer but in the interval $[i+5,\infty)$, the parity check sub-symbols of $\cC_1$ are of indices $i+5$ and later. Using the fact that $(B_1,T_1)$ SCo code has a memory of $T_1$ symbols, it can be easily shown that these parity check sub-symbols combine only source sub-symbols of time $i$ and later which are not erased and thus can be computed as well(c.f. \eqref{eq:Code3579}). 
\item In steps (a) and (b), we show that all the parity check sub-symbols of $\cC_1$ in layer (3) can be computed except for the interval $[i+2,i+4]$. Let us mark the uncomputed parity check sub-symbols as erased source sub-symbols with two rows and three columns.
\item Moreover, the parity check sub-symbols of $\cC_2$ in layer (4) spanning the last three rows of parity check sub-symbols in the interval $[i,i+1]$ are again unerased source sub-symbols and thus can be cancelled and the corresponding parity check sub-symbols of $\cC_3$ can be computed in this interval.
\item $\cC_3$ is a concatenation of $\cC_{3,1} = (2,2)$ repetition code producing two parity-check sub-symbols $(p_0^3[.],p_1^3[.])$ and a $\cC_{3,2} = (1,2)$ SCo code producing a single parity-check sub-symbol $p_2^3[.]$. At time $i$ and $i+1$, the parity checks of $\cC_{3,1}$,
\begin{align*}
\left(
\begin{array}{cc}
\tilde{p}_0^3[i]\big{|}_{i} \\
\tilde{p}_1^3[i]\big{|}_{i}
 \end{array}
\right) =
\left(
\begin{array}{cc}
\tilde{p}_1[i+2]\big{|}_i \\
\tilde{p}_2[i+2]\big{|}_i
 \end{array}
\right),
\end{align*}
thus, $\tilde{p}_1[i+2]\big{|}_i$ and $\tilde{p}_2[i+2]\big{|}_i$ can be directly recovered, while their corresponding non-causal parts can be computed before time $i+2$. Similarly, $\tilde{p}_1[i+3]\big{|}_i$ and $\tilde{p}_2[i+3]\big{|}_i$ can be recovered at time $i+1$ and their corresponding non-causal parts can be retrieved before $i+3$. The remaining column, $(\tilde{p}_1[i+4]\big{|}_i,\tilde{p}_2[i+4]\big{|}_i)^T$ can be recovered using the parity checks of $\cC_{3,2} = (1,2)$ SCo code at time $i$ and $i+1$, $p_2^3[i]$ and $p_2^3[i+1]$ in a similar way used in the previous example.
\end{enumerate}

After recovering these parity check sub-symbols of $\cC_1$, the decoder can cancel their effect in the second and third rows of parity checks (layer (3)) at times $i+2$, $i+3$ and $i+4$. Moreover, in the same rows and starting at time $i+5$ all parity checks of code $\cC_1$ combine only unerased source symbols (c.f. \eqref{eq:Code3579}) and thus can be cancelled as well. 
\item  Remove interference in layer (4) starting at time $i+2$.\\
The parity check sub-symbols of $\cC_3$ interfering in the last two rows (layer (4)) starting at time $i+2$ are of indices $i+4$ and later which are either recovered in Step (1) or can be calculated as they combine unerased source sub-symbols (c.f.~\eqref{eq:Code3579}).
\item  Use the parity-checks in layer (3) and (4) to recover $\bs[i-7],\dots,\bs[i-1]$.\\

According to Step (3) and (4), the parity checks of $\cC_2$ in layers (3) and (4) are now free of any interference starting at time $i+2$ and thus, the decoder of user 2 is capable of recovering the erased source sub-symbols.
\end{itemize}
\end{itemize}

\section{Proof of Lemma.~\ref{lem:step1}}
\label{app:step1}

\begin{figure*}
\centering
\resizebox{\textwidth}{!}{\includegraphics[scale=1]{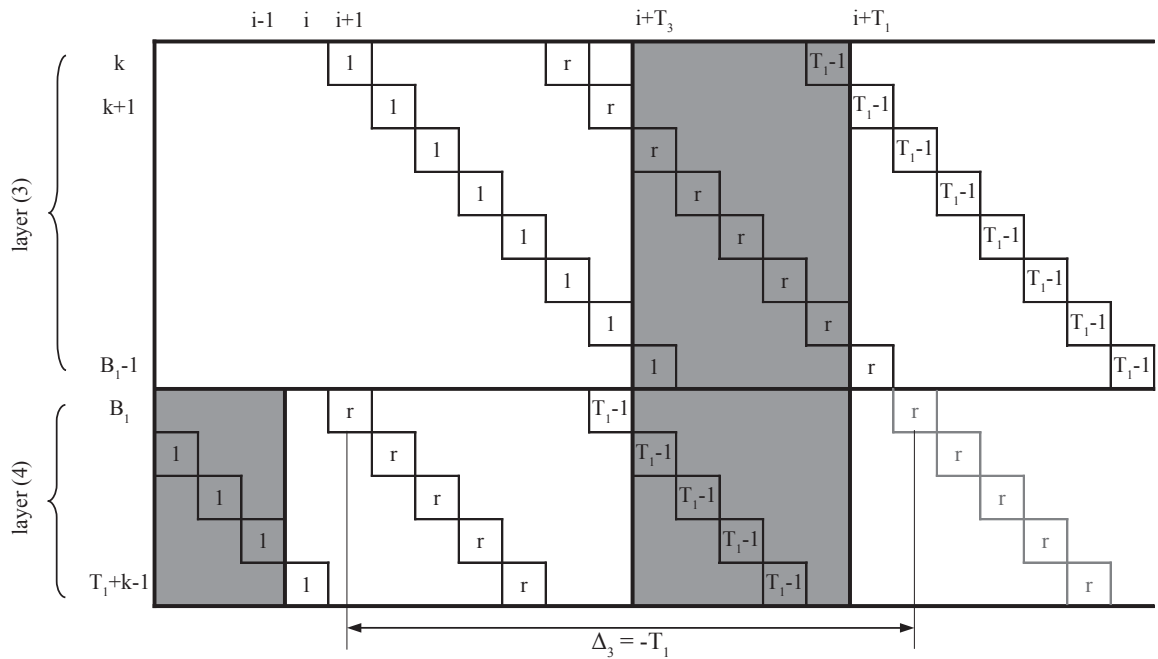}}
\caption{Diagonal Embedding of parity checks for the construction in section~\ref{sec:Region_e_CC}. The parity checks $p^3[\cdot]$ are applied using a $(T_3, B_3)$ SCo code onto the last $B_1-k$ parity checks  of $p^I[\cdot]$ in layer 3.  The parity checks $p^3[\cdot]$ are shifted back by $T_1$ units as discussed before.}
\label{fig:step1}
\end{figure*}


The parity-check sub-symbols of $\cC_2$ in the interval $[t_2,t_3-1] = [i,i-B_2+T_2-1]$ are source sub-symbols in the interval $[t_2-T_2, t_3-T_2-1] = [i-T_2,i-B_2-1]$ which are not erased. Thus, they can be computed and subtracted to recover the combined parity-check sub-symbols of $\cC_1$ and the causal part of that of $\cC_3$ in layers (3) and (4), respectively. More specifically, the parity-check sub-symbols $p_{j_1}^{\rm{I}}[\cdot]$ for $j_1 \in \{ k,\dots,B_1-1 \}$ and $\tilde{p}_{j_2}^{3}[\cdot]$ for $j_2 \in \{0,\dots,B_3-1 \}$ are recovered.

Recall that $\cC_3$ is a $(B_3, T_3)$ is a SCo code applied by taking the last $B_1-k$ parity check sub-symbols of $\cC_1$ as source sub-symbols.

Let us define the parity-check symbols that need to be recovered as
\begin{align}
\bw[t] = (w_0[t],\dots,w_{T_3-1}[t]) = (p_{k}^{\rm{I}}[t],\dots,p_{B_1-1}^{\rm{I}}[t]).
\end{align}

We first consider case (A) i.e., when $T_1 \le 2(B_1-k)$. Since $\cC_3$ is an SCo which involves diagonal interleaving of Low Delay - Burst Erasure Block Codes (LD-BEBC), the diagonals that span the sub-symbols of interest are as follows:

\begin{multline}
\bar{\bd}_r = (w_0[i+r],\dots,w_{T_3-1}[i+r+T_3-1],p_{0}^{3}[i+r+T_3],\dots,p_{B_3-1}^{3}[i+r+T_3+B_3-1]), \\ r \in {1,\dots,T_3+B_3-1} \label{eq:bdr}
\end{multline}

Since the parity check sub-symbols of $\cC_3$ are shifted back by $T_1 = T_3+B_3$ keeping only their causal part, the corresponding diagonals of interest are
\begin{multline}
\label{eq:bdr2}
\bd_r = (w_0[i+r],\dots,w_{T_3-1}[i+r+T_3-1], \tl{p}_{0}^{3}[i+r+ T_3]\big{|}_{i+r-B_3},  \\ \dots, \tl{p}_{B_3-1}^{3}[i+r+T_3+B_3-1]\big{|}_{i+r-1}).
\end{multline}
where recall that $\tilde{p}_j[t_1]\big{|}_{t_2}$ denotes the causal part of the parity check $p_j[t_1]$ w.r.t.\ $t_2$ (c.f.~\eqref{eq:p-causal}).

With every parity check sub-symbol projected to a different time instant, one can clearly see that $\bd_r$ is no more a code-word of an LD-BEBC code.

The following conditions are sufficient to establish Lemma.~\ref{lem:step1},
\begin{enumerate}[c1]
\item The diagonals $d_r$ for $r \in \{1,\dots,T_3+B_3-1\}$ span all the parity-check sub-symbols that need to be recovered, i.e., $p_j^{\rm{I}}[\cdot]$ for $j \in \{k,\dots,B_1-1\}$ in the interval $[t_3,t_4-1] = [i+T_2-B_2,i+T_1-1]$.
\item The decoder can compute the non-causal part of each parity-check $p_j^{3}[\cdot]$ in the interval $[t_2,t_3-1]$ and reduce~\eqref{eq:bdr2} to~\eqref{eq:bdr}. This step should not violate the zero-delay constraint for any erased symbol on the diagonal i.e. the non-causal part of the parity-check sub-symbol $p_{j_1}^3[t_x]$ responsible for the  recovery of a given parity check $w_{j_2}[t_y]$ should combine source sub-symbols $\bs[.]$ which are both, not erased and from time earlier than $t_y$.
\item Each diagonal $d_r$ should have no more than $B_3$ erased sub-symbols.
\end{enumerate}


For c1, we note that the diagonal $\bd_1$ covers $w_{T_3-1}[i+T_3] = p_{B_1-1}^{\rm{I}}[i+T_3]$ which is the lower left most sub-symbol that needs to be recovered.
At $r=T_3+B_3-1$, one can see that $\bd_r$ combines $w_{0}[i+T_3+B_3-1] = p_k^{\rm{I}}[i+T_3+B_3-1]$ which is the upper right most sub-symbol that needs to be recovered. 
Fig.~\ref{fig:step1} easily illustrates that the diagonal $\bd_r$ for $r \in [1, T_3+B_3-1]$ cover all of the erased sub-symbols. 

For c2, we note that all elements of a diagonal $\bd_r$ combine source symbols $\bs[\cdot]$ from time $i+r-1$ and earlier according to the diagonal interleaving property of SCo codes. Thus, one can conclude that the non-causal part of any parity-check sub-symbol $p_{j}^{3}[i+r+ T_3+j]\big{|}_{i+r- B_3+j}$ for $j \in {0,\dots,B_3-1}$ in $\bd_r$ is just a combination of source symbols in the interval $[i+r-B_3+j,i+r-1]$. Thus the entire non-causal part of each parity check is available before time $i+r$ and the reduction to~\eqref{eq:bdr} is possible for each $\bd_r$.

Finally note that the zero delay constraint also requires that the  symbols $w_j[t]$ with $t \ge i+T_1$ in $\bd_r$ be made available before time ${t = i+r}$. Since each $w_j[t]$ for $t \ge i+T_1$ only consists of combinations of source symbols in $[i, i+r-1]$ these symbols can be explicitly computed by the decoder by time $i+r-1$ and c2 follows.

For c3, we divide the values of $r$ into three intervals.
\begin{itemize}
\item $\bd_r$ for $r \in \{1,\dots,T_1-T_3\}$\\
In this range, one can see that the following symbols are available,
\begin{align}
&(w_0[i+r],\dots,w_{T_3-r-1}[i+T_3-1],\tl{p}_{B_3-r}^{3}[i+T_3+B_3],\dots,\tl{p}_{B_3-1}^{3}[i+r+T_3+B_3-1]), \nonumber
\end{align}
which are a total of $T_3$ sub-symbols in the beginning and the end of the diagonals $\bd_r$ which contains $T_3+B_3$ sub-symbols. In other words, each such diagonal has $B_3$ erased sub-symbols happening in a burst.
\item $\bd_r$ for $r \in \{T_1-T_3+1,\dots,T_3\}$\\
In these diagonals, the following symbols are available,
\begin{align}
&(w_0[i+r],\dots,w_{T_3-r-1}[i+T_3-1],w_{T_1-r}[i+T_1], \dots, w_{T_3-1}[i+r+T_3-1],\tl{p}_0^{3}[i+r+T_3] \nonumber \\
&\quad ,\dots,\tl{p}_{B_3-1}^{3}[i+r+T_3+B_3-1]), \nonumber
\end{align}
The first group is a total of $T_3-r$ consecutive sub-symbols, while the other two groups are a total of $r$ consecutive sub-symbols. This implies that each such diagonal $\bd_r$ has $B_3$ erased sub-symbols in a burst.
\item $\bd_r$ for $r \in \{T_3+1,\dots,T_3+B_3-1\}$\\
The available sub-symbols in these diagonals are,
\begin{align}
&(w_{T_1-r}[i+T_1],\dots,w_{T_3-1}[i+r+T_3-1],\tl{p}_0^{3}[i+r+T_3],\dots,\tl{p}_{T_3+B_3-r-1}^{3}[i+2T_3+B_3-1]), \nonumber
\end{align}
which are again a total of $T_3$ consecutive sub-symbols  which implies that the considered diagonals $\bd_r$ has $B_3$ erased sub-symbols in a burst and the c3 follows.
We note that LD-BEBC codes are capable of recovering wrap-around burst which may start at the end of the block and wrap around to the beginning of that block.
\end{itemize}

When $T_1 > 2(B_1-k)$ note that $\cC_3$ is a concatenation of $r+1$ codes, the first $r$ of which are repetition codes with parity check sub-symbols given by~\eqref{eq:C_B}. These parity-check sub-symbols in the interval $[i,i+(B_1-k)-1]$ can be used to recover the causal part of the parity-check sub-symbols $(p^{\rm{I}}_{k}[t_1],\dots,p^{\rm{I}}_{B_1-1}[t_1])$ for $t_1 \in \{ i+(B_1-k),\dots,i+(r+1)(B_1-k)-1 \} = \{ i+T_2-B_2,\dots,i+T_1-q-1 \}$. The non-causal part of these parity-check sub-symbols  combine source sub-symbols in the interval $[i,t_1-1]$ which are not erased and thus can be recovered.

The remaining $q$ columns of parity-check sub-symbols $(p^{\rm{I}}_{k}[t_2],\dots,p^{\rm{I}}_{B_1-1}[t_2])$ for $t_2 \in \{ i+(r+1)(B_1-k), \dots, i+(r+1)(B_1-k)+q-1 \} = \{ i+T_1-q,\dots,i+T_1-1 \}$ can be recovered using the parity-check sub-symbols of $\cC_{3,r+1} = (q,B_1-k)$. This step is similar to that of recovering the $T_1-(B_1-k)$ columns of parity-check sub-symbols of $\cC_1$ using $\cC_3 = (T_1-(B_1-k),B_1-k)$ done above, except that $B_3 = T_1 - (B_1-k)$ is replaced by $B_{3,r+1} = q$.

\section{Proof Of~\eqref{eq:eRegion_a} and~\eqref{eq:eRegion_b}}
\label{app:eRegion}

One can get the result in~\eqref{eq:eRegion_a} through the following steps,
{\allowdisplaybreaks
\begin{align}
	H&(W_0^{k}) \geq H(V_0^{k-1}) + H\Bigl(W_k \Big| V_0^{k-1} \bx\sqstack{0}{kd-1}\Bigr) \nonumber \\
	&= H(V_0^{k-1}) + H\Bigl(\bs\sqstack{kd}{kd+a-1} W_{k} \Big| V_0^{k-1} \bx\sqstack{0}{kd-1}\Bigr) - H\Bigl(\bs\sqstack{kd}{kd+a-1} \Big| V_0^{k-1} \bx\sqstack{0}{kd-1} W_{k}\Bigr) \nonumber \\
	&\overset{(a)}{=} H(V_0^{k-1}) + H\Bigl(\bs\sqstack{kd}{kd+a-1} W_{k} \Big| V_0^{k-1} \bx\sqstack{0}{kd-1}\Bigr) \nonumber \\
	&= H(V_0^{k-1}) + H\Bigl(\bs\sqstack{kd}{kd+a-1} \Big| V_0^{k-1} \bx\sqstack{0}{kd-1}\Bigr) + H\Bigl(W_{k} \Big| V_0^{k-1} \bs\sqstack{kd}{kd+a-1} \bx\sqstack{0}{kd-1}\Bigr) \nonumber \\
	&\overset{(b)}{=} H\Bigl(V_0^{k-1} \bs\sqstack{kd}{kd+a-1}\Bigr) + H\Bigl(W_{k} \Big| V_0^{k-1} \bs\sqstack{kd}{kd+a-1} \bx\sqstack{0}{kd-1}\Bigr) \nonumber \\
	&\geq H\Bigl(V_0^{k-1} \bs\sqstack{kd}{kd+a-1}\Bigr) + H\Bigl(W_{k} \Big| V_0^{k-1} \bs\sqstack{kd}{kd+a-1} \bx\sqstack{0}{kd+b-1}\Bigr) \nonumber \\
	&= H\Bigl(V_0^{k-1} \bs\sqstack{kd}{kd+a-1}\Bigr) + H\Bigl(\bs\sqstack{kd+b}{kd+c-1} W_{k} \Big| V_0^{k-1} \bs\sqstack{kd}{kd+a-1} \bx\sqstack{0}{kd+b-1}\Bigr) \nonumber \\
		&\quad - H\Bigl(\bs\sqstack{kd+b}{kd+c-1} \Big| V_0^{k-1} \bs\sqstack{kd}{kd+a-1} \bx\sqstack{0}{kd+b-1} W_{k}\Bigr) \nonumber \\
	&\overset{(c)}{=} H\Bigl(V_0^{k-1} \bs\sqstack{kd}{kd+a-1}\Bigr) + H\Bigl(\bs\sqstack{kd+b}{kd+c-1} W_{k} \Big| V_0^{k-1} \bs\sqstack{kd}{kd+a-1} \bx\sqstack{0}{kd+b-1}\Bigr) \nonumber \\
	&= H\Bigl(V_0^{k-1} \bs\sqstack{kd}{kd+a-1}\Bigr) + H\Bigl(\bs\sqstack{kd+b}{kd+c-1} \Big| V_0^{k-1} \bs\sqstack{kd}{kd+a-1} \bx\sqstack{0}{kd+b-1}\Bigr) \nonumber \\
		&\quad + H\Bigl(W_{k} \Big| V_0^{k-1} \bs\sqstack{kd}{kd+a-1} \bs\sqstack{kd+b}{kd+c-1} \bx\sqstack{0}{kd+b-1}\Bigr) \nonumber \\
	&\overset{(d)}{=} H\Bigl(V_0^{k-1} \bs\sqstack{kd}{kd+a-1} \bs\sqstack{kd+b}{kd+c-1}\Bigr) + H\Bigl(W_{k} \Big| V_0^{k-1} \bs\sqstack{kd}{kd+a-1} \bs\sqstack{kd+b}{kd+c-1} \bx\sqstack{0}{kd+b-1}\Bigr) \nonumber \\
	&\geq H\Bigl(V_0^{k-1} \bs\sqstack{kd}{kd+a-1} \bs\sqstack{kd+b}{kd+c-1}\Bigr) + H\Bigl(W_{k} \Big| V_0^{k-1} \bs\sqstack{kd}{kd+a-1} \bs\sqstack{kd+b}{kd+c-1} \bx\sqstack{0}{kd+c-1}\Bigr)
\end{align}
}%
We use \eqref{eq:dldr_e_region_rule1} to remove the negative term before step (a). Similarly, we remove the negative term before step (c) using \eqref{eq:dldr_e_region_rule2}. Steps (b) and (d) use the fact that source packets are independent of each other and of previous channel packets.

The following steps help finding the result in~\eqref{eq:eRegion_b},
{\allowdisplaybreaks
\begin{align}
	&H(W_0^{k+1}) \nonumber \\
	&\geq H\Bigl(V_0^{k-1} \bs\sqstack{kd}{kd+a-1} \bs\sqstack{kd+b}{kd+c-1}\Bigr) + H\Bigl(W_{k} \Big| V_0^{k-1} \bs\sqstack{kd}{kd+a-1} \bs\sqstack{kd+b}{kd+c-1} \bx\sqstack{0}{kd+c-1}\Bigr) + H(W_{k+1} | W_0^k) \nonumber \\
	&\geq H\Bigl(V_0^{k-1} \bs\sqstack{kd}{kd+a-1} \bs\sqstack{kd+b}{kd+c-1}\Bigr) + H\Bigl(W_k^{k+1} \Big| V_0^{k-1} \bs\sqstack{kd}{kd+a-1} \bs\sqstack{kd+b}{kd+c-1} \bx\sqstack{0}{kd+c-1}\Bigr) \nonumber \\
	&= H\Bigl(V_0^{k-1} \bs\sqstack{kd}{kd+a-1} \bs\sqstack{kd+b}{kd+c-1}\Bigr) + H\Bigl(\bs\sqstack{kd+c}{(k+1)d-1} W_{k}^{k+1} \Big| V_0^{k-1} \bs\sqstack{kd}{kd+a-1} \bs\sqstack{kd+b}{kd+c-1} \bx\sqstack{0}{kd+c-1}\Bigr) \nonumber \\
		&\quad - H\Bigl(\bs\sqstack{kd+c}{(k+1)d-1} \Big| V_0^{k-1} \bs\sqstack{kd}{kd+a-1} \bs\sqstack{kd+b}{kd+c-1} \bx\sqstack{0}{kd+c-1} W_{k}^{k+1}\Bigr) \nonumber \\
	&\overset{(e)}{=} H\Bigl(V_0^{k-1} \bs\sqstack{kd}{kd+a-1} \bs\sqstack{kd+b}{kd+c-1}\Bigr) + H\Bigl(\bs\sqstack{kd+c}{(k+1)d-1} W_{k}^{k+1} \Big| V_0^{k-1} \bs\sqstack{kd}{kd+a-1} \bs\sqstack{kd+b}{kd+c-1} \bx\sqstack{0}{kd+c-1}\Bigr) \nonumber \\
	&= H\Bigl(V_0^{k-1} \bs\sqstack{kd}{kd+a-1} \bs\sqstack{kd+b}{kd+c-1}\Bigr) + H\Bigl(\bs\sqstack{kd+c}{(k+1)d-1} \Big| V_0^{k-1} \bs\sqstack{kd}{kd+a-1} \bs\sqstack{kd+b}{kd+c-1} \bx\sqstack{0}{kd+c-1}\Bigr) \nonumber \\
		&\quad + H\Bigl(W_{k}^{k+1} \Big| V_0^{k-1} \bs\sqstack{kd}{kd+a-1} \bs\sqstack{kd+b}{(k+1)d-1} \bx\sqstack{0}{kd+c-1}\Bigr) \nonumber \\
	&\overset{(f)}{=} H\Bigl(V_0^{k-1} \bs\sqstack{kd}{kd+a-1} \bs\sqstack{kd+b}{(k+1)d-1}\Bigr) + H\Bigl(W_{k}^{k+1} \Big| V_0^{k-1} \bs\sqstack{kd}{kd+a-1} \bs\sqstack{kd+b}{(k+1)d-1} \bx\sqstack{0}{kd+c-1}\Bigr) \nonumber \\
	&= H(V_0^{k}) + H\Bigl(W_{k}^{k+1} \Big| V_0^{k} \bx\sqstack{0}{kd+c-1}\Bigr) \nonumber \\
	&= H(V_0^{k}) + H\Bigl(W_{k} \Big| V_0^{k} \bx\sqstack{0}{kd+c-1}\Bigr) + H\Bigl(W_{k+1} \Big| V_0^{k} \bx\sqstack{0}{kd+c-1} W_{k}\Bigr) \nonumber \\
	&\geq H(V_0^{k}) + H\Bigl(W_{k+1} \Big| V_0^{k} \bx\sqstack{0}{(k+1)d-1}\Bigr) \label{eq:dldr_e_region_ind_step}
\end{align}
}%
Once again, we remove the negative term before step (e) using and \eqref{eq:dldr_e_region_rule3}. Steps (f) uses the fact that each source packet is independent of each other.

\section{Proof Of~\eqref{eq:f1Region_a} and~\eqref{eq:f1Region_b}}
\label{app:f1Region}

For the result in~\eqref{eq:f1Region_a}, we walk through the following steps,
{\allowdisplaybreaks
\begin{align}
	&H(W_0^{k}) \geq H(V_0^{k-1}) + H\Bigl(W_{k}  \Big|  V_0^{k-1} \bx\sqstack{0}{kd-1}\Bigr) \nonumber \\
	&\overset{(a)}{=} H(V_0^{k-1}) + H\Bigl(\bs\sqstack{kd}{kd+a-1} W_{k}  \Big|  V_0^{k-1} \bx\sqstack{0}{kd-1}\Bigr) - H\Bigl(\bs\sqstack{kd}{kd+a-1}  \Big|  V_0^{k-1} \bx\sqstack{0}{kd-1} W_{k} \Bigr) \nonumber \\
	&= H(V_0^{k-1}) + H\Bigl(\bs\sqstack{kd}{kd+a-1} W_{k}  \Big|  V_0^{k-1} \bx\sqstack{0}{kd-1}\Bigr) \nonumber \\
	&= H(V_0^{k-1}) + H\Bigl(\bs\sqstack{kd}{kd+a-1}  \Big|  V_0^{k-1} \bx\sqstack{0}{kd-1}\Bigr) + H\Bigl(W_{k}  \Big|  V_0^{k-1} \bs\sqstack{kd}{kd+a-1} \bx\sqstack{0}{kd-1}\Bigr) \nonumber \\
	&= H\Bigl(V_0^{k-1} \bs\sqstack{kd}{kd+a-1}\Bigr) + H\Bigl(W_{k}  \Big|  V_0^{k-1} \bs\sqstack{kd}{kd+a-1} \bx\sqstack{0}{kd-1}\Bigr) \nonumber \\
	&\geq H\Bigl(V_0^{k-1} \bs\sqstack{kd}{kd+a-1}\Bigr) + H\Bigl(W_{k}  \Big|  V_0^{k-1} \bs\sqstack{kd}{kd+a-1} \bx\sqstack{0}{kd+a-1}\Bigr) \nonumber \\
	&= H\Bigl(V_0^{k-1} \bs\sqstack{kd}{kd+a-1}\Bigr) + H\Bigl(\bs\sqstack{kd+a}{kd+b-1} W_{k}  \Big|  V_0^{k-1} \bs\sqstack{kd}{kd+a-1} \bx\sqstack{0}{kd+a-1}\Bigr) \nonumber \\
	    &\quad - H\Bigl(\bs\sqstack{kd+a}{kd+b-1}  \Big|  V_0^{k-1} \bs\sqstack{kd}{kd+a-1} \bx\sqstack{0}{kd+a-1} W_{k} \Bigr) \nonumber \\
	&\overset{(b)}{=} H\Bigl(V_0^{k-1} \bs\sqstack{kd}{kd+a-1}\Bigr) + H\Bigl(\bs\sqstack{kd+a}{kd+b-1}  \Big|  V_0^{k-1} \bs\sqstack{kd}{kd+a-1} \bx\sqstack{0}{kd+a-1}\Bigr) \nonumber \\
	    &\quad + H\Bigl(W_{k}  \Big|  V_0^{k-1} \bs\sqstack{kd}{kd+b-1} \bx\sqstack{0}{kd+a-1}\Bigr) \nonumber \\
	&\geq H\Bigl(V_0^{k-1} \bs\sqstack{kd}{kd+b-1}\Bigr) + H\Bigl(W_{k}  \Big|  V_0^{k-1} \bs\sqstack{kd}{kd+b-1} \bx\sqstack{0}{kd+c-1}\Bigr)
	\end{align}
}
The negative terms in (a) and (b) are removed using \eqref{eq:e_rule1} and \eqref{eq:e_rule2} respectively.

Next, we start by adding $H(W_{k+1}|W_0^k)$ to both sides of~\eqref{eq:f1Region_a} to find~\eqref{eq:f1Region_b} as follows:
{\allowdisplaybreaks
\begin{align}
&H(W_0^{k+1}) \nonumber \\
&\geq H\Bigl(V_0^{k-1} \bs\sqstack{kd}{kd+b-1}\Bigr) + H\Bigl(W_{k}  \Big|  V_0^{k-1} \bs\sqstack{kd}{kd+b-1} \bx\sqstack{0}{kd+c-1}\Bigr) + H(W_{k+1}|W_0^k) \nonumber \\
&\geq H\Bigl(V_0^{k-1} \bs\sqstack{kd}{kd+b-1}\Bigr) + H\Bigl(W_{k}  \Big|  V_0^{k-1} \bs\sqstack{kd}{kd+b-1} \bx\sqstack{0}{kd+c-1}\Bigr) \nonumber \\
		&\quad + H(W_{k+1}|V_0^{k-1} \bs\sqstack{kd}{kd+b-1} \bx\sqstack{0}{kd+c-1} W_0^k) \nonumber \\
&\geq H\Bigl(V_0^{k-1} \bs\sqstack{kd}{kd+b-1}\Bigr) + H\Bigl(W_{k}^{k+1}  \Big|  V_0^{k-1} \bs\sqstack{kd}{kd+b-1} \bx\sqstack{0}{kd+c-1}\Bigr) \nonumber \\
&= H\Bigl(V_0^{k-1} \bs\sqstack{kd}{kd+b-1}\Bigr) + H\Bigl(\bs\sqstack{kd+c}{(k+1)d-1} W_{k}^{k+1}  \Big|  V_0^{k-1} \bs\sqstack{kd}{kd+b-1} \bx\sqstack{0}{kd+c-1}\Bigr) \nonumber \\
	    &\quad - H\Bigl(\bs\sqstack{kd+c}{(k+1)d-1}  \Big|  V_0^{k-1} \bs\sqstack{kd}{kd+b-1} \bx\sqstack{0}{kd+c-1} W_{k}^{k+1}\Bigr) \nonumber \\
	&\overset{(c)}{=} H\Bigl(V_0^{k-1} \bs\sqstack{kd}{kd+b-1}\Bigr) + H\Bigl(\bs\sqstack{kd+c}{(k+1)d-1}  \Big|  V_0^{k-1} \bs\sqstack{kd}{kd+b-1} \bx\sqstack{0}{kd+c-1}\Bigr) \nonumber \\
	    &\quad + H\Bigl(W_{k}^{k+1}  \Big|  V_0^{k-1} \bs\sqstack{kd}{kd+b-1} \bs\sqstack{kd+c}{(k+1)d-1} \bx\sqstack{0}{kd+c-1}\Bigr) \nonumber \\
	&\overset{(d)}{=} H(V_0^{k}) + H\Bigl(W_{k}^{k+1}  \Big|  V_0^{k} \bx\sqstack{0}{kd+c-1}\Bigr) \nonumber \\
	&= H(V_0^{k}) + H\Bigl(W_{k}  \Big|  V_0^{k} \bx\sqstack{0}{kd+c-1}\Bigl) + H\Bigl(W_{k+1}  \Big|  V_0^{k} \bx\sqstack{0}{kd+c-1} W_{k}\Bigr) \nonumber \\
	&\geq H(V_0^{k}) + H\Bigl(W_{k+1}  \Big|  V_0^{k} \bx\sqstack{0}{(k+1)d-1}\Bigr)
\end{align}
}
The negative in (c) is removed using~\eqref{eq:e_rule3}. Step (d) follows from the fact that source symbols are independent and~\eqref{eq:f1Region_b} follows.

\section{Proof Of~\eqref{eq:fRegion_a} and~\eqref{eq:fRegion_b}}
\label{app:fRegion}

The steps to get~\eqref{eq:fRegion_a} are,
{\allowdisplaybreaks
\begin{align}
	&\sum_{i=B_2}^{j+1} H(\bx[i]) \nonumber \\
	&\overset{(a)}{\geq} H\Bigl(\bs\sqstack{0}{j-B_2}\Bigr) + H\Bigl(\bs\sqstack{B_2-B_1}{j-T_1}\Bigr) + H\Bigl(\bx\sqstack{B_2}{j} \Big| \bs\sqstack{0}{j-B_2} \bs\sqstack{B_2-B_1}{j-T_1} \bx\sqstack{0}{j-B_2}\Bigr) \nonumber \\
	&\quad + H(\bs[j+1-B_2]) + H\Bigl(\bx[j+1]\Big| \bs[j+1-B_2] \bx\sqstack{0}{j+1-B_2}\Bigr) \nonumber \\
	&\geq H\Bigl(\bs\sqstack{0}{j+1-B_2}\Bigr) + H\Bigl(\bs\sqstack{B_2-B_1}{j-T_1}\Bigr) + H\Bigl(\bx\sqstack{B_2}{j+1} \Big| \bs\sqstack{0}{j+1-B_2} \bs\sqstack{B_2-B_1}{j-T_1} \bx\sqstack{0}{j+1-B_2}\Bigr) \nonumber \\
	&= H\Bigl(\bs\sqstack{0}{j+1-B_2}\Bigr) + H\Bigl(\bs\sqstack{B_2-B_1}{j-T_1}\Bigr) + H\Bigl(\bs[j+1-T_1] \bx\sqstack{B_2}{j+1} \Big| \bs\sqstack{0}{j+1-B_2} \bs\sqstack{B_2-B_1}{j-T_1} \bx\sqstack{0}{j+1-B_2}\Bigr) \nonumber \\
		&\quad - H\Bigl(\bs[j+1-T_1] \Big| \bs\sqstack{0}{j+1-B_2} \bs\sqstack{B_2-B_1}{j-T_1} \bx\sqstack{B_2}{j+1} \bx\sqstack{0}{j+1-B_2} \Bigr) \nonumber \\
	&\overset{(b)}{=} H\Bigl(\bs\sqstack{0}{j+1-B_2}\Bigr) + H\Bigl(\bs\sqstack{B_2-B_1}{j-T_1}\Bigr) + H\Bigl(\bs[j+1-T_1] \bx\sqstack{B_2}{j+1} \Big| \bs\sqstack{0}{j+1-B_2} \bs\sqstack{B_2-B_1}{j-T_1} \bx\sqstack{0}{j+1-B_2}\Bigr) \nonumber \\
	&= H\Bigl(\bs\sqstack{0}{j+1-B_2}\Bigr) + H\Bigl(\bs\sqstack{B_2-B_1}{j-T_1}\Bigr) + H\Bigl(\bs[j+1-T_1] \Big| \bs\sqstack{0}{j+1-B_2} \bs\sqstack{B_2-B_1}{j-T_1} \bx\sqstack{0}{j+1-B_2}\Bigr) \nonumber \\
		&\quad + H\Bigl(\bx\sqstack{B_2}{j+1} \Big| \bs\sqstack{0}{j+1-B_2} \bs\sqstack{B_2-B_1}{j+1-T_1} \bx\sqstack{0}{j+1-B_2}\Bigr) \nonumber \\
	&\overset{(c)}{\geq} H\Bigl(\bs\sqstack{0}{j+1-B_2}\Bigr) + H\Bigl(\bs\sqstack{B_2-B_1}{j+1-T_1}\Bigr) + H\Bigl(\bx\sqstack{B_2}{j+1} \Big| \bs\sqstack{0}{j+1-B_2} \bs\sqstack{B_2-B_1}{j+1-T_1} \bx\sqstack{0}{j+1-B_2}\Bigr)
\end{align}
}%
Step (a) is the addition of \eqref{eq:dldr_f_region_relation2b} and \eqref{eq:dldr_f_region_target_ineq}, step (b) uses \eqref{eq:dldr_f_region_relation1} to remove the negative term in the previous step, and step (c) uses the fact that the source packets are independent of each other. The result is the form \eqref{eq:dldr_f_region_target_ineq} for $m=l+1$.

While the working out to get~\eqref{eq:fRegion_b} is as follows,
{\allowdisplaybreaks
\begin{align}
	&\sum_{i=B_2}^{B_2+T_1-1} H(\bx[i]) \nonumber \\
	&\geq H\Bigl(\bs\sqstack{0}{T_1-1}\Bigr) + H\Bigl(\bs\sqstack{B_2-B_1}{B_2-1}\Bigr) + H\Bigl(\bx\sqstack{B_2}{B_2+T_1-1} \Big| \bs\sqstack{0}{T_1-1} \bs\sqstack{B_2-B_1}{B_2-1} \bx\sqstack{0}{T_1-1}\Bigr) \nonumber \\
 	&\geq H\Bigl(\bs\sqstack{0}{T_1-1}\Bigr) + H\Bigl(\bs\sqstack{B_2-B_1}{B_2-1}\Bigr) + H\Bigl(\bx\sqstack{B_2}{B_2+T_1-1} \Big| \bs\sqstack{0}{T_1-1} \bs\sqstack{B_2-B_1}{B_2-1} \bx\sqstack{0}{B_2-1}\Bigr) \nonumber \\
	&= H\Bigl(\bs\sqstack{0}{T_1-1}\Bigr) + H\Bigl(\bs\sqstack{B_2-B_1}{B_2-1}\Bigr) + H\Bigl(\bs\sqstack{B_2}{B_2+T_1-1} \bx\sqstack{B_2}{B_2+T_1-1} \Big| \bs\sqstack{0}{T_1-1} \bs\sqstack{B_2-B_1}{B_2-1} \bx\sqstack{0}{B_2-1}\Bigr) \nonumber \\
		&\quad - H\Bigl(\bs\sqstack{B_2}{B_2+T_1-1} \Big| \bs\sqstack{0}{T_1-1} \bs\sqstack{B_2-B_1}{B_2-1} \bx\sqstack{0}{B_2+T_1-1}\Bigr) \nonumber \\
	&\overset{(d)}{=} H\Bigl(\bs\sqstack{0}{T_1-1}\Bigr) + H\Bigl(\bs\sqstack{B_2-B_1}{B_2-1}\Bigr) + H\Bigl(\bs\sqstack{B_2}{B_2+T_1-1} \bx\sqstack{B_2}{B_2+T_1-1} \Big| \bs\sqstack{0}{T_1-1} \bs\sqstack{B_2-B_1}{B_2-1} \bx\sqstack{0}{B_2-1}\Bigr) \nonumber \\
	&= H\Bigl(\bs\sqstack{0}{T_1-1}\Bigr) + H\Bigl(\bs\sqstack{B_2-B_1}{B_2-1}\Bigr) + H\Bigl(\bs\sqstack{B_2}{B_2+T_1-1} \Big| \bs\sqstack{0}{T_1-1} \bs\sqstack{B_2-B_1}{B_2-1} \bx\sqstack{0}{B_2-1}\Bigr) \nonumber \\
		&\quad + H\Bigl(\bx\sqstack{B_2}{B_2+T_1-1} \Big| \bs\sqstack{0}{T_1-1} \bs\sqstack{B_2-B_1}{B_2+T_1-1} \bx\sqstack{0}{B_2-1}\Bigr) \nonumber \\
	&= H\Bigl(\bs\sqstack{0}{T_1-1}\Bigr) + H\Bigl(\bs\sqstack{B_2-B_1}{B_2+T_1-1}\Bigr) + H\Bigl(\bx\sqstack{B_2}{B_2+T_1-1} \Big| \bs\sqstack{0}{T_1-1} \bs\sqstack{B_2-B_1}{B_2+T_1-1} \bx\sqstack{0}{B_2-1}\Bigr) \nonumber \\
	&\geq H\Bigl(\bs\sqstack{0}{T_1-1}\Bigr) + H\Bigl(\bs\sqstack{B_2-B_1}{B_2+T_1-1}\Bigr)
\end{align}
}%
where step (d) makes use of \eqref{eq:dldr_f_region_step3b}.



\end{document}